\documentclass[journal=jacsat,manuscript=article]{achemso}
\SectionNumbersOn

\usepackage[version=3]{mhchem} 
\usepackage{adjustbox}
\usepackage{threeparttable}
\usepackage{physics}
\usepackage{amssymb}

\usepackage{xcolor}
\usepackage{siunitx}
\usepackage[normalem]{ulem}

\usepackage{lineno} 

\author{Jan Elsner}
\affiliation{Department of Physics and Astronomy and Thomas Young Centre, University College London, London WC1E 6BT, UK.}
\author{Yucheng Xu}
\affiliation{University of Cambridge, Cavendish Laboratory, Cambridge, UK CB3 0HE, UK.}
\author{Elliot D. Goldberg}
\affiliation{University of Cambridge, Cavendish Laboratory, Cambridge, UK CB3 0HE, UK.}
\author{Filip Ivanovic}
\affiliation{Department of Physics and Astronomy and Thomas Young Centre, University College London, London WC1E 6BT, UK.}
\author{Aaron Dines}
\affiliation{Department of Physics and Astronomy and Thomas Young Centre, University College London, London WC1E 6BT, UK.}
\author{Samuele Giannini}
\affiliation{Department of Physics and Astronomy and Thomas Young Centre, University College London, London WC1E 6BT, UK.}
\alsoaffiliation{Institute for the Chemistry of OrganoMetallic Compounds, National Research Council (ICCOM-CNR), I-56124 Pisa, Italy.}
\author{Henning Sirringhaus}
\affiliation{University of Cambridge, Cavendish Laboratory, Cambridge, UK CB3 0HE, UK.}
\author{Jochen Blumberger}
\affiliation{Department of Physics and Astronomy and Thomas Young Centre, University College London, London WC1E 6BT, UK.}
\email{j.blumberger@ucl.ac.uk}

\title
{Thermoelectric transport in molecular crystals driven by gradients of thermal electronic disorder}

\begin{document}

\newpage

\begin{abstract}
Thermoelectric materials convert a temperature gradient into a voltage. This phenomenon is relatively well understood for 
inorganic materials, but much less so for organic semiconductors (OSs).  
These materials present a challenge because the strong thermal fluctuations of electronic coupling between the molecules 
result in partially delocalized charge carriers that cannot be treated with traditional theories for thermoelectricity.    
Here we develop a novel quantum dynamical simulation approach revealing in atomistic detail how the charge carrier wavefunction 
moves along a temperature gradient in an organic molecular crystal. We find that the 
wavefunction propagates from hot to cold in agreement with experiment and we obtain a Seebeck coefficient 
in good agreement with values obtained from experimental measurements that are also reported in this work. 
Detailed analysis of the dynamics reveals that the directional charge carrier motion is due to the gradient in 
thermal electronic disorder, more specifically in the spatial gradient of thermal fluctuations of electronic couplings. 
It causes an increase in the density of thermally accessible electronic states, the delocalization of states and 
the non-adiabatic coupling between states with decreasing temperature.  As a result, the carrier wavefunction transitions with 
higher probability to a neighbouring electronic state towards the cold side compared to the hot side generating a thermoelectric current. 
Our dynamical perspective of thermoelectricity suggests that the temperature dependence of electronic disorder plays an important role in 
determining the magnitude of the Seebeck coefficient in this class of materials, 
opening new avenues for design of OSs with improved Seebeck coefficients. 
\end{abstract}

\newpage 

\section{Introduction} \label{section:Introduction}

Organic semiconductors (OSs) have emerged as promising materials for thermoelectric applications\cite{lu2016review, zhang2020exploring, venkateshvaran2014approaching}. 
Recent studies have shown that relatively high ZT figure of merit values are achievable ($\text{ZT}=T \alpha^2\sigma/\kappa$, $T$ is temperature, $\alpha$ is the Seebeck coefficient, $\sigma$ is the electronic conductivity and 
$\kappa$ is the thermal conductivity), for example, ZT $= 0.42$ has been measured in the doped conducting polymer poly(3,4-ethylenedioxythiophene) (PEDOT:PSS) at room temperature\cite{kim2013engineered}.  The combination of good thermoelectric properties with intrinsic mechanical flexibility opens up a range of new possibilities, for example in wearable devices, 
a rapidly growing industry where the need for batteries or external charging could be eliminated \cite{masoumi2022organic}. As such, there is a need for a detailed fundamental understanding of thermoelectric transport in OSs that can aid the interpretation of experiments and inform the design of improved organic thermoelectric materials. 

Here we aim to establish a molecular-scale understanding of thermoelectricity and the Seebeck coefficient, $\alpha$, in high-mobility OSs. 
The Seebeck coefficient quantifies the open-circuit voltage, $\Delta V_{oc}$, developed in response to an applied temperature difference, $\Delta T$, 
$\alpha = - \Delta V_{oc}/\Delta T$. For wide-band inorganic semiconductors, $\alpha$ is usually computed in the framework of coherent transport theories, e.g., the Boltzmann transport equation\cite{Wang12pccp,wang2018first,wu2020contributions} or Landauer theory\cite{zevalkink2018practical, lundstrom2012near}. For narrow-band semiconductors, where charge transport occurs via incoherent hopping of localized charge carriers, Heikes formula\cite{austin1969polarons,chaikin1976thermopower} or Emin's theory\cite{emin1975thermoelectric,emin1999enhanced,emin2014seebeck} may be used instead. It is now well established, however, that the transport scenario for ordered OSs is in between these two limiting extremes\cite{Troisi06,fratini2016transient,fratini2017map,Fratini20,Few15,Wang13jpcl,Wang15pccp,Giannini23,Heck15,Xie20,Roosta22,Taylor18,Balzer21,Willson23,Sneyd21,Sneyd22,giannini2018crossover,
giannini2019quantum,giannini2020flickering,giannini2022charge}. Charge carriers partially delocalize over the molecular units of these materials due to sizable electronic couplings whilst their delocalization is limited by intermolecular electron-phonon couplings. Transient localisation theory has been derived specifically for this intermediate regime and has been successful in predicting charge carrier mobilities, as well as in providing new design rules for this class of materials \cite{fratini2017map, fratini2016transient}. Yet, a theoretical description of thermoelectricity in this regime is still relatively unexplored. 

Computer simulations, specifically atomistic mixed quantum-classical molecular dynamics such as fragment orbital-based surface hopping (FOB-SH)
\cite{giannini2018crossover,giannini2019quantum,giannini2020flickering,giannini2022charge,Giannini23} and similar implementations\cite{Heck15,Xie20,Roosta22}, 
have given important complementary molecular-level insight 
into the charge transport process in ordered OSs, largely supporting the assertions of transient localization theory. In FOB-SH, the quantum dynamics of an excess charge 
carrier in the valence or conduction band of the semiconductor is propagated under the influence of time-dependent classical nuclei according to Tully's fewest switches 
surface hopping\cite{tully1990molecular}. This method is particularly well suited for the simulation of charge transport in the difficult intermediate 
transport regime where relevant transport parameters, notably electronic coupling and reorganization energy, are on the same order of magnitude, as is the case for 
high mobility OS. Applications of FOB-SH to ordered OSs have shown that thermal electronic excitations give rise to expansion and contraction events of the charge 
carrier wavefunction, also denoted ``transient delocalizations", which result in charge displacement, diffusion and mobility. However, it remains unknown how the important 
effect of transient delocalizations play out in a system subject to a temperature gradient. What is the microscopic origin that drives charge carriers across a temperature 
gradient in an OS? How can the directional charge flow and the Seebeck coefficient be enhanced?   

To investigate these questions we abandon the open-circuit condition, i.e. zero charge flow, for which Seebeck coefficients are usually measured or computed, and simulate 
the real-time propagation of the charge carrier wavefunction using FOB-SH in an OS subjected to a temperature gradient, representing short-circuit conditions in experiment. 
We choose single crystalline rubrene for this purpose, an experimentally well characterised high-mobility OS where hole transport is thought to occur in the transient 
delocalization regime\cite{Troisi06,giannini2019quantum,giannini2020flickering,giannini2022charge}. We observe a net migration of the hole wavefunction from hot to cold indicative of the Seebeck effect and in agreement with experiment. We show that the directional motion of the hole wavefunction is due to the higher density and non-adiabatic coupling of neighbouring low-energy 
electronic states towards the cold side compared to the hot side. This causes the carrier wavefunction to transition with higher probability to a neighbouring electronic state 
on the cold side compared to the hot side resulting in the movement from hot to cold. Hence, our simulations show that gradients in thermal electronic disorder, specifically the decreasing off-diagonal electronic disorder with decreasing temperature, are an important consideration in understanding thermoelectricity in this class of materials.

Our results are analyzed in terms of the general expression for current density in the presence of gradients in temperature, chemical potential and electrical potential 
(see equation~\ref{eq:current_density} below). In this framework, the Seebeck coefficient, $\alpha$, can be written as the sum of three contributions, a kinetic contribution due to the 
temperature gradient-induced current or drift velocity, $\alpha_v$, a thermodynamic contribution due to the temperature gradient-induced change in chemical potential, $\alpha_c$, 
and an electric field contribution, $\alpha_e$, $\alpha\!=\!\alpha_v+\alpha_c+\alpha_e$. Under short-circuit conditions, all terms contribute in general, whereas under open-circuit conditions only the thermodynamic and electric field terms contribute and the kinetic term is zero. 
$\alpha_c$ and $\alpha_e$ are independent of transport mechanism and have been the sole focus of most computational studies\cite{emin1999enhanced, wang2018first, apertet2016note}, whilst the kinetic term depends on the charge transport mechanism and, to our best knowledge, has eluded a rigorous calculation so far owing to the complex nature of the transient delocalization mechanism described above. 

Simulations without an external electric field ($\alpha_e\!=\!0$), representing short-circuit conditions in experiment, show that the kinetic contribution to the Seebeck coefficient is significant ($\alpha_v > \frac{k_B}{e}$), on the same order of magnitude as, albeit somewhat smaller than, the thermodynamic contribution $\alpha_c$. These results are consolidated by simulations with an electric field that is chosen to cancel the kinetic contribution, representing the open-circuit condition in experiment, reproducing the Seebeck coefficient obtained without an external electric field. Based on these results, we discuss viable strategies for increasing the Seebeck coefficient of thermoelectric materials in the regime of transient delocalization.

\section{Results} \label{section:Results}

\begin{figure}[H]
\centering
\includegraphics[width=0.95\textwidth]{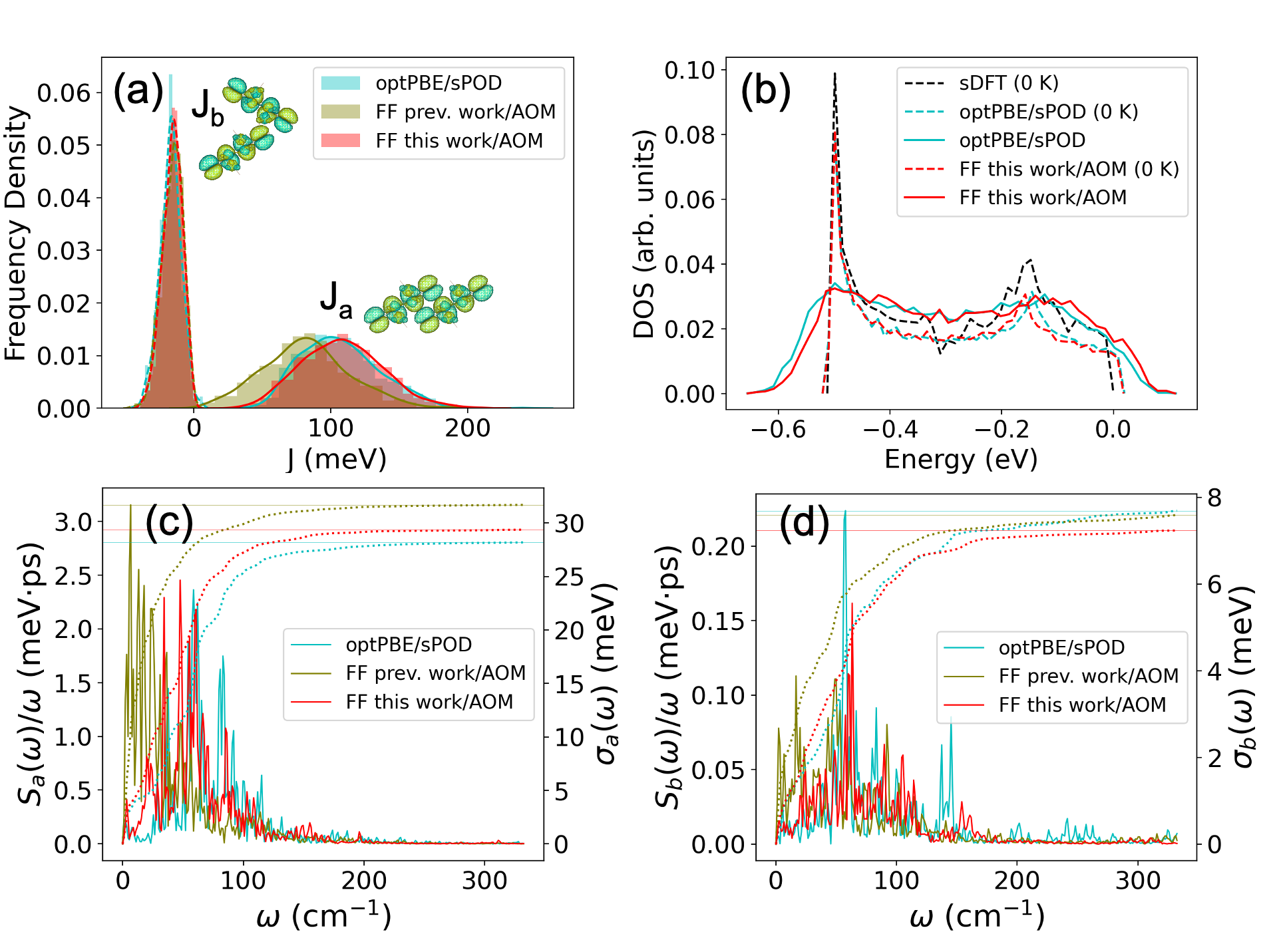}
  \caption{Validation of optimized force field (FF this work) and fast electronic coupling calculation (AOM) employed in FOB-SH simulations of rubrene against ab-initio calculations. 
  (a) Thermal distributions of electronic couplings along crystallographic directions $a$ and $b$ of rubrene, $J_a$ and $J_b$, at 300 K.  optPBE/sPOD denotes ab-initio molecular dynamics simulation of rubrene crystal using the optPBE functional\cite{klimevs2009chemical} with electronic couplings calculated along the trajectory according to the scaled projector operator-based diabatization (sPOD)\cite{futera2017electronic} method using PBE. FF prev.\ work/AOM and FF this work/AOM denote classical molecular dynamics simulation of the rubrene crystal with dihedral parameters used in Ref. \cite{giannini2019quantum,giannini2020flickering} 
   and re-optimized for this work (see Supplementary Note 1), respectively, with electronic couplings calculated along the trajectory using the AOM\cite{gajdos2014ultrafast, ziogos2021ultrafast}. 
  (b) Normalised density of states (DOS) from the Kohn-Sham Hamiltonian at PBE level (sDFT, black) and from the valence band Hamiltonian, equation~\ref{eq:Hamiltonian}, with electronic couplings obtained from optPBE/sPOD (blue) or FF this work/AOM (red). Site energies were sampled from a Gaussian distribution centred on zero with variance $\sigma_E^2 = k_B T \lambda$, $\lambda = 0.152$ eV is the DFT reorganisation energy for hole transfer\cite{giannini2019quantum}. Dashed lines correspond to optimized structures (0 K) and solid lines correspond to sampled configurations at 300 K. In each case, the peak of the DOS was aligned with the peak of the sDFT DOS at -0.499 eV relative to the top of the valence band.
  (c) and (d) Spectral density functions from the cosine transform of the autocorrelation function of the $J_a$ and $J_b$ time series (solid), and accumulated frequency resolved root-mean-square-fluctuations of $J_a$ and $J_b$, $\sigma_a$ and $\sigma_b$, respectively\cite{elsner2021mechanoelectric}.}
  \label{fig:FF_validation}
\end{figure}

{\bf Molecular model} \\
In the FOB-SH method\cite{spencer2016fob, carof2019calculate, giannini2021atomic} that will be used for simulation of thermoelectric transport, the electronic Hamiltonian for hole transport is constructed in a site basis of highest occupied molecular orbitals with site energies obtained from force fields and electronic couplings from an ultrafast coupling estimator denoted analytic overlap method (AOM)\cite{gajdos2014ultrafast, ziogos2021ultrafast}, see {\it Methods} for details. Using such an approximate but computationally efficient scheme, it is vital to demonstrate that properties governing charge transport are well captured when compared to higher-level electronic structure methods. We have used a force field with optimized dihedral parameters for rubrene, alongside optimized AOM electronic coupling parameters (denoted FF this work/AOM). We validate their performance by comparison with results from ab-initio molecular dynamics simulation (using the optPBE van-der-Waals density functional\cite{klimevs2009chemical}) and scaled projector operator-based diabatization method (sPOD)\cite{futera2017electronic} for electronic couplings (denoted optPBE/sPOD in Figure~\ref{fig:FF_validation}). We find that the electronic coupling distributions (Fig.~\ref{fig:FF_validation}a) and the density of valence band states at $T\!=\!0$ and 300 K (Fig.~\ref{fig:FF_validation}b) are very well reproduced. Moreover, the spectral density functions of the electronic couplings related to off-diagonal electron-phonon coupling compare well with the ab-initio MD results both in terms of frequency (0-200 cm$^{-1}$) and intensity (panels c-d). The force field we have previously used for rubrene\cite{giannini2019quantum,giannini2020flickering} (denoted FF prev.\ work/AOM) underestimates couplings along the high mobility direction ($a$) and gives red-shifted spectral density functions. Both deficiencies are cured with the improved dihedral parameters. We report numerical values for couplings and coupling fluctuations obtained using the different approaches in Supplementary Table 2. A description of the new force field parameters and a more detailed discussion of results shown in Figure~\ref{fig:FF_validation} are presented in Supplementary Notes~1 and~2, respectively. 
%
%

\begin{figure}[!htb]
\centering
\includegraphics[width=1.0\textwidth]{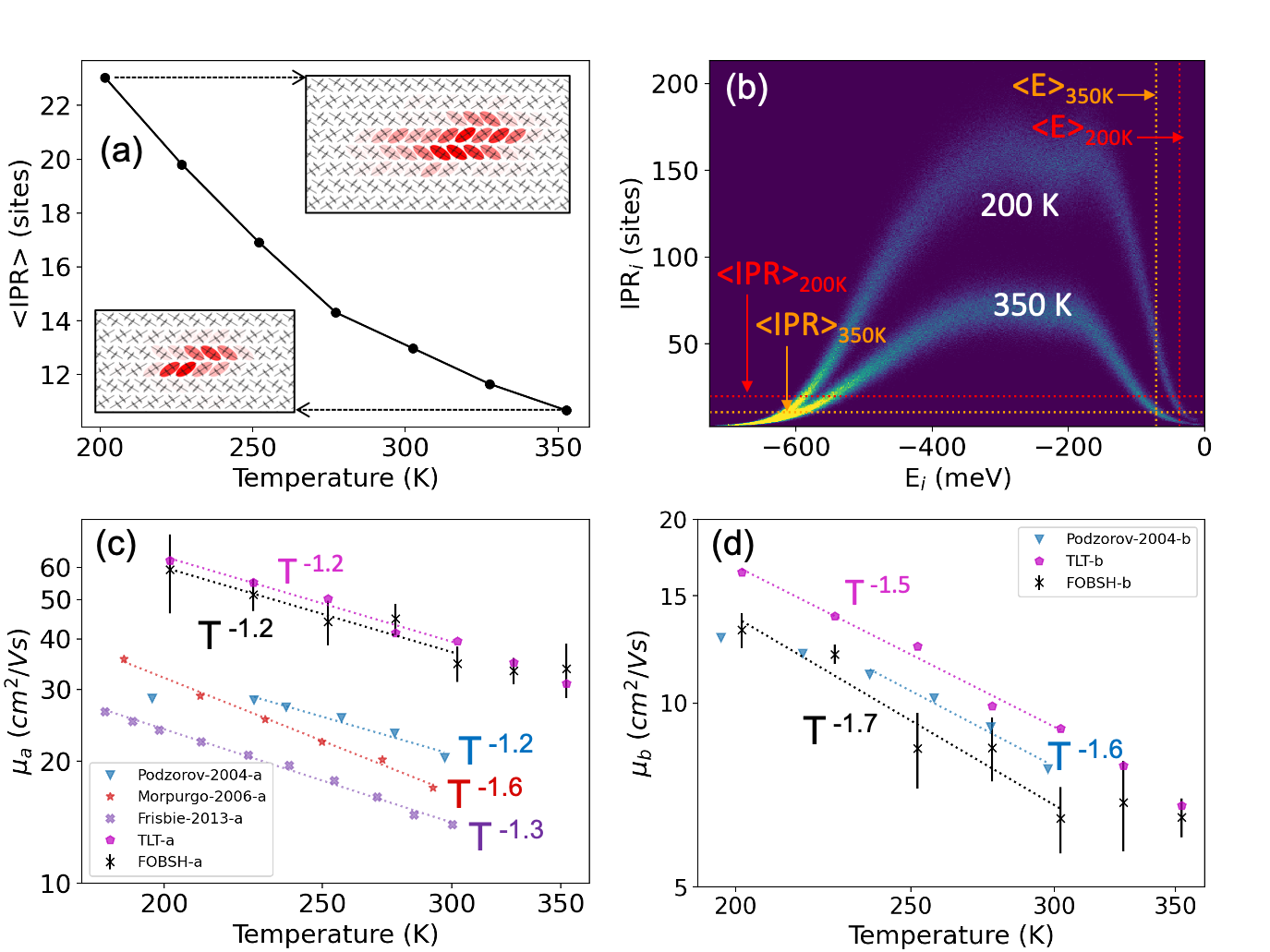}
\caption{Temperature dependences of inverse participation ratio, density of valence band states and hole mobility in rubrene. (a) Time-averaged IPR of the hole wavefunction $\Psi (t)$ (Eq.~\ref{eq:IPR_WFN}), $\langle \text{IPR} \rangle$, showing increasing localization with increasing temperature. 
 (b) Density of valence band states (increasing from dark blue to yellow) with respect to their IPR (Eq.~\ref{eq:IPR_adiabat}) and 
 energy ($E_k^{\text{ad}}$, eigenvalues of Eq.~\ref{eq:Hamiltonian}) relative to the energy at the valence band maximum, for $T = 200$ K and $T = 350$ K. Data were averaged over configurations from MD simulations run at the specified temperature. Mean values for the IPR of the hole wavefunction (Eq.~\ref{eq:IPR_WFN}) and for the potential energy of the hole ($\sum_k |u_k|^2 E_k$) from FOB-SH simulations are shown in dashed lines.
 (c), (d) Hole mobility along the $a$ and $b$ directions, $\mu_a$ and $\mu_b$ respectively, from FOB-SH simulation and~Eq.~\ref{eq:Einstein_relation} (black), 
 and experimental data from Refs. \cite{podzorov2004intrinsic} (blue), \cite{hulea2006tunable} (red) and \cite{xie2013high} (purple). 
 Hole mobilities from transient localisation theory, Eq.~S6 in Supplementary Note 8, are in magenta.
 Best fits $\mu \propto T^{-n}$ are indicated in dashed lines. Error bars were calculated by partitioning the total number of trajectories into 5 blocks ($\approx 130$ trajectories per block) and taking the standard deviation of the block-averaged mobilities.
  }
  \label{fig:RUB_TEMP}
\end{figure}

\noindent{\bf Temperature dependence of hole transport} \\
To set the scene for hole transport subject to a temperature gradient, we first present results for simulations at constant temperatures
in the range 200--350\,K. The lattice parameters were adjusted for each temperature to account for the small thermal expansion 
of the crystal as observed in experiment\cite{jurchescu2006low}. The FOB-SH simulations of hole transport follow a protocol very similar to the 
one established in previous works\cite{giannini2020flickering, giannini2019quantum} except that the improved force field is used, 
see section {\it Methods} for details.  The results obtained are presented in Figure~\ref{fig:RUB_TEMP} and numerical values are 
summarized in Table~\ref{table:FOBSH_electronic_couplings_tau}. 

Figure \ref{fig:RUB_TEMP}(a) shows that the hole wavefunction propagated by FOB-SH becomes increasingly localised as temperature increases. 
The average inverse participation ratio (IPR) of the wavefunction decreases from 23 molecules at 200 K to 13 and 11 molecules at 300 and 350 K, respectively.  
 The same trend can be seen in the IPR-resolved density of electronic valence band states that make up the hole wavefunction, shown for 200 K and 350 K in Figure~\ref{fig:RUB_TEMP}(b). The average delocalization of valence band states at a given energy is significantly smaller at 350 K than at 200 K due to 
 the increased dynamic disorder of the electronic Hamiltonian. Indeed, the width of the thermal distribution of electronic couplings, i.e., off-diagonal electron-phonon coupling, 
 $\sigma_{a}$ and $\sigma_{b}$, are about 25\% higher at 350 K than at 200 K whilst the magnitudes of the mean values of electronic 
 couplings slightly decrease due to the small thermal expansion of the crystal (see Table~\ref{table:FOBSH_electronic_couplings_tau} and Supplementary Note 3). 
 The maximum of the IPR is shifted towards the top of the valence band at both temperatures, a consequence of the overall positive 
 electronic coupling sign combination in the rubrene crystal, sgn($P$,$T_1$,$T_2$) = sgn($J_a$, $J_b$, $J_b$) = (+, -, -)\cite{fratini2017map, giannini2022charge}, 
 but the height of the maximum is markedly reduced at the higher temperature. Although the hole occupies more highly excited (= lower energy) valence band states at the higher temperature, the average IPR is significantly lower than at the lower temperature 
 (compare horizontal dashed lines in Figure~\ref{fig:RUB_TEMP}(b)). The same holds for the hole wavefunction propagated by FOB-SH (panel a) as it closely samples a Boltzmann distribution of the valence band states in the long-time limit\cite{Giannini23}.    

The hole mobility obtained from the time-dependent mean-square displacement of the hole wavefunction (see {\it Methods}) is shown in Figure~\ref{fig:RUB_TEMP}(c)-(d) as a function of temperature, 
with fits to $\mu \sim T^{-n}$.  Good agreement of the theoretical and experimental data to the power law decay fits indicates a band-like temperature dependence of mobility. 
We obtain exponents $n=-1.2$ for mobility along $a$ and $n=-1.7$ for mobility along $b$, which is within the relatively narrow experimental range of estimates. 
Our absolute mobilities are in good agreement with results from transient localisation theory (see Supplementary Note~8 for details), but about a factor of 1.5-2 higher than the experimental values. 
For instance, our predicted room temperature mobility along the high-mobility direction $a$ is $34.8 \pm 3.5$ $\text{cm}^2\,\text{V}^{-1}\,\text{s}^{-1}$ compared to $\sim 20$ $\text{cm}^2\,\text{V}^{-1}\,\text{s}^{-1}$ 
reported by Podzorov {\it et al.}\cite{podzorov2004intrinsic}. This difference may be due to factors not accounted for in the FOB-SH simulations, for example remaining 
structural disorder or chemical impurities in the crystalline samples or surface and finite carrier concentration effects.  

\begin{table}
  \caption{Summary of temperature dependent properties of rubrene obtained from FOB-SH simulations at constant temperature unless noted otherwise.    
    }
  \label{table:FOBSH_electronic_couplings_tau}
  \begin{tabular}{lllllllll}
    \hline
    T (K) & $\langle J_a \rangle^a$ &  $\sigma_a^a$ & $\langle J_b \rangle^a$ & $\sigma_b^a$ & $\langle \text{IPR} \rangle^b$ & $\sigma_\text{IPR}^b$ &$\tau_r^c$ & $t_r^c$ \\
    \hline
    200 & 113.4 & 23.7 & -17.6 & 5.7 & 23.0 & 16.2 & 88.4 & 11.9 \\
    225 & 112.0 & 24.9  & -17.2 & 6.0 & 19.8 & 14.1 & 83.6 & 11.0 \\
    250 & 110.6 & 26.1  & -16.7 & 6.2 & 16.9 & 12.6 & 81.2 & 10.4 \\
    275 & 108.5 &  27.2 & -16.1 & 6.5 & 14.3 & 10.7 & 76.4 & 9.8 \\
    276$^{d}$  &  107.6        &   27.3       &  -15.9         &   6.4   &  13.7      & 10.0      &         &        \\
    300 & 106.8 &  28.2 & -15.6 & 6.7 & 13.0 & 9.6 & 72.8 & 9.1 \\
    300$^{d}$       &   106.8        &   28.2       &   -15.6        &  6.6    &  12.5      & 9.4       &         &  \\
    325 & 105.2 & 29.2  & -15.2 & 6.8 & 11.6 & 8.8 & 71.5 & 9.0 \\
    325$^{d}$       & 106.1         &   29.0       &  -15.4         &   6.8    &  11.7      &  8.8       &         &   \\ 
    350 & 103.5 & 30.1  & -14.7 & 7.0 & 10.7 & 8.1 & 69.1 & 8.6 \\
    \hline
  \end{tabular}
  \begin{flushleft}
   $^a$ Mean value, $\langle J_{a(b)} \rangle$, and root-mean-square fluctuations, $\sigma_{a(b)}$, of electronic coupling 
   in meV. \\
   $^b$ Average inverse participation ratio, $\langle$IPR$\rangle$ (Eq.~\ref{eq:IPR_WFN}), and corresponding root-mean-square fluctuations, $\sigma_\text{IPR}$, of the hole wavefunction. \\
   $^c$ Average time between two transient delocalization events, $\tau_r$ (fs), defined as configurations where the IPR of a trajectory is greater than $\langle$IPR$\rangle+\sigma_\text{IPR}$ and average duration of a transient delocalization event, $t_r$ (fs). \\
   $^d$ From FOB-SH simulation under a temperature gradient $\partial_x T\!=\!2.8$ K/nm for bins of length 2.9 nm centred at the indicated temperature (same binning procedure as in Figure~\ref{fig:v_s_ipr_cell}).
 \end{flushleft}
\end{table}

The mechanism of charge transport is transient delocalization (TD) at all temperatures, see our previous work for a detailed description\cite{giannini2020flickering}.  
We define a transient delocalisation event as a period inf time where the IPR is larger than a threshold given by $\langle$IPR$\rangle + \sigma_{\text{IPR}}$, 
where $\langle$IPR$\rangle$ and  $\sigma_\text{IPR}$ are the mean and the width of the thermal distribution of the IPR\cite{giannini2019quantum}. We find that both the average 
duration of a TD event, $t_r$, and the average time between two subsequent TD events, $\tau_r$, decrease with increasing temperature 
(Table~\ref{table:FOBSH_electronic_couplings_tau}) but the effect is rather small. Such behaviour is expected because TD events are associated 
with transitions (surface hops) between electronic valence band states and the frequency of such transitions increases with increasing kinetic 
energy. The small reduction of $\tau_r$ would lead to an increase in mobility with increasing temperature but this effect is smaller than the decrease in mobility due to increasing hole localization. 

\noindent{\bf Thermoelectric transport} \\
To study thermoelectric transport using FOB-SH, we require a computational protocol for simulating a uniform temperature gradient. 
This was achieved by defining local heat baths at specified temperatures, maintained through thermostatting \cite{li2019influence}. 
Thermal bath regions were defined at $T$ = 250 K and $T$ = 350 K, separated by 36 nm along the $a$-direction (50 unit cells), resulting in a stable and linear temperature profile, see Supplementary Figures~S8-9. We note that such a temperature gradient is about 3 orders of magnitude larger than realisable 
in experiment, though necessary in our simulations to ensure convergence of statistical sampling of drift velocity, see below. 
The hole wavefunction in FOB-SH simulations was initialised on a single molecule at 5 different initial positions, 
uniformly spaced along the $x$-direction of the active region spanning the ``hot" and the ``cold" end of the simulation cell
(400 trajectories each). All properties obtained from FOB-SH, including drift velocity, were averaged over all sets of starting positions. We also performed a control simulation using the same setup as above except that the temperature of both thermal baths were set equal to 300 K, i.e. constant room temperature simulation. Further details regarding the computational set-up are provided in {\it Methods} and Supplementary Note 10. 

\begin{figure}[H]
\centering
\includegraphics[width=0.75\textwidth]{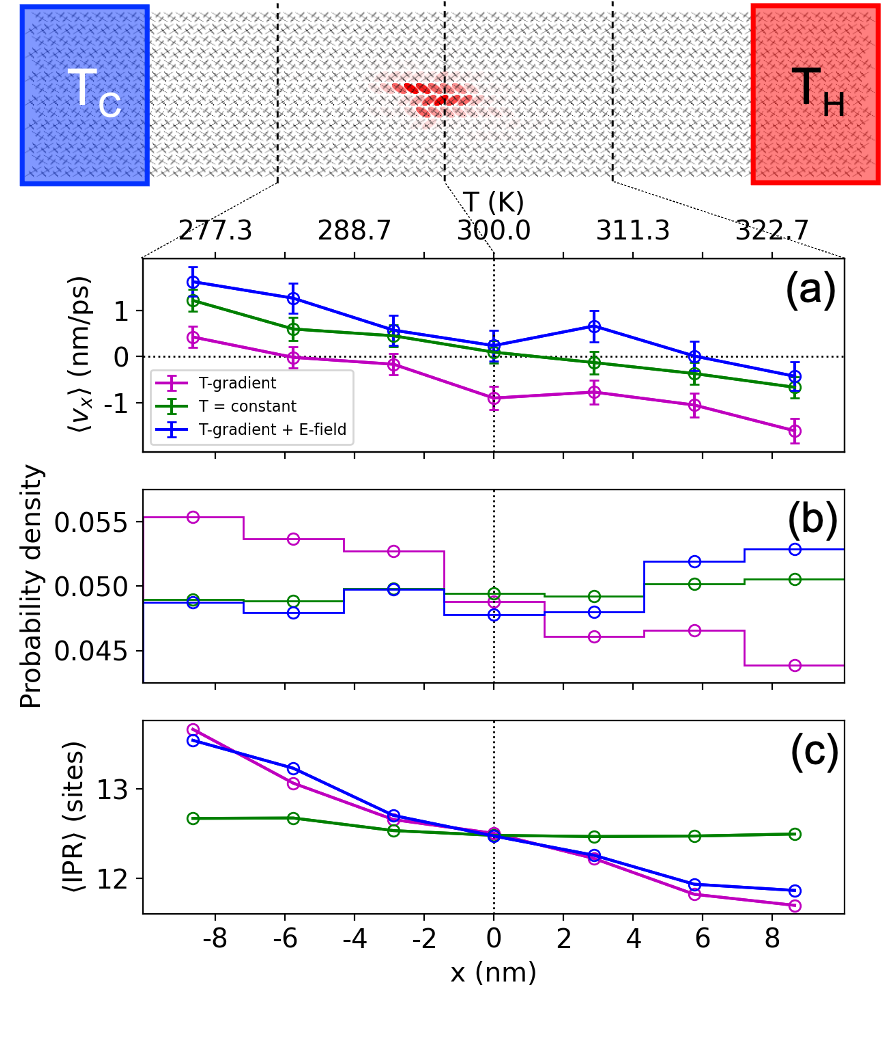}
  \caption{ Thermoelectric hole transport in rubrene from FOB-SH. The graphic at the top shows the simulation box with heat baths on 
  the cold and hot sides indicated. A snapshot of the hole wavefunction $\Psi$ as it moves from the hot to the cold end is depicted by superposing red ellipses on each molecular site with opaqueness proportional to the site population $|u_k|^2$.
  Position-resolved quantities obtained from FOB-SH simulations under a temperature gradient ($\partial_x T\!=\!2.8$ K/nm, magenta), under a temperature gradient and an external electric field ($\partial_x T\!=\!2.8$ K/nm and $-\partial_x \phi\!=\!3 \times 10^3$ V/cm, blue) and at constant 
temperature ($T\!=\!300$ K, green) are shown in panels (a)-(c) for the central 20 nm of the simulation cell centred around 300 K (region 
indicated in black dashed lines). The data were binned using bin widths of 2.9 nm and averaged over all time steps and FOB-SH trajectories.      
(a) Drift velocity of the COC (Eq.~\ref{eq:COC_wfn}) of the hole wavefunction $\Psi$, $\langle v_x \rangle$, see {\it Methods} for details of calculation. Error bars correspond to the standard error of the velocity distribution, $\sigma_v(x)/\sqrt{N(x)}$, where $\sigma_v(x)$ is the standard deviation of velocities observed in bin $x$ and $N(x)$ is the number of data points. (b) Probability density of the COC of the hole wavefunction.   
(c) Average IPR of the hole wavefunction (Eq.~\ref{eq:IPR_WFN}), $\langle\text{IPR}\rangle$.}
\label{fig:v_s_ipr_cell}
\end{figure}

Figure \ref{fig:v_s_ipr_cell} shows the position-resolved average drift velocity of the centre-of-charge (COC) of the hole wavefunction (panel a), probability density of the COC (panel b) and average IPR (panel c) for positions within $\pm 10$ nm with respect to the middle of the simulation cell (at position $x$ = 0, $T$ = 300 K) spanning temperatures between 275-325 K. In the control simulations at constant temperature (data in green), the probability density of COC and the average IPR are approximately the same at all positions. The average drift velocity at $x = 0$ is zero, as expected. In fact, the average drift velocity should vanish at all positions for a constant temperature profile, but in practice this is not the case. We observe a small linear change in position-dependent drift velocity as one approaches the hard boundaries at the hot and the cold side of the simulation cell. At the boundaries the charge carrier gets reflected and this results in a boundary force, leading to a small drift velocity pointing towards the middle of the simulation cell. In the centre of the simulation cell the artificial boundary effects cancel and the average drift velocity vanishes.

Remarkably, in the presence of the temperature gradient (data in magenta) the drift velocities shift fairly uniformly to more negative values  
compared to the constant temperature simulations. At the centre of the cell where no artificial boundary effects are present, 
we obtain $\langle v_x \rangle = -0.89 \pm 0.25$ nm$\,\text{ps}^{-1}$, corresponding to a net motion of the charge carrier from the hot to the cold region, i.e. the thermoelectric effect. 
We emphasize that the thermoelectric effect obtained from FOB-SH is not due to skewed initial conditions - 
the same number of trajectories are initialised from hot and cold regions and from the middle, see above. Rather, the hole wavefunction 
moves, on average, faster from hot to cold than from cold to hot giving rise to a net current and an increase in probability density of the hole 
on the cold side (panel b). 

\begin{figure}[H]
\centering
\includegraphics[width=0.85\textwidth]
{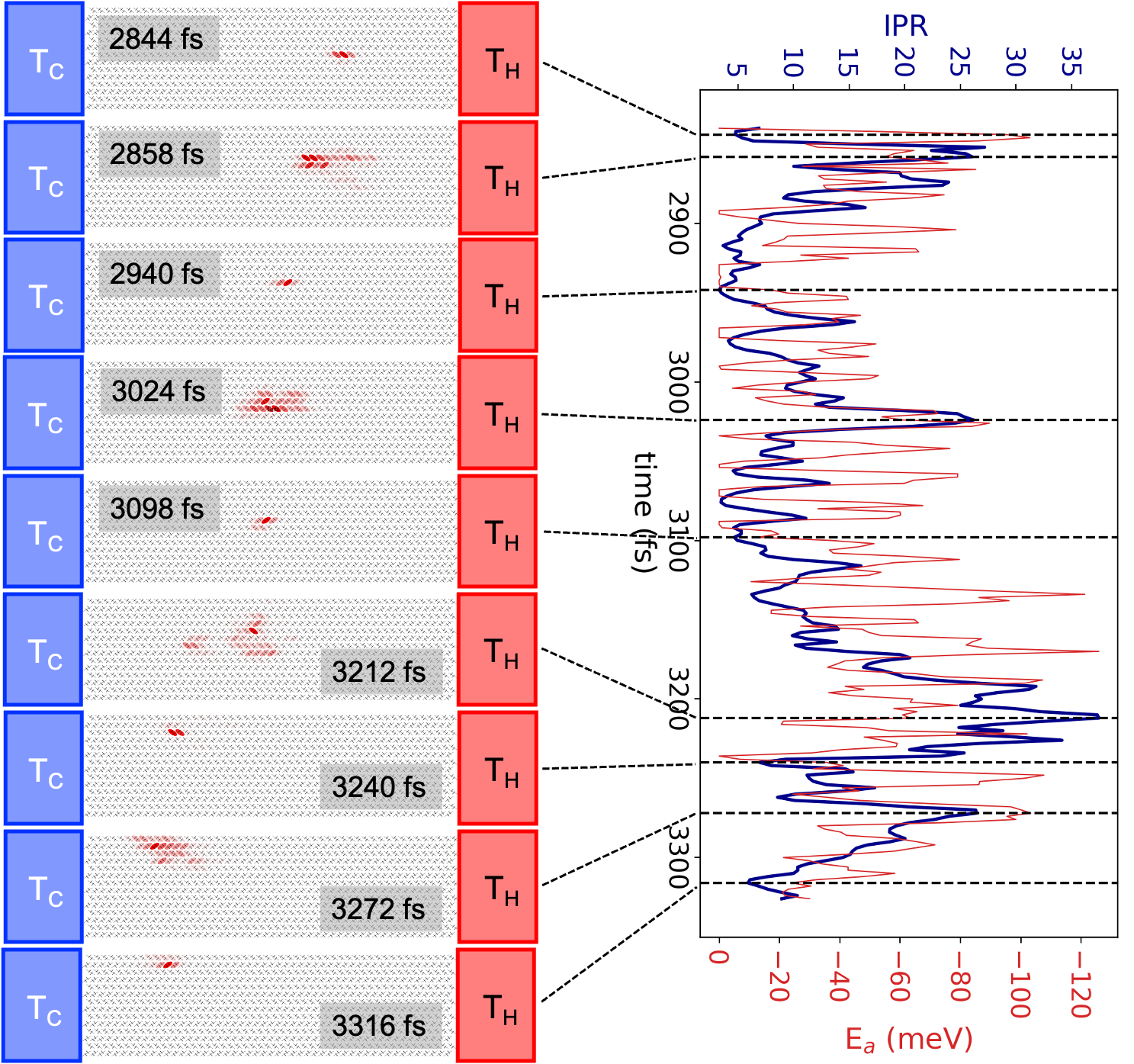}
  \caption{Snapshots of the charge carrier wavefunction, $\Psi$ (Eq.~\ref{eq:wavefunction}), along a single FOB-SH trajectory as it moves from hot to cold. $\Psi$ is represented by superposing red ellipses onto each molecular site with opaqueness proportional to the site population, $|u_k|^2$. The IPR of the carrier wavefunction (Eq.~\ref{eq:IPR_WFN}) 
  and the energy of the active adiabatic state, $E_a$, are plotted on the right hand side in dark blue and red, respectively.}
  \label{fig:thermoelectric_transport}
\end{figure}
 
A representative FOB-SH trajectory of a hole injected in the hot region and moving towards the cold region is shown 
in Figure \ref{fig:thermoelectric_transport}. We find that the transport occurs via a series of transient delocalization events, similarly to the case of 
constant temperature.  However, on average, the hole wavefunction steadily expands as it moves from hot to cold, see the position-resolved 
average IPR in Figure \ref{fig:v_s_ipr_cell}c (data in magenta). In fact, the average IPR at a given temperature in the temperature 
gradient simulations is very similar to the corresponding value obtained in a constant temperature simulation at this temperature. 
This is due to electronic couplings and root-mean-square fluctuations (i.e. off-diagonal electronic disorder) 
along the temperature gradient being very similar too, see Table~\ref{table:FOBSH_electronic_couplings_tau}. Hence, the trends discussed above 
for temperature-dependent delocalization and electronic disorder carry over, in a quasi-continuous manner, to the system under a temperature gradient. 
The spatial heterogeneity of wavefunction delocalization under a temperature gradient is a feature of the transient delocalisation regime in contrast to the standard assumption that spatial variation caused by the temperature gradient is related only to the spreading-out of the Fermi-Dirac distribution i.e transport level of carriers\cite{ioffe1959semiconductor, lundstrom2012near}.

\begin{figure}[H]
\centering
\includegraphics[width=1.0\textwidth]{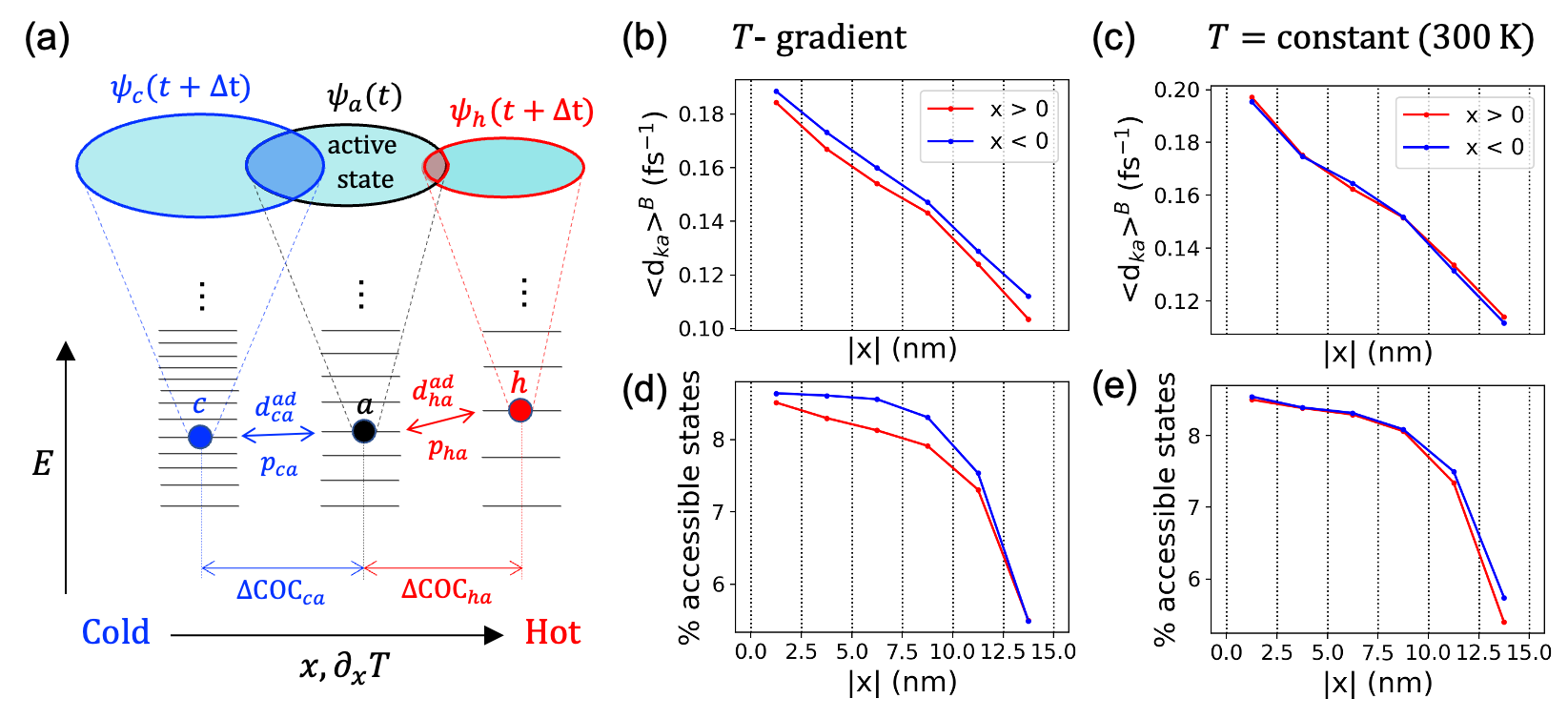}
  \caption{Interpretation of the net drift velocity from the hot to the cold side observed in FOB-SH simulations with a temperature gradient. 
  (a) Energy levels are schematically shown for valence band (adiabatic) states with their COC located in the bin centred at 300 K (black), and in adjacent 
        bins towards lower (blue) and higher temperature (red). The active valence band state $\psi_a$ governing nuclear dynamics at time $t$ and  
        example adiabatic states $\psi_c$ and $\psi_h$ at time $t+\Delta t$, which are a distance $\Delta$COC$_{ca}$ and $\Delta$COC$_{ha}$ displaced 
        towards the cold and hot sides, respectively, are shown as ellipses, with increasing size from hot to cold to reflect their increasing IPR. 
        NACEs ($d_{ka}^{\text{ad}}$) and surface hopping probabilities ($p_{ka}$) between $\psi_a$ and $\psi_c$ ($d_{ca}^{\text{ad}}$, $p_{ca}$) 
        and between $\psi_a$ and $\psi_h$ ($d_{ha}^{\text{ad}}$, $p_{ha}$) are indicated.
 (b, c) Position-resolved thermally averaged NACE, Eq.~\ref{eq:Boltz_av} where $P_k\!=\!d_{ka}^{\text{ad}}$, towards the cold side (data in blue) and hot side (data in red) for simulations under 
          (b) a temperature gradient $\partial_x T\!=\!2.8$ K/nm and (c) at constant temperature of 300 K. 
 (d, e) Position-resolved percentage of thermally accessible states, $N^\text{acc}(x_i) / \sum_j N^\text{acc}(x_j)\times 100\%$ where $N^\text{acc}(x_i)$ is given by Eq.~\ref{eq:N_accessible},     
           towards the cold side (data in blue) and hot side (data in red) for simulations under (d) a temperature gradient $\partial_x T\!=\!2.8$ K/nm and (e) at constant temperature of 300 K.}
  \label{fig:NACE_av_d_n_states}
\end{figure}

What causes the directional motion of the hole wavefunction from hot to cold? Our mechanistic proposal based on FOB-SH simulations
is illustrated in Figure~\ref{fig:NACE_av_d_n_states}(a). We assume that the hole wavefunction $\psi_a$ at a given time $t$ 
is located in the central bin (shown schematically as an ellipse in the middle of Figure~\ref{fig:NACE_av_d_n_states}(a)) and we analyse the likelihood for transitions 
from $\psi_a$ to one of the states towards the cold side (denoted ``cold states", e.g., $\psi_c$ indicated by an ellipse to the left) or to one of the 
states towards the hot side (denoted ``hot states", e.g., $\psi_h$ indicated by an ellipse to the right). In other words, we aim to explain why, on average,  
$\psi_a$ is more likely to transition to cold than to hot states thereby generating the negative drift velocity $\langle v_x \rangle$ obtained in the simulations.       
We find that major displacements of the hole wavefunction contributing to drift velocity are typically induced by thermally induced electronic transitions (surface hops)
from  $\psi_a$ to other electronic states. The probability for transitions to occur is proportional to (i) the density of thermally accessible states, i.e., density of states 
that are within a few $k_{\text{B}}T$ of $\psi_a$ and (ii) the magnitude of the non-adiabatic coupling element between 
$\psi_a$ and these states. We find that the Boltzmann-averaged (i.e., thermally accessible) density of cold states 
(Fig.~\ref{fig:NACE_av_d_n_states}(d), data in blue) is higher than for hot states (data in red), and this is the case for all distances between the 
COCs of $\psi_a$  and the states on the cold or hot side ($x=\Delta\text{COC}_{ka}$, $k=c$ or $h$). 
Moreover, we find that the Boltzmann-averaged non-adiabatic coupling between $\psi_a$ and the cold states is somewhat larger than between 
$\psi_a$ and the hot states (Fig.~\ref{fig:NACE_av_d_n_states}(b)). Thus, according to this analysis of our simulation data, the negative drift velocity 
is due to the increasing density of thermally accessible states and the increasing non-adiabatic coupling between states with decreasing temperature. 
Both effects are a consequence of the decreasing electronic coupling fluctuations along the temperature gradient from hot to cold.    
In the control simulations at constant temperature there are no such gradients for these quantities, as expected, see Figure~\ref{fig:NACE_av_d_n_states}(e) and (c).

The gradients in thermally accessible density of electronic states and non-adiabatic coupling are not unrelated. 
This is because the non-adiabatic coupling is inversely proportional to the energy difference between the coupled states, 
$d^{\text{ad}}_{ja} \propto 1/\Delta E_{ja}$\cite{baer2002non, domcke2004conical} (see Supplementary Note S11). Indeed we find good correlation 
between these two quantities, Supplementary Figure~S14 (upper row). Thus, the larger density of 
energetically-close cold states (small $\Delta E_{ja}$) compared to hot states contributes to the larger average non-adiabatic coupling 
for transition to cold states. Moreover, noting that the non-adiabatic coupling is given by $\langle \psi_j | \dot{\psi}_a\rangle$, an additional factor potentially 
further enhancing the average non-adiabatic coupling to cold states could be their greater delocalization compared to hot states.  
Indeed we find some correlation between non-adiabatic coupling and delocalization of electronic states, Supplementary Figure~S14 (bottom row), though the scatter 
is relatively large and the correlation is not as strong as for the energy difference above.

\noindent{\bf Seebeck coefficient} \\
To link our simulations to the theory of thermoelectricity and the Seebeck coefficient, $\alpha$, 
we consider the general expression for the current density along direction $x$ ($J_x$) in the presence of gradients of
temperature ($T$), chemical potential ($\mu_c$) and electrical potential ($\phi$)\cite{callen1948application, emin2014seebeck},  
\begin{equation}
  J_x = - \sigma \alpha \partial_x T - \frac{\sigma}{q} \partial_x \mu_c - \sigma \partial_x \phi , \label{eq:current_density}  
\end{equation}
where $\sigma$ is the electrical conductivity and $q$ the charge of the carriers. (Note, $\alpha$ and $\sigma$ are in general tensorial quantities but for ease of notation we omit indices).  
The second and third terms on the right hand side (RHS) of equation~\ref{eq:current_density} can alternatively be expressed in terms of the electrochemical potential $\bar{\mu} = \mu_c + q \phi$. 
The current density obtained from FOB-SH, $J_x = q n \langle v_x \rangle$, where $n$ formally corresponds to the concentration of the single charge carrier in our simulation cell, 
is the result of the first two terms on the RHS of equation~\ref{eq:current_density}.    
The first term, directly proportional to the Seebeck coefficient and the temperature gradient, drives the charge carrier from hot to cold,  
whilst the second term, proportional to the chemical potential gradient that is induced by the temperature gradient, drives the carriers in the opposite 
direction from cold to hot. The third term on the right hand side (RHS) of equation~\ref{eq:current_density} is zero in the above simulations since no electrical 
potential gradients were applied. This differs from the usual experimental setup, 
where the electrical potential energy gradient, $\partial_x \phi$, is adjusted by an external electric field so that the net current arising from the first two contributions is 
compensated, i.e. $J_x\!=\!0$, and the resulting voltage is the open-circuit voltage, $\Delta V_{oc} =  \Delta \mu_c/q + \Delta \phi$. 

Rearranging terms in equation~\ref{eq:current_density} one finds
\begin{equation}
\alpha =  -\frac{q}{e}\frac{\langle v_x \rangle}{\mu \partial_x T}   -\frac{1}{q} \frac{\partial_x \mu_c}{\partial_x T} - \frac{\partial_x \phi}{\partial_x T} = 
\alpha_v + \alpha_c + \alpha_e, \label{eq:alpha_FOBSH_contributions} 
\end{equation}
where $\mu = \sigma/(e n)$ is the charge mobility along $x$ and $e$ the elementary charge. 
Hence, the Seebeck coefficient is comprised of three terms. The first term, $\alpha_v$, is proportional to the net current flow in the system. 
This contribution is of kinetic origin and depends on the charge transport mechanism. 
The second and third terms, $\alpha_c$ and $\alpha_e$, are due to the gradient in chemical potential and external electric field, respectively, hence 
they are of thermodynamic origin. We emphasize that the separation of the Seebeck coefficient in kinetic and thermodynamic contributions
rigorously follows from Eq.~\ref{eq:current_density}\cite{callen1948application, emin2014seebeck} without assuming any specific transport mechanism. 
Emin carried out a similar separation for systems that are in the phonon-assisted tunneling (hopping) regime by writing the total Seebeck coefficient in terms 
of a contribution of kinetic origin, $\alpha_{\text{transport}}$, and a contribution of thermodynamic origin, $\alpha_{\text{presence}}$\cite{emin1975thermoelectric,emin1999enhanced}. 
Experiments are typically carried out in 
open-circuit conditions where the external electric field is adjusted such that the net current flow, and therefore the kinetic contribution $\alpha_v$, 
is equal to zero. Thus, in experiment the total Seebeck coefficient arises from the thermodynamic contributions only. Generally, the kinetic and 
thermodynamic contributions to the Seebeck coefficient will depend on the applied external electric field.
  
All three terms to the Seebeck coefficient Eq.~\ref{eq:alpha_FOBSH_contributions} can be obtained from computation. Using the average drift velocity obtained from 
FOB-SH simulation under a temperature gradient without external field 
(Figure 2a, data in magenta, $\langle v_x \rangle = -0.89$ nm$\,\text{ps}^{-1}$ at $x\!=\!0$), we obtain a kinetic contribution to the 
Seebeck coefficient $\alpha_v\!=\!97.4 \pm 27.5$~$\mu$V\,K$^{-1}$ at 300~K.
Here, a charge mobility value of $\mu = 33.0$~$\text{cm}^2\,\text{V}^{-1}\,\text{s}^{-1}$  
was used, as obtained from FOB-SH simulation at constant temperature $T\!=\!300$\,K employing the same electronically active region size 
($50\times 7$ unit cells) as in the simulations with a temperature gradient. (Notice, this value differs slightly from the one quoted above, 
$34.8 \pm 3.5$ $\text{cm}^2\,\text{V}^{-1}\,\text{s}^{-1}$, which was obtained for a larger electronically active region.) 
The thermodynamic contribution is given by $\alpha_c$ only since $\alpha_e\!=\!0$ at zero external field. $\alpha_c$ has previously been evaluated for the case 
of a non-degenerate semiconductor with parabolic bands, where a simple expression exists for $\mu_c$ \cite{ioffe1959semiconductor, mahan2000density}. 
Such approximations should be regarded with caution in the case of rubrene, where the effect of thermal disorder on the band states is very strong. 
Here, we calculate $\frac{\partial \mu_c}{\partial T}$ explicitly by noting that the chemical potential $\mu_c$ is related to the free energy change upon 
insertion of a hole (electron) in the valence (conduction) band, equations~\ref{eq:mu_c_ref} and~\ref{eq:mu_c}, see 
{\it Methods} and Supplementary Note~13 for details. At carrier density $n= 2.74 \times 10^{15}$~m$^{-2}$, corresponding to a single carrier 
in the active region area of $50\times7$ unit cells, we obtain $\alpha_c\!=\!331 \pm 6$~$\mu$V\,K$^{-1}$. The total Seebeck coefficient 
at $T = 300$~K and $n= 2.74 \times 10^{15}$~m$^{-2}$ is thus $\alpha\!=\! \alpha_v + \alpha_c \!=\!429 \pm 28$~$\mu$V\,K$^{-1}$.

We verify that the Seebeck coefficient computed from Eq.~\ref{eq:alpha_FOBSH_contributions} is independent of the chosen external electric field by carrying out FOB-SH simulations subject to a temperature gradient but this time under open-circuit conditions, like in experiment. 
To do so we apply an external electric field in our simulation such that the average drift velocity and the resultant kinetic contribution to the Seebeck coefficient, $\alpha_v$, 
vanish. This should be the case for an external electric field 
of $\approx 3 \times 10^3$~V$\,$cm$^{-1}$ pointing in the direction from cold to hot. 
The results are shown in Figure~\ref{fig:v_s_ipr_cell} (data in blue). Indeed, the average drift velocity in the central bin is now zero within the statistical uncertainty, 
$\langle v_x \rangle = 0.24 \pm 0.33$~nm$\,\text{ps}^{-1}$ (panel a) and the overall Seebeck coefficient, $\alpha \!=\! \alpha_v + \alpha_c + \alpha_e \!=\! 397 \pm 37 $~$\mu$V\,K$^{-1}$, encompasses the value obtained without the field within the statistical uncertainty.
Thus, the two simulation approaches (i.e., with and without an external electric field) for calculating the Seebeck coefficient within the framework of Eq.~\ref{eq:alpha_FOBSH_contributions} yield consistent results. 
Note that the hole wavefunction is still very dynamic, frequently moving towards the cold or the hot side but at about equal amounts of time, thus averaging close to zero. We also find that the position-resolved IPR remains virtually unchanged when compared to the results without the electric field (Fig.~\ref{fig:v_s_ipr_cell}(c) magenta vs blue). This is expected because the above field strength corresponds to an electrostatic 
site energy difference between adjacent rubrene molecules along the $a$-direction of only about 0.2~meV which is much smaller than the electronic coupling (100~meV), thus has very little impact on band structure and delocalization of valence band states. 

\noindent{\bf Comparison to experiment} \\ 
To validate our simulations, we carried out experimental measurements of the Seebeck coefficient 
in rubrene single crystals as a function of carrier concentration ($n$). These are measured under temperature differences of typically 5--10~K 
across a channel length of \SI{420}{\micro m} and under open circuit conditions, see {\it Methods} and Supplementary Note~14 for further details.

\begin{figure}[H]
\centering
\includegraphics[width=0.8\textwidth]{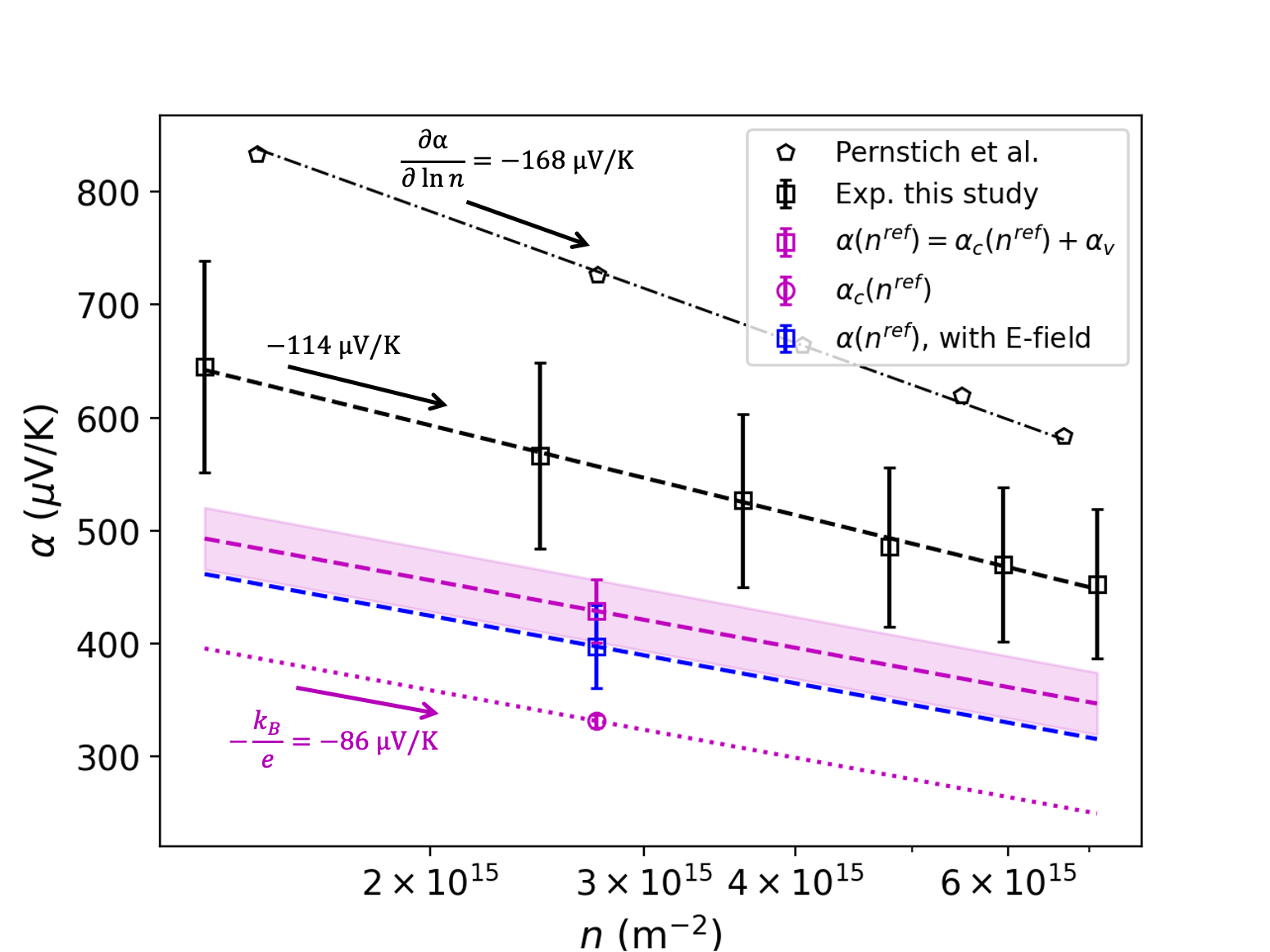}
  \caption{ Experimental and calculated Seebeck coefficient $\alpha$ as a function of carrier concentration $n$. Black squares and pentagons correspond to experimental data from this study and Ref.~\citenum{pernstich2008field}, respectively, with best fits indicated with black dashed lines. 
  The chemical potential contribution to the Seebeck coefficient, $\alpha_{c}$ (Eq.~\ref{eq:alpha_c}), computed at a reference carrier concentration $n^\text{ref} = 2.7\times 10^{15}$\,m$^{-2}$, is indicated by the dotted magenta line, $\alpha_c(n^\text{ref}=2.7\times 10^{15}\text{ m}^{-2}) = 331 \pm 6$~$\mu$V\,K$^{-1}$.
  The total calculated Seebeck coefficient (for simulations with no external electric field, $\alpha_e = 0$) is obtained by adding the kinetic contribution $\alpha_v\!=\!97.4\pm 27.5$~$\mu$V\,K$^{-1}$, $\alpha(n^\text{ref}) = \alpha_{c}(n^\text{ref}) + \alpha_v = 429 \pm 28$~$\mu$V\,K$^{-1}$(magenta dashed line). 
  The shaded area indicates $\pm$ the standard error in the calculation of $\alpha$ at $n^\text{ref}$. 
  For simulations employing an external electric field, the total Seebeck coefficient is plotted with the blue dashed line.
  }
  \label{fig:alpha_vs_n}
\end{figure}

Figure~\ref{fig:alpha_vs_n} shows the experimental Seebeck coefficients obtained in the present study (black squares) 
and those from Ref.~\citenum{pernstich2008field} (black pentagons). The best fits of experimental data points to $A + B\ln(n)$, where $A$ and $B$ 
are optimization parameters, are indicated with black dashed lines. The total computed Seebeck coefficient, $\alpha = \alpha_c + \alpha_v + \alpha_e$, 
obtained without and with an external electric field in FOB-SH simulations are indicated
at $n= 2.7 \times 10^{15}$~m$^{-2}$ corresponding to the carrier density present in FOB-SH simulations (squares in magenta and blue, respectively), 
along with $\alpha_c$ (circle in magenta).  
The concentration dependence of computed $\alpha$ and $\alpha_c$ (lines in magenta and blue) is given by equation~\ref{eq:alpha_c}. 
The uncertainty in $\alpha$ due to the statistical error from FOB-SH simulations is indicated over the 
entire concentration range by the magenta shaded region. 

The computed total $\alpha \!=\! 429 \pm 28$~$\mu$V\,K$^{-1}$ and $\alpha \!=\! 397 \pm 37$~$\mu$V\,K$^{-1}$ obtained from FOB-SH simulations without and 
with external electric field, respectively, compares favourably with the experimental estimate at the same carrier density ($n= 2.7 \times 10^{15}$~m$^{-2}$), $557 \pm 82$~$\mu$V\,K$^{-1}$, 
with experimental and computational error bars nearly overlapping.   
The experimental slope $\partial \alpha/\partial \ln(n)=-114.4$~$\mu$V\,K$^{-1}$ is also in reasonable agreement with the slope predicted by theory, $\partial \alpha/\partial \ln(n)=-k_B/e=-86.2$~$\mu$V\,K$^{-1}$.
The remaining discrepancy could be caused by model assumptions, for instance, that the thermoelectric effect remains linear over the 1000-fold larger temperature interval applied in the simulations when compared with experiment. 
Another source for the remaining discrepancy could be the 
presence of shallow trap states in experimental field effect transistors (the existence of which are confirmed by observing the experimental threshold voltage shift as a function of temperature), which are known to cause an enhancement of the Seebeck coefficient and the slope $\partial \alpha/\partial \ln(n)$, as well as depress the experimental mobility\cite{venkateshvaran2014approaching}. Therefore, the experimentally measured Seebeck coefficients are expected to be larger than the simulation where intrinsic valence band states are considered. 

In Figure~\ref{fig:alpha_vs_n} we also plot data taken from Pernstich \textit{et. al} to highlight the sample-to-sample variation present in measurements\cite{pernstich2008field}. This data shows a significantly larger magnitude of the Seebeck coefficient and slope, $\partial \alpha/\partial \ln(n) = -168$~$\mu$V\,K$^{-1}$. Experimental variations are likely due to differences in crystal quality, especially at the dielectric interface, as well as differences in the exact experimental set-up and measurement methods. We highlight these differences in order to contextualise the comparison between the simulation and experimental data, where a similar magnitude of discrepancy is seen within different experimental determinations of the Seebeck coefficient. Additionally, we note that although it is generally believed that a 2D model describes very well the charge transport situation in a field effect transistor, the carriers in the experiment are not confined in the third dimension, which may also have a small effect on the charge carrier compared to the simulation. Even in a 2D model, transport is anisotropic so any misalignment of the high-mobility crystal axis with the temperature gradient will also result in some discrepancy.

\section{Discussion} \label{section:conclusions}
In this work we have simulated the quantum dynamics of a charge carrier in an OS subject to a temperature gradient providing a dynamical perspective and understanding of a transport problem that is typically approached from a purely static perspective. At the heart of our dynamical perspective of thermoelectricity is the gradient in the thermal fluctuations of electronic couplings that gives rise to gradients in the spatially resolved density of electronic states and in non-adiabatic coupling causing the carrier wavefunction to move from hot to cold. The explicit time-dependent carrier simulation presented here 
has also allowed us to understand the Seebeck coefficient in terms of a kinetic ($\alpha_v$) and thermodynamic contribution ($\alpha_c+\alpha_e$). 
We have shown that simulations with no external electric field, where the carrier migrates from hot to cold, and those with a field to cancel such motion yield consistent Seebeck coefficients. 
Whilst in the current simulations at low carrier density the Seebeck coefficient was dominated by the thermodynamic contribution, at high carrier density relevant in practical situations the thermodynamic contribution will strongly decrease and the kinetic contribution is expected to become very important if not dominating.    
 
The mechanistic insight we have obtained from our simulations opens up a new avenue for the design of organic semiconductors with improved Seebeck coefficients.  
Previously, focus has been placed on the chemical potential contribution to the Seebeck coefficient, $\alpha_c$, sometimes referred to as the entropy of mixing. 
Indeed, increasing the entropy of the charge carrier in the valence or conduction band by increasing the density of thermally accessible valence band states 
will lead to an increase in $\alpha_c$. Yet, for ordered organic semiconductors where charge transport occurs in the transient delocalization regime we propose an 
additional route aimed at increasing the kinetic contribution to the Seebeck coefficient, $\alpha_v$, or equivalently the electric field contribution, $\alpha_e$, 
compensating $\alpha_v$ in open-circuit conditions. 

Recall that thermoelectric transport in OS is driven by the gradient in thermal electronic disorder and that a sensitive probe 
of the disorder is the average delocalization of the carrier wavefunction (IPR in Fig.~\ref{fig:v_s_ipr_cell}(c)). Thus we expect that systems with increasing 
sensitivity of thermal electronic disorder and, concurrently, charge carrier delocalization to changes in temperature will exhibit increasing values of 
$\alpha_v$. Hence, we assert that $\alpha_v \propto - \Delta L_{x} / \Delta T$, where $L_{x}$ is the localization length of the charge carrier (related to IPR). Assuming validity of 
transient localization theory ($\mu_x \!=\! (e/(k_{\text{B}}T)) L_{x}^2/(2 \tau)$) and temperature dependences of charge mobility 
$\mu_x \propto T^{-n}$ and localization time $\tau \propto T^{-m}$,  $\alpha_v \propto - \Delta L_{x} / \Delta T \propto (n+m-1)$. 
Hence the kinetic contribution to the Seebeck coefficient is predicted to increase with increasing exponents of the temperature 
dependences of $\mu$ and $\tau$. This hypothesis could be tested in future work on a range of systems exhibiting varying temperature 
dependences of localization length or mobility.

\section{Methods} \label{section:Methods}

{\bf Fragment orbital-based surface hopping (FOB-SH)} \\
Fragment orbital-based surface hopping (FOB-SH) is a mixed quantum-classical dynamics method based on fewest switches surface hopping which allows for simulation of charge 
transport on the true nanoscale (10-100 nm)\cite{spencer2016fob, carof2019calculate, giannini2021atomic}. In this method a single excess charge carrier (in this work, hole carrier) is propagated in space and time according to the time-dependent Schr{\"o}dinger equation under the influence of time-dependent classical nuclear motion. The single-particle time-dependent 
wavefunction, $\Psi(t)$ is expanded in a basis of orthogonalised time-dependent frontier orbitals which mediate charge transport, 
\begin{equation}
 \ket{\Psi(t)} = \sum_{l=1}^M u_l(t) \ket{\phi_l(\mathbf{R}(t))}, \label{eq:wavefunction}
 \end{equation}
where $\mathbf{R}(t)$ denotes time-dependent nuclear positions. In the case of hole (electron) transport, the basis functions $\phi_l$ are the orthogonalised HOMOs (LUMOs) of the molecules which constitute the crystal lattice. The valence band in which the excess hole is propagated is described by the following Hamiltonian:
\begin{equation}
H = \sum_{k}\epsilon_k\ket{\phi_k}\bra{\phi_k} + \sum_{k\neq l}H_{kl}\ket{\phi_k}\bra{\phi_l}, \label{eq:Hamiltonian}
\end{equation}
where $\epsilon_k = \bra{\phi_k}H\ket{\phi_k}$ are the site energies, i.e. the potential energy of the system when the excess charge is localised on molecule $k$, and $H_{kl} = \bra{\phi_k}H\ket{\phi_l}$ are the electronic couplings. To facilitate simulations on large systems over long time scales, we have developed a parameterized approach which avoids explicit electronic structure calculations. Site energies $\epsilon_k$ are calculated using a classical force field where molecule $k$ is charged and all other molecules are neutral. Electronic couplings $H_{kl}$ are calculated using the efficient analytic overlap method 
(AOM)\cite{gajdos2014ultrafast, ziogos2021ultrafast}, which assumes a linear relationship between electronic coupling and orbital overlap
$H_{kl} = \bar{C} \bar{S}_{kl}$ where $\bar{S}_{kl}$ is the overlap between fragment orbitals $k$ and $l$ projected into a minimal Slater-type orbital basis 
and $\bar{C}$ is a fitting parameter obtained by correlating $\bar{S}_{kl}$ with reference $H_{kl}$ computed from DFT, specifically the projector operator-based 
diabatization (POD) method \cite{futera2017electronic}. We note that for rubrene this provides a very good approximation, with mean absolute errors (mean relative unsigned errors) of
5.8 meV (5.5 \%) and 3.2 meV (23.6 \%) for $a$ and $b$--direction AOM couplings with respect to DFT, $J_a$ and $J_b$, respectively \cite{hafizi2023ultrafast}.
Inserting equation~\ref{eq:wavefunction} into the time-dependent Sch{\"o}dinger equation yields
\begin{equation}
 i\hbar\dot{u}_k(t) = \sum_{l=1}^M u_l(t) (H_{kl}(\mathbf{R}(t)) - i\hbar d_{kl}(\mathbf{R}(t))), \label{eq:TDSE} 
 \end{equation}
 where $d_{kl} = \bra{\phi_k}\ket{\dot{\phi_l}}$ are the NACEs in the quasi-diabatic site basis, which are usually close to zero, in contrast to the NACEs in the adiabatic basis $d^{\text{ad}}_{kl} = \bra{\psi_k}\ket{\dot{\psi_l}}$. At any given time, the nuclear dynamics is propagated classically on one of the adiabatic states that results from diagonalising the electronic Hamiltonian. 
This adiabatic state is denoted the active surface, $E_a(\mathbf{R}(t))$. At each time step, the Tully surface hopping probability for a surface hop from state $a$ to another adiabatic state $j$, $p_{ja}$, is calculated\cite{tully1990molecular},
\begin{equation} 
p_{ja}=-\frac{\mathbb{R}(c_j^* c_a d^{\text{ad}}_{ja})}{|c_a|^2}\Delta t,
\label{eq:Tully_hop} 
\end{equation} 
where $c_j$ and $c_a$ are the wavefunction expansion coefficients of adiabats $j$ and $a$, respectively, and $\Delta t$ is the nuclear time step.
The probability to remain on the active surface is given by $p_{aa}=1-\sum_{j\neq a} p_{ja}$. 
A random uniform number is drawn to stochastically decide whether to attempt hop to a new surface $j$. Energy conservation after a successful hop is 
enforced according to the standard procedure of rescaling the velocity components in the direction of the non-adiabatic coupling vector (NACV). 
If the nuclear kinetic energy is not sufficient to fulfill energy conservation, the hop is rejected and the the nuclear velocity components along the 
direction of the NACV are reversed \cite{carof2019calculate, carof2017detailed}. In addition to the standard prescription of surface hopping, 
three extensions to the original algorithm are required to ensure convergence of mobility with system size, detailed balance and good internal consistency. 
These extensions include a decoherence correction, trivial crossing detection and elimination of decoherence-induced spurious long-range charge transfer. 
We refer to Refs. \citenum{carof2019calculate} and \citenum{carof2017detailed} for a detailed discussion of these important additions to the method.
A drawback of the surface hopping method is that nuclei are treated as classical particles which means that certain nuclear quantum effects including 
zero-point energy and tunneling are not included. We do not think this is a major problem for the current system because these effects typically become important 
for systems characterised by high energetic barriers, i.e., in the small polaron hopping regime.

\noindent{\bf FOB-SH simulations at constant temperature} \\
Initial atomic positions for rubrene were taken from the Cambridge Crystallographic Data Centre (CCDC) structure with identifier QQQCIG01. 
Thermal expansion of the unit cell was accounted for by using temperature dependent lattice parameters determined by a linear fit to the experimental  
temperature-dependent lattice parameters of Ref. \cite{jurchescu2006low}, see Supplementary Note 3 for details. 
The unit cell was optimised for each temperature under the constraint of fixed lattice parameters corresponding to that temperature.
The force field used was based on the GAFF parameters, with selected parameters re-optimized to better reproduce results from ab-initio MD simulations as reported in Ref.~\cite{elsner2021mechanoelectric}, see Supplementary Note 1 for details. 
Supercells composed of $54\times13\times1$ unit cells were prepared and equilibrated to 200, 225 and 250~K, 
and slightly smaller supercells composed of  $50\times13\times1$ unit cells were equilibrated to 275, 300, 325 and 350~K
for 200~ps in the NVT ensemble applying periodic boundary conditions, a Nos\'e-Hoover thermostat and a MD time step of 1 fs. This was followed by 
at least 325 ps MD simulation in the NVE ensemble.    
Initial positions and velocities for the swarm of FOB-SH trajectories were drawn from snapshots separated by 0.5 ps from the equilibrated NVE trajectory. For each trajectory, the wavefunction was initialised on a single molecular site (i.e. diabatic state) located at the corner of the simulation box and propagated for 900 fs in the NVE ensemble using an MD time step of 0.05 fs and an electronic time step of 0.01 fs for integration of equation~\ref{eq:TDSE} employing the Runge-Kutta algorithm to 4th order.  At least 650 FOB-SH trajectories were run for each temperature.   
The initial 200 fs of dynamics, corresponding to quantum relaxation from the initial diabatic state, was neglected in all analysis. 
The diffusion tensor for each temperature was obtained from a linear best fit of the mean squared displacement against time, equation~\ref{eq:Diffusion_tensor}, 
between 200 fs to 900 fs (see Supplementary Fig.~5). Mobility was calculated using the Einstein relation, equation~\ref{eq:Einstein_relation}. 
The  mobility values reported herein are well converged with system size, even for the low temperatures where hole delocalization is extensive.  
A detailed analysis of the convergence is given in Supplementary Note~4 and Tables~S4-5.  

\noindent{\bf FOB-SH simulations with a temperature gradient} \\ 
A supercell of size $120\times7\times1$ (using the 300 K lattice parameters) was defined in periodic boundary conditions. A saw-tooth temperature profile was achieved by defining two thermal bath regions of size $10\times7\times1$ unit cells at temperatures 250 K and 350 K, separated by 50 unit cells along the $a$ direction. The temperature in the thermal bath regions was constrained to the target temperatures through a velocity rescaling procedure, as implemented in CP2K \cite{kuhne2020cp2k} 
(see Supplementary Note 9 for the detailed procedure). A set of 400 positions and velocities sampled every 0.5 ps from a 200 ps non-equilibrium run under the temperature gradient were used as initial coordinates for subsequent FOB-SH runs. The (non-periodic) electronically active region in FOB-SH simulations was defined over one of the linear portions of the temperature profile (see Supplementary Fig.~8, top panel). In order to eliminate any possible boundary effects associated with initialising the wavefunction at a given position, the hole wavefunction was initialised on a single molecule at 5 different initial positions, uniformly spaced along the $x$-direction of the FOB-SH active region (at 250, 275, 300, 325 and 350 K). Overall, 2000 trajectories of length 5 ps were run ($400 \times 5$ initial wavefunction positions). 
In all production runs, velocity rescaling to the target temperature was only applied within the thermal bath regions. The dynamics in the electronically active region evolved according to the standard FOB-SH algorithm, i.e., 
Newtonian dynamics on one of the adiabatic potential energy surfaces and rescaling of the velocity component parallel to the NACV after successful surface hops. The settings and integration time steps were the same as described above for FOB-SH simulation at constant temperature. Further simulation details are presented in Supplementary Note 9.

\noindent{\bf FOB-SH simulations with a temperature gradient and an external electric field} \\ 
The above FOB-SH simulations with a temperature gradient were repeated under the presence of a constant external electric field along $x$, $E_x$.
The effect of the field was simply modelled by a linear change of the site energies of the electronic Hamiltonian, equation~\ref{eq:Hamiltonian}, 
$\epsilon_k ({\bf R}) \rightarrow  \epsilon_k ({\bf R}) - q E_x x_k $, where $q\!=\!+e$, $x_k$ is the centre of mass of molecule $k$ and  
$E_x\!=\!-\partial_x \phi = 2.557 \times 10^{3}$ V$\,$cm$^{-1}$. The corresponding additional forces on the nuclei have been accounted for but are 
negligibly small for the small fields applied. The same simulation protocol was followed as for the simulations without external field 
except that trajectories were run to 3~ps rather than 5~ps due to our finding that trajectories of length 3~ps are sufficiently converged, 
see Supplementary Figure~S15. All simulations were carried out with our in-house implementation of FOB-SH 
in the CP2K simulation package \cite{kuhne2020cp2k}.

\noindent{\bf Calculation of charge mobility and inverse participation ratio} \\
Hole mobilities (Fig.~2(c),(d)) were obtained from FOB-SH trajectories run at constant temperature. 
The mean-square-displacement (MSD)  of the hole wavefunction is calculated as follows,
\begin{align}
\text{MSD}_{\alpha\beta} &= \frac{1}{N_\text{traj}}\sum_{n=1}^{N_\text{traj}} \bra{\Psi_n(t)}(\alpha - \alpha_{0,n})(\beta - \beta_{0,n})\ket{\Psi_n(t)}, \label{eq:MSD1}  \\
                            &\approx \frac{1}{N_\text{traj}}\sum_{n=1}^{N_\text{traj}} \left( \sum_{k=1}^{M}  | u_{k,n} (t) |^2 (\alpha_{k,n} (t) -  \alpha_{0,n} ) (\beta_{k,n} (t) -  \beta_{0,n} ) \right )   \label{eq:MSD2}  
\end{align}
where $\Psi_n(t)$ is the hole wavefunction in FOB-SH trajectory $n$, $\alpha\!=\!x,y$, $\beta\!=\!x,y$ are Cartesian coordinates along the crystallographic directions $a,b$,
$\alpha_{0,n} (\beta_{0,n})$ are the initial positions of the COC in trajectory $n$, $\alpha_{0,n}\!=\!\langle \Psi_n(0) | \alpha | \Psi_n (0) \rangle$, 
and $N_\text{traj}$ is the number of FOB-SH trajectories. In equation~\ref{eq:MSD2} the coordinates of the hole are discretized and replaced by the centre of mass of molecule $k$ in trajectory $n$, $\alpha_{k,n}$, and $\alpha_{0,n}\!=\!\langle \alpha_n \rangle (0)$ where $\langle \alpha_n \rangle (t)$ is the $\alpha$-coordinate of the COC at time $t$ in trajectory $n$, $\langle \alpha_n \rangle (t) \!=\! \sum_{k=1}^{M} | u_{k,n}(t) |^2 \alpha_{k,n} (t)$ and  $| u_{k,n}(t) |^2$ is the hole population of site $k$ in trajectory $n$.   
The diffusion tensor $D_{\alpha\beta}$ is given by half of the slope of MSD$_{\alpha\beta}$, 
\begin{equation}
D_{\alpha\beta} = \frac{1}{2}\lim_{t \rightarrow \infty} \frac{\text{d}\text{MSD}_{\alpha\beta}(t)}{\text{d}t}, \label{eq:Diffusion_tensor} 
\end{equation}
which allows the charge mobility $\mu_{\alpha\beta}$ to be calculated from the Einstein relation, 
\begin{equation}
\mu_{\alpha\beta} = \frac{e D_{\alpha\beta}}{k_BT}, \label{eq:Einstein_relation} 
\end{equation}
where $e$ is the elementary charge, $k_B$ is the Boltzmann constant and $T$ is temperature.
Delocalisation of the hole wavefunction $\Psi (t)$ (Fig.~2(a)) is quantified using the inverse participation ratio (IPR), 
\begin{equation}
\text{IPR}(t) = \frac{1}{N_\text{traj}}\sum_{n=1}^{N_\text{traj}} \frac{1}{\sum_{k=1}^{M}|u_{k,n}(t)|^4}, \label{eq:IPR_WFN}  
\end{equation}
where $u_{k,n}(t)$ are the expansion coefficients of $\Psi_n (t)$ in the diabatic or site basis $\phi_{k,n}$ in trajectory $n$
and $M$ is the number of rubrene molecules in the electronically active region. 
The IPR of a given eigenstate or valence band state of the electronic Hamiltonian equation~\ref{eq:Hamiltonian}, $\psi_i$ (Fig.~2(b)), 
is given by:
\begin{equation}
\text{IPR}_i(t) = \frac{1}{N_\text{traj}}\sum_{n=1}^{N_\text{traj}} \frac{1}{\sum_{k=1}^{M}|U_{ki}(t)|^4}, \label{eq:IPR_adiabat} 
\end{equation}
where $U_{ki}(t)$ are the expansion coefficients of eigenstate $\psi_i$ in the site basis $\phi_k$. 

\noindent{\bf Calculation of drift velocity, $\langle v_x \rangle$} \\
Position-resolved drift velocities along the $a$-crystallographic direction (Fig.~\ref{fig:v_s_ipr_cell}(a)) 
were obtained as follows. The COC of the hole wavefunction along $x$ (crystallographic direction $a$), 

\begin{equation}
\langle x_n \rangle (t) = \sum_{k=1}^M |u_{k,n}(t)|^2 x_{k,n} (t), \label{eq:COC_wfn} 
\end{equation}
was calculated for each FOB-SH trajectory $n$ every 2 fs. The drift velocity was calculated from the finite difference time 
derivative of the COC, $v_{x,n}(t) = [\langle x_n \rangle (t + \delta t) - \langle x_n \rangle (t)] / \delta t$
with $\delta t = 2$ fs. Position bins of length 4 unit cells (2.9 nm) along the $x$-direction were defined and a given velocity was associated with the bin containing the COC at time $t$. The velocity at each time step over all trajectories was binned this 
way and the data in each bin was averaged over giving $\langle v_x \rangle$. The same binning procedure was used for the probability density of COC (Fig.~\ref{fig:v_s_ipr_cell}(b)), obtained by counting the number of occurrences of the COC in each bin, and for the IPR of the hole wavefunction, equation~\ref{eq:IPR_WFN} (Fig.~\ref{fig:v_s_ipr_cell}(c)). 
We note that using a smaller $\delta t$ for the calculation of COC drift velocity makes little difference to the average drift velocity in the central bin compared to the statistical uncertainty. This was checked for the simulations at constant temperature and those employing both a temperature gradient and an external electric field, where the charge carrier wavefunction was printed every 0.5~fs (i.e. four times more frequently than for simulations with a temperature gradient and no external electric field). For the constant temperature simulations, using $\delta t = 0.5$~fs results in a mean drift velocity in the central position bin of $\langle v_x \rangle = 0.15$~nm$\,\text{ps}^{-1}$, whereas using $\delta t = 2$~fs results in $\langle v_x \rangle = 0.10 \pm 0.24$~nm$\,\text{ps}^{-1}$. For simulations with both a temperature gradient and an external electric field, using $\delta t = 0.5$~fs results in $\langle v_x \rangle = 0.28$~nm$\,\text{ps}^{-1}$ compared to $\langle v_x \rangle = 0.24 \pm 0.33$ using $\delta t = 2$~fs. In both cases, the values obtained using $\delta t = 0.5$~fs are well within the statistical uncertainty of the values computed using $\delta t = 2$~fs.
Details concerning the distribution of drift velocities in the case of constant temperature and with temperature gradient are given in Supplementary Note 10 and Figure~11. Convergence of drift velocity with number of trajectories and trajectory length is demonstrated in Supplementary Figure~15.

\noindent{\bf Analysis of thermoelectric motion} \\
For the analysis shown in Figure \ref{fig:NACE_av_d_n_states} (b), (d) we used 800 FOB-SH trajectories of length 2 ps 
with the temperature gradient applied. All configurations where a successful surface hop occured and the 
COC of the active adiabatic state (index $a$), $\langle x_{a} \rangle$, 
was located within the central 10 units cells (7.2 nm) along the $a$-direction were included 
($\langle x_{a} \rangle  = \sum_{l=1}^M |U_{l,a}(t)|^2 x_{l} (t)$ with $x_{l}$ the centre of mass of molecule $l$). This amounted to 
around 60,000 configurations overall. For each configuration, there are 700 adiabatic states $\psi_k$, including the active state $\psi_a$. 
Each adiabat $k \neq a$ was binned according to the difference in the COC position of that state and the COC position of the active state, 
$\Delta \text{COC}_{ka}\!=\!\langle x_{k} \rangle - \langle x_{a} \rangle$, using bin widths of 2.5 nm. The centre of each bin is denoted $x_i$
in the following.  
To account for finite thermal accessibility of states $k$ from state $a$, i.e. the fact that surface hops to adiabatic states deep inside the 
valence band may be energetically forbidden, a Boltzmann weight was assigned to each state according to its energy relative to the active 
adiabatic state energy, $E_k - E_a$, 
\begin{equation}
w^\text{B}_k = \text{min} \left[ \exp(\beta(E_k-E_a)), 1 \right]. 
\end{equation}  
Defining the number of thermally accessible states within a given bin $x_i$ as the sum of 
the Boltzmann weights of all states $k$ within this bin, 
\begin{equation}
N^\text{acc}(x_i) = \sum_{k \in x_i} w^\text{B}_k, \label{eq:N_accessible} 
\end{equation} 
%
the thermally averaged property $P_k$ over adiabatic states in bin $x_i$, $P_k\!=\! d_{ka}^{\text{ad}}, p_{ka}$, IPR$_k$, is given by
\begin{equation} 
\langle P_k \rangle^{\text{B}} (x_i) = \frac{\sum_{k \in x_i} P_k w^\text{B}_k}{ N^\text{acc}(x_i)}.
\label{eq:Boltz_av} 
\end{equation}
The NACE, Tully hopping probability and IPR thermally averaged over adiabatic states in a distance bin $x_i$, $\langle d_{ka}^{\text{ad}} \rangle^{\text{B}}$, 
$\langle p_{ka} \rangle^{\text{B}}$ and $\langle \text{IPR}_k \rangle^{\text{B}}$, respectively, are shown for all distance bins
in Figure~4(b) and Supplementary Figure~12(c) and (a). The percentage of thermally accessible states in a distance bin $x_i$, 
$N^\text{acc}(x_i) / \sum_j N^\text{acc}(x_j)\times 100\%$, is shown for all distance bins in Figure~4(d).   
Moreover, the total thermally weighted probability for a surface hop to any state in bin $x_i$, $N^\text{acc}(x_i) \langle p_{ka}\rangle^\text{B}(x_i)$, 
is shown in Supplementary Figure~12(f). 
To facilitate a like-for-like comparison with the 300 K constant temperature simulation (Fig.~\ref{fig:NACE_av_d_n_states}(c) and (e) and Supplementary Fig.~13), 
the same procedure was carried out at constant temperature of 300 K using an identical active region size of $50\times 7$ unit cells and 
identical initial conditions.  

\noindent{\bf Chemical potential contribution to Seebeck coefficient, $\alpha_c$} \\
The chemical potential of holes in the valence band is equal to the free energy change upon hole insertion into the band. It depends on 
hole density $n$ and temperature $T$. For a given reference hole density, $n^\text{ref}$, and $T$ it is given by
\begin{align}
    \mu_c^\text{ref}(T, n^\text{ref}) &= F_\text{hole}(T, n^\text{ref}) - F_\text{neutral}(T, n^\text{ref})  \\
    &= -k_B T \ln \Big \langle \sum_i^\text{vb} e^{\beta [E_i(\mathbf{R}) + E_\text{neutral}(\mathbf{R})}  \Big \rangle_{E_\text{neutral}(\mathbf{R})}^{n^{\text{ref}}},  \label{eq:mu_c_ref}
\end{align} where $F$ denotes free energy, $E_i(\mathbf{R})$ is the $i^\text{th}$  
valence band (vb) state at nuclear positions $\mathbf{R}$ (i.e. eigenstate of the electronic Hamiltonian equation~\ref{eq:Hamiltonian}, where the index $i$ runs from the top to 
the bottom of the valence band) and $E_\text{neutral} (\mathbf{R})$ is the energy of the neutral system at nuclear positions $\mathbf{R}$. 
The brackets denote taking the ensemble average over configurations sampled from a molecular dynamics trajectory of the neutral system at carrier density $n^\text{ref}$. 
Note that $E_i$ are electronic energy levels, i.e., $E_i$ decreases with increasing hole excitation energy (see also Fig.~\ref{fig:RUB_TEMP}(b)), 
thus the positive sign in the exponent of the Boltzmann weight. A derivation of equation~\ref{eq:mu_c_ref} is presented in Supplementary Note 13. 
The chemical potential at a general carrier density $n = 1 / A$, where $A$ is the area of the electronically active region within the $a\!-\!b$ plane, is given by
\begin{equation}
\mu_c(T, n) = \mu_c^\text{ref}(T, n^\text{ref}) + k_B T \ln\frac{n}{n^\text{ref}}. \label{eq:mu_c} 
\end{equation}
The chemical potential contribution to the Seebeck coefficient, that is the second term on the RHS of equation~\ref{eq:alpha_FOBSH_contributions}, is then given by
\begin{equation}
\alpha_c = -\frac{1}{q}\frac{\partial \mu_c}{\partial T} = -\frac{1}{q}\Bigg[ \frac{\partial \mu_c^\text{ref}}{\partial T} + k_B \ln\frac{n}{n^\text{ref}}\Bigg]. \label{eq:alpha_c} 
\end{equation}
The first term on the RHS of equation~\ref{eq:alpha_c}, can be obtained by calculating $\mu_c^\text{ref}(T, n^\text{ref})$ at different temperatures around 300 K and taking the slope. 
Details on these calculations are presented in Supplementary Note 13 where we also show that the numerical results are virtually independent of the chosen reference concentration ($n^\text{ref}$), 
Supplementary Table~S8 and Figure~S16. 

\noindent{\bf Experimental Details} \\
Rubrene single crystals were grown via physical vapour transport under Ar flow from $\geq$ 98~\% pure rubrene powder (Sigma Aldrich – used as purchased). The devices were made on \SI{175}{\micro\metre} PET (Melinex\textsuperscript{$\circledR$} ST504, DuPont Teijin Films). After sonic cleaning in acetone and isopropyl alcohol, the gate and heater electrodes were deposited via shadow mask with a 3~nm Cr adhesion layer followed by 20~nm Au. The gate dielectric is formed of a 500~nm layer of CYTOP, deposited by spin coating followed by thermal annealing at \SI{90}{\degree C}. The source and drain contacts (also 20~nm Au) were then evaporated, followed by manual placement of the grown single crystals aligning along the high mobility direction. 

The device architecture and the Seebeck measurement are carried out in the similar way as our previous study\cite{staz2020seebeck}.
All mobility and Seebeck measurements were performed in a Lake Shore CRX-4K cryogenic probe station using Keithley SMU models 2612B, 6430 and 2182 nanovoltmeters. For more details on the Seebeck measurement see SI.

\subsection*{Author Contributions}
Conceptualization: JE, HS, JB \\
Software: AD, SG \\
Methodology: JE, YX, HS, JB \\
Investigation: JE, YX, EDG, FI \\
Visualization: JE, EDG \\
Supervision: HS, JB \\
Writing—original draft: JE, JB \\
Writing—review \& editing: JE, YX, EDG, FI, AD, SG, HS, JB


\subsection*{Data and code availability}
The full data for this study total several terabytes and are in cold storage accessible by the corresponding authors and available upon reasonable request. The custom FOB-SH code for non-adiabatic molecular dynamics, the python code used for the analysis and other post-processing tools used for this study are available from the corresponding authors upon request.

\begin{acknowledgement}
J.E. was supported by a departmental Ph.D. studentship, F.I., A.D. were supported by EPSRC DTP studentships (EP/W524335/1)
and S.G. was supported by the European Research Council (ERC) under the European Union, Horizon 2020 research 
and innovation programme (grant agreement no. 682539/SOFTCHARGE). 
Via our membership of the UK's HEC Materials Chemistry Consortium, which is funded by EPSRC (EP/L000202, EP/R029431), 
this work used the ARCHER UK National Supercomputing Service (http://www.archer.ac.uk) as well as the UK Materials and Molecular 
Modelling (MMM) Hub, which is partially funded by EPSRC (EP/P020194). We also acknowledge the use of the UCL Kathleen High 
Performance Computing Facility. Y.X. acknowledges support from the Cambridge Commonwealth, European \& International Trust 
and Chinese Scholarship Council. E.D.G. was supported by an EPSRC DTP Studentship provided by the Department of Physics, 
University of Cambridge. For the experimental work we acknowledge financial support from the Royal Society (RP/R1/201082), the European Research Council (101020872) and the Engineering and Physical Sciences Research Council (EP/W017091/1).

\end{acknowledgement}

\nocite{wang2004development, bulgarovskaya1983, ziogos2021hab79, nematiaram2020modeling, ciuchi2011transient, nematiaram2019practical, chui1992temperature}

\vspace{1cm}

\providecommand{\latin}[1]{#1}
\makeatletter
\providecommand{\doi}
  {\begingroup\let\do\@makeother\dospecials
  \catcode`\{=1 \catcode`\}=2 \doi@aux}
\providecommand{\doi@aux}[1]{\endgroup\texttt{#1}}
\makeatother
\providecommand*\mcitethebibliography{\thebibliography}
\csname @ifundefined\endcsname{endmcitethebibliography}  {\let\endmcitethebibliography\endthebibliography}{}

\newpage
\newcommand{\SItitlepage}{
    \clearpage
    \begin{center}
        \vspace*{1cm}
        {\titlefont \LARGE Supplementary Information: \\ Thermoelectric transport in molecular crystals driven by gradients of thermal electronic disorder}\\[1.5cm]
        {\authorfont \large Jan Elsner,\textsuperscript{\dag} Yucheng Xu,\textsuperscript{\ddag} Elliot D. Goldberg,\textsuperscript{\ddag} Filip Ivanovic,\textsuperscript{\dag} Aaron Dines,\textsuperscript{\dag} Samuele Giannini,\textsuperscript{\dag,\P} Henning Sirringhaus,\textsuperscript{\ddag} and Jochen Blumberger\textsuperscript{*,\dag}}\\[0.5cm]
        {\affilfont \normalsize
        \textsuperscript{\textnormal{\dag}}Department of Physics and Astronomy and Thomas Young Centre, University College London, London WC1E 6BT, UK.\\
        \textsuperscript{\textnormal{\ddag}}University of Cambridge, Cavendish Laboratory, Cambridge, UK CB3 0HE, UK.\\
        \textsuperscript{\textnormal{\P}}Institute for the Chemistry of OrganoMetallic Compounds, National Research Council (ICCOM-CNR), I-56124 Pisa, Italy.\\}
        {\authorfont Email: j.blumberger@ucl.ac.uk
        }
    \end{center}
    \newpage
}

\newcommand{\beginsupplement}{
    \setcounter{section}{0}
    \renewcommand{\thesection}{S\arabic{section}}
    \setcounter{table}{0}
    \renewcommand{\thetable}{S\arabic{table}}
    \setcounter{figure}{0}
    \renewcommand{\thefigure}{S\arabic{figure}}
    \setcounter{equation}{0}
    \renewcommand{\theequation}{S\arabic{equation}}
}

\beginsupplement
\SItitlepage

\section{Force field parameterization} \label{section:FF_param}

\begin{figure}[!htb]
\centering
\includegraphics[width=1.0\textwidth]{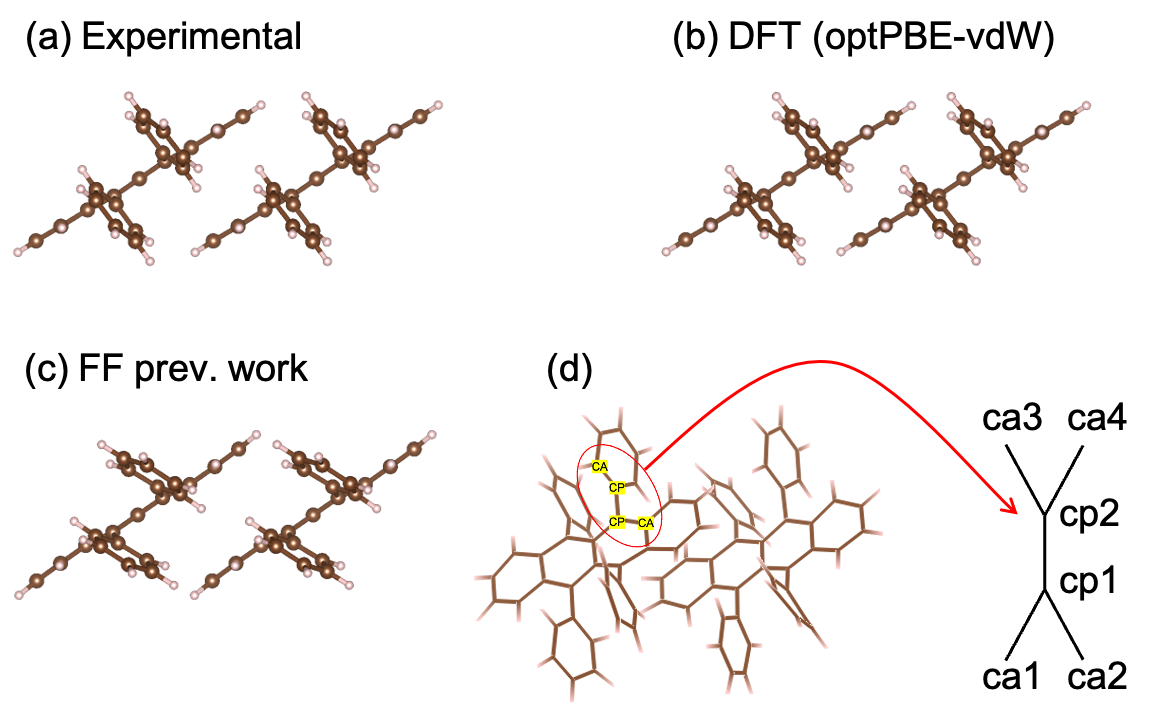}
  \caption{(a) Experimental dimer, (b) dimer extracted from a unit cell optimised with DFT using the optPBE-vdW density functional and (c) dimer extracted from a unit cell optimised using the previously employed force field. Optimisation using the latter results in a misorientation of phenyl side groups relative to the tetractene backbone, quantified by the pertinent connecting dihedral angles illustrated in (d). }
  \label{fig:FF_dimers}
\end{figure}

FOB-SH simulations employ force fields for the calculation of diagonal Hamiltonian matrix elements (i.e. site energies). The matrix element $H_{kk}$ corresponds to the energy of the system with molecule $k$ charged and all other molecules neutral. The force field for the neutral system is based on the general AMBER force field (GAFF) \cite{wang2004development} and parameters for the charged state were obtained by displacing the equilibrium bond lengths of the molecule with respect to the neutral state in order to reproduce the DFT reorganization energy, $\lambda = 0.152$~eV (see refs\cite{giannini2019quantum, giannini2020flickering} for details). The 300 K distributions of electronic couplings obtained using the force field for rubrene employed in previous work\cite{giannini2019quantum, giannini2020flickering} deviate slightly from those obtained using ab initio molecular dynamics with the optPBE-vdW density functional\cite{elsner2021mechanoelectric} (Prev. FF in Fig. 1 of the main text). This discrepancy stems from the use of an incorrect dihedral angle term linking phenyl side chains to the tetracene backbone such that parameters were set to match the dihedral terms describing in-plane interactions.
This resulted in a slight misalignment of the phenyl side chains of optimized structures, visually apparent in Figure~\ref{fig:FF_dimers}, which shows dimers extracted from optimized unit cells using the previously employed FF and optPBE-vdW DFT, as well as an experimental structure (CCDC database identifier QQQCIG01) \cite{bulgarovskaya1983}. While this oversight has little effect on the conclusions drawn in previous studies, we have now updated the relevant dihedral terms so that the dihedral angles of optimized structures and the 300 K distributions of electronic couplings obtained from force field molecular dynamics better reproduce those obtained from ab-initio molecular dynamics.

The orientation of the phenyl side chains relative to the tetracene backbone can be quantified through the relevant dihedral angles, as indicated in Figure \ref{fig:FF_dimers}(d). For each phenyl side chain, there are 4 dihedral angles to consider: $\phi_1 = \phi(\text{ca1-cp1-cp2-ca3})$, $\phi_2 = \phi(\text{ca1-cp1-cp2-ca4})$, $\phi_3 = \phi(\text{ca2-cp1-cp2-ca3})$ and $\phi_4 = \phi(\text{ca2-cp1-cp2-ca4})$. The dihedral energy term has the form

\begin{equation}
E_\phi(\phi) = K_\phi (1 + \cos{(2\phi - \phi_0)}), \label{eq:dihedral}  
\end{equation}
where the parameters used previously are $K_\phi = 3.625$ kcal/mol/rad$^2$ and $\phi_0 = 180$ deg. 
This has an energy minimum at $\phi = 0$ deg, resulting in incorrect geometries for this system and an underestimation of the $a$ direction electronic coupling, $J_a$. Table~\ref{table:dihedrals} lists dihedral angles $\phi_1$, $\phi_2$, $\phi_3$ and $\phi_4$, of the experimental structure and of 0 K optimised structures using DFT and force fields with different values of $K_\phi$. Optimisation using the previously employed force field (Prev. FF in Table~\ref{table:dihedrals}) yields dihedral angles that are either too large or too small. Setting $K_\phi = 0$ (i.e. turning off entirely) for dihedral terms linking the atoms depicted in Figure~\ref{fig:FF_dimers}(d) results in much better agreement with optPBE-vdW and the experimental structure (FF$_0$ in Table~\ref{table:dihedrals}). Further improvement can be obtained by using a small value of $K_\phi = 0.6000$ kcal/mol/rad$^2$. We refer to this as FF$_\text{opt}$ in Table~\ref{table:dihedrals}, which is the force field we have used throughout the present study. We note that the parameters for all other dihedral angles (i.e. those which do not connect the tetracene backbone to the phenyl side chains) and all other interactions are the same as in previous work\cite{giannini2019quantum, giannini2020flickering}.

\begin{table}
  \caption{ Dihedral angles for the experimental dimers and dimers extracted from optimised unit cells using potential energy surfaces. }
  \label{table:dihedrals}
  \begin{tabular}{lllllll}
      \hline
     & $K_\phi^a$ & $\phi_0^a$ & $\phi_1^b$ & $\phi_2^b$ & $\phi_3^b$  & $\phi_4^b$ \\
    \hline
    Experimental & & & 75.9 & 111.8 & 96.6 & 75.7 \\
    optPBE-vdW & & & 78.0 & 109.6 & 94.8 & 77.6 \\
    Prev. FF & 3.625 & 180 & 49.5 & 129.1 & 116.4 & 65.0 \\
    FF$_0$ & 0.000 & 180 & 79.6 & 108.1 & 92.9 & 79.3 \\
    FF$_\text{opt}$ & 0.600 & 180 & 77.2 & 110.1 & 94.6 & 78.1 \\

    \hline
  \end{tabular}
  \begin{flushleft} 
 $^a$ Parameters for the dihedral energy term corresponding to atoms in Fig. \ref{fig:FF_dimers}(d), Eq. \ref{eq:dihedral}, for force field optimizations, in kcal/mol/rad$^2$. \\
 $^b$ Dihedral angles for the atoms shown in Fig. \ref{fig:FF_dimers}(d), in degrees.
\end{flushleft}
\end{table}

\begin{table}
  \caption{ Summary of electronic couplings and timescales of electronic coupling fluctuations using DFT and the different force fields described in the text. }
\label{table:coupling_distr_optimise_dih}
  \begin{tabular}{llllllllll}
      \hline
     & $K_\phi$ & $J_a^a$ & $\langle J_a \rangle^b$  & $\sigma_a^b$ & $J_b^a$ & $\langle J_b \rangle^b$ & $\sigma_b^b$ & $\tau_a^c$ & $\tau_b^c$ \\
    \hline
    optPBE-vdW/sPOD & & 110.2 & 107.1 & 28.2 & -20.0 & -17.0 & 7.7 & 78.5 & 96.3 \\
    Prev. FF/AOM & 3.625 & 73.4 & 82.3 & 31.8 & -18.5 & -15.7 & 7.6 & 43.0 & 70.0  \\
    FF$_0$/AOM & 0.000 & 108.8 & 111.1 & 31.0 & -22.6 & -16.0 & 9.0 & 64.8 & 79.9  \\
    FF$_\text{opt}$/AOM & 0.6000 & 106.8 & 108.3 & 29.5 & -21.9 & -15.8 & 7.3 & 66.0 & 82.3 \\
    \hline
  \end{tabular}
\begin{flushleft} 
 $^a$ Electronic couplings of dimers taken from optimized cells along the $a$ and $b$ crystallographic directions, $J_a$ and $J_b$, respectively, in meV. \\
 $^b$ Mean $\langle J_{a(b)} \rangle$ and root-mean-square fluctuations $\sigma_{a(b)}$ of the distribution of electronic couplings obtained by sampling dimers from a molecular dynamics trajectory at 300 K, in meV. \\
 $^c$ Average timescale of coupling fluctuations, $\tau_{a(b)} \!=\!\int \text{d}\omega \, \omega (S_{a(b)}(\omega)/\omega) / \int \text{d}\omega (S_{a(b)}(\omega)/\omega)$, in cm$^{-1}$.    
\end{flushleft}
\end{table}

Electronic couplings of 0K optimized structures, as well as mean and root-mean-square fluctuations of the 300 K distributions, using DFT and the various force fields are listed in Table~\ref{table:coupling_distr_optimise_dih}. The distributions of electronic couplings were obtained by sampling dimers every 50 fs from molecular dynamics trajectories of length 15 ps (timestep 1 fs) at 300 K. The force field used in previous work (Prev. FF) underestimates $J_a$ for the optimized structure, as well as the mean value of the 300 K distribution, $\langle J_a \rangle$. This is remedied by reducing the force constant of the relevant dihedral terms. Best agreement with respect to ab initio molecular dynamics is obtained using $K_\phi = 0.6000$ kcal/mol/rad$^2$ (FF$_\text{opt}$). The timescales of electronic coupling fluctuations, $\tau_a$ and $\tau_b$ in Table~\ref{table:coupling_distr_optimise_dih}, are also much improved using FF$_\text{opt}$; see also Figure~1 of the main text.

Electronic couplings for structures obtained from optPBE-vdW MD were calculated using the projector operator-based diabatization (POD) method \cite{futera2017electronic} in combination with the PBE density functional and uniform scaling by 1.325 (sPOD) \cite{ziogos2021hab79}, as outlined in Ref.~\citenum{elsner2021mechanoelectric}. For structures obtained from force field MD, the analytic overlap method (AOM) \cite{gajdos2014ultrafast, ziogos2021ultrafast} was used, as employed in FOB-SH simulations.

 \clearpage
 
\section{Detailed description of results in Figure 1 main text}
Figure~1(a) in the main text shows the distributions of electronic couplings $J_a$ and $J_b$ over trajectories at 300 K obtained with three different methods: ab-initio MD with the optPBE-vdW functional, classical MD using the force field employed in previous studies\cite{giannini2019quantum, giannini2020flickering} and classical MD using the optimized force field employed in this work, FFopt (SI section 1). In the first case, we utilise the DFT-based sPOD method for calculation of electronic couplings\cite{futera2017electronic}, whereas in the latter two cases we utilise the analytic overlap method (AOM), as employed in FOB-SH \cite{ziogos2021ultrafast}. We note significantly better agreement in the distributions of electronic couplings when using the optimized force field FFopt compared to the force field used previously (see SI section 1). 

Figure~1(b) shows the normalised density of states (DOS) obtained using the different methods. The DFT/PBE density of states of the optimized structure is shown in the black dashed line. We note that the bandwidth has been scaled by a factor of 1.325 (hence the label sDFT) due to a tendency for the PBE functional to underestimate electronic couplings \cite{ziogos2021hab79}. The energy of the top of the valence band was set to 0. All other lines are obtained using the valence band Hamiltonian (Eq.~4 in the main text) with matrix elements sampled from the calculated 
distributions of electronic couplings. Dashed lines are for optimized structures with no electronic disorder i.e. all diagonal elements set to zero and the distributions of $J_a$ and $J_b$ are given by delta functions. Solid lines are for averages over 50 Hamiltonians with different realisations of disorder, obtained by sampling the 300 K distributions of electronic couplings. In each case, the peak of the DOS was aligned in energy with the peak of the sDFT/PBE DOS at -0.499 eV relative to the top of the valence band. We find that our valence band Hamiltonians achieve good accuracy compared to the full DFT DOS for the optimized structures, indicating that such valence band Hamiltonian accurately describes the valence band electronic structure \cite{nematiaram2020modeling}. Excellent agreement is achieved with the FFopt Hamiltonians compared to the optPBE-vdW Hamiltonians (red vs cyan). The effect of finite temperature is to smear out density of states due to electronic disorder i.e. the spread in the electronic couplings matrix elements, as indicated by the solid compared to dashed lines. 

Figure~1(c) and (d) show the spectral density functions of the electronic coupling time series for $J_a$ and $J_b$. The running integral of the spectral density yields the cumulative disorder including all frequencies up to $\omega$, $\sigma_{\alpha}(\omega)$, allowing us to quantify the relative contribution of each mode, $\sigma_{\alpha}(\omega) = \Big[ \frac{8}{\beta \pi} \int_{0}^{\omega} d\omega' \,\,  \frac{S_{\alpha}(\omega')}{\omega'} \Big]^{\frac{1}{2}}$, where $S_{\alpha}(\omega)$ is the spectral density for time series $J_{\alpha}$ ($\alpha = a, b$) and $\beta=k_BT$. Including all frequencies $\omega \rightarrow \infty$ returns the root-mean-square fluctuation of the time series, which we note is similar using FFopt dynamics compared to optPBE-vdW dynamics (red vs cyan). 
 
\clearpage
 
\section{Temperature dependence of lattice parameters} \label{section:T_dependent_lattice_params}

\begin{figure}[!htb]
\centering
\includegraphics[width=1.0\textwidth]{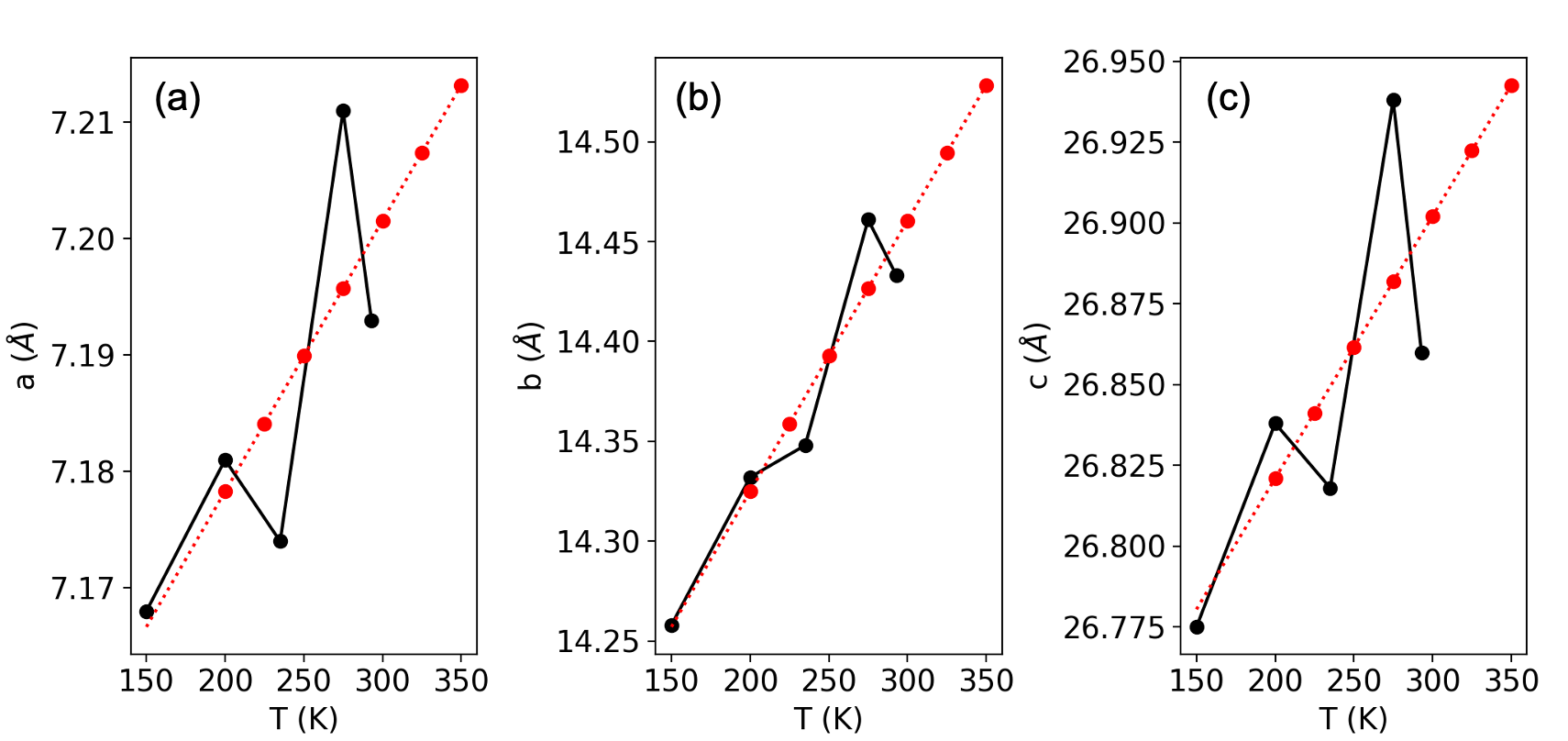}
  \caption{Experimental temperature-dependent lattice parameters of rubrene for the $a$, $b$ and $c$ directions from Ref.~\citenum{jurchescu2006low} in black. Best fit lines are shown in red (dashed) and values employed in FOB-SH simulations are marked by red circles.}
  \label{fig:lattice_params}
\end{figure}

All FOB-SH simulations were carried out at fixed cell volume. In the case of the constant temperature simulations over a range of temperatures, thermal expansion of the lattice was taken into account using a linear interpolation of the experimental temperature-dependent lattice parameters \cite{jurchescu2006low}. The best fit line was extrapolated to yield lattice parameters for the higher temperatures (325 K and 350 K) not measured by experiment. For simulations under a temperature gradient from 250 K to 350 K centred at 300 K, linear expansion with temperature was assumed and the 300 K lattice parameters were used. 
Figure \ref{fig:lattice_params} shows the experimental lattice parameters along the $a$, $b$ and $c$ crystallographic directions (black lines), with linear fits in red dashed lines. The temperature-dependent lattice parameters employed in FOB-SH simulations are marked by red circles. Numerical values are listed in Table~\ref{table:lattice_params}.

\begin{table}[H]
  \caption{ Temperature dependent lattice parameters of rubrene from experiment \cite{jurchescu2006low} and those employed in FOB-SH using a linear fit to the experimental values. }
  \label{table:lattice_params}
  \begin{tabular}{lllll}
    \hline
    Temperature (K) &  & $a$ (\AA) & $b$ (\AA) & $c$ (\AA) \\
    \hline
    150 & Exp & 7.168 & 14.258 & 26.775 \\
    200 & Exp & 7.181 & 14.332 & 26.838 \\
    235 & Exp & 7.174 & 14.348 & 26.818 \\
    275 & Exp & 7.211 & 14.461 & 26.938 \\
    293 & Exp & 7.193 & 14.433 & 26.860 \\
    \hline
    200 & Fit & 7.178 & 14.325 & 26.821 \\
    225 & Fit & 7.184 & 14.359 & 26.841 \\
    250 & Fit & 7.190 & 14.393 & 26.862 \\
    275 & Fit & 7.196 & 14.427 & 26.882 \\
    300 & Fit & 7.202 & 14.461 & 26.902 \\
    325 & Fit & 7.207 & 14.494 & 26.922 \\
    350 & Fit & 7.213 & 14.528 & 26.943 \\
    \hline
  \end{tabular}
\end{table}

 \clearpage
 
\section{Convergence of hole mobility with respect to system size} \label{section:Supercell_convergence}

\begin{figure}[!htb]
\centering
\includegraphics[width=0.5\textwidth]{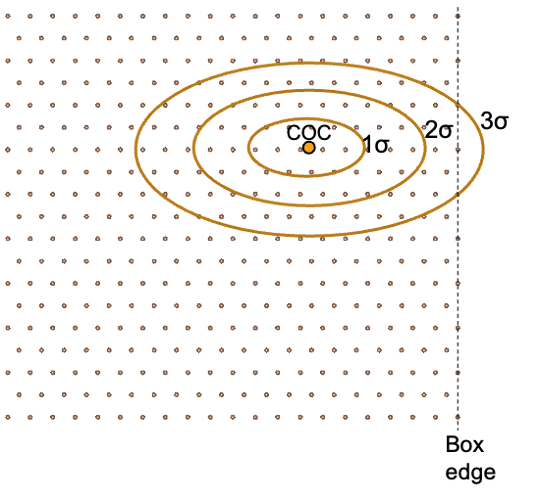}
  \caption{ Illustration of how hits are counted. If the position of the centre-of-charge (COC) comes within $d_\text{hit} = n \sigma$ of the boundary, a hit is counted. The fraction of trajectories containing a hit gives an indication of whether the cell size is large enough. Too many hits indicate that the charge is significantly restricted by the boundary. In the present illustration a hit would be counted using $n \ge 3$. }
  \label{fig:hits}
\end{figure}

Due to the increase in diffusivity with decreasing temperature, simulations at the lowest temperature, $T = 200$ K, are the hardest to converge with FOB-SH active region size. FOB-SH runs at different supercell sizes are required to ensure that mobility is converged with system size. If the cell is too small in a particular direction, the charge carrier wavefunction will be restricted in that direction once it reaches the vicinity of the boundary, leading to an underestimation in the slope of mean-squared-displacement with time. We can quantify the boundary effect by counting the fraction of trajectories where the wavefunction `hits' the far boundary. A `hit' refers to the centre-of-charge of the wavefunction coming within some defined distance $d_\text{hit}$ of the boundary. This is illustrated in Figure \ref{fig:hits}. 

We use two definitions of $d_\text{hit}$. In the first case, we define $d^{n}_\text{av, x}$ to be $n\times$ the standard deviation of the wavefunction projected along the~$x$ direction, averaged over all time steps: 

\begin{equation}
d^{n}_{\text{av}, x} = n \bar{\sigma}_x = n \bigg \langle \sqrt{\langle x^2 \rangle - \langle x \rangle^2} 
\bigg \rangle_{t} = n \Bigg \langle \sqrt{\sum_{k}|u_k|^2x_k^2 - \Big( \sum_{k}|u_k|^2 x_k \Big )^2 } 
\Bigg \rangle_{t}, \label{eq:d_hit_av}    
\end{equation}
where $u_k$ are the wavefunction coefficients in the diabatic (site) basis, $x_k$ denotes the position of site $k$ along direction $x$ and the average is taken over all time steps. An alternative choice is to define $d^{n}_{\text{inst}, x}(t)$ to be $n\times$ the instantaneous standard deviation of the wavefunction at time $t$, projected along the $x$ direction.

\begin{equation}
 d^{n}_{\text{inst}, x}(t) = n \sigma_x(t) = n  \sqrt{\langle x(t)^2 \rangle - \langle x(t) \rangle^2}  = n  \sqrt{\sum_{k}|u_k(t)|^2x_k^2(t) - \Big( \sum_{k}|u_k(t)|^2 x_k(t) \Big )^2 }. \label{eq:d_hit_inst}   
\end{equation}

Equation \ref{eq:d_hit_inst} provides a more stringent criterion than equation \ref{eq:d_hit_av} in the scenario where the wavefunction is located close to the boundary and undergoing a transient delocalisation event, $\sigma_x(t) \gg \bar{\sigma}_x$. In this case, a hit may be counted by equation~\ref{eq:d_hit_inst} but not by equation~\ref{eq:d_hit_av}. 

The rubrene unit cell comprises four molecules (280 atoms) and spans two distinct high-mobility $a-b$ planes perpendicular to the out-of-plane $c$ direction. The overall 3D periodic MD cell used in FOB-SH simulations includes both $a-b$ layers to ensure structural integrity, however the FOB-SH active region (where the FOB-SH electronic Hamiltonian is defined) includes only a single high-mobility layer, representing half the unit cell along the $c$ direction. This is justified by the fact that electronic couplings along the $c$ direction are orders of magnitude smaller than electronic couplings within the $a-b$ plane. In the following we denote supercell sizes by $N_a \times N_b$ and omit the dimension along the c-axis which is always $1/2$ except where indicated otherwise. 

For simulations at 200 K, we considered four supercell sizes with FOB-SH active regions of $32\times18$, $41\times14$, $50\times13$ and $54\times13$ unit cells. We note that while the overall MD cell is 3D periodic, the FOB-SH active region which defines the FOB-SH Hamiltonian is not. 
For simulations with the $32\times18$ active cell, a nuclear time step of 0.1 fs was used, whereas a smaller time step of 0.05 fs was used for all other cells. Such small nuclear time steps are necessary to avoid trivial crossings \cite{carof2019calculate}. The percentage of hits within 900 fs along the $a$ and $b$ directions are listed in Table~\ref{table:hits_200K_av_std} and Table \ref{table:hits_200K_inst_std} using the definitions in equations \ref{eq:d_hit_av} and \ref{eq:d_hit_inst}, respectively, for different choices of~$n$.

\begin{table}
  \caption{ 
Percentage of hits in the $a$ and $b$ directions for FOB-SH simulations of length 900 fs at 200 K using different active cell sizes. $N_\text{traj}$ denotes the total number of FOB-SH trajectories. Hits are counted using $d^{n}_{\text{av}, x}$ i.e. equation \ref{eq:d_hit_av}. }
  \label{table:hits_200K_av_std}
  \begin{tabular}{lllllllll}
    \hline
    \multicolumn{3}{c}{} & \multicolumn{6}{c}{\% hits} \\
    \cline{4-9}
    Active cell & Dimensions $(\text{nm} \times \text{nm})$ & N$_{\text{traj}}$ &  $d^{n=1}_{\text{av}, a}$ & $d^{n=1}_{\text{av}, b}$ &  $d^{n=2}_{\text{av}, a}$ &   $d^{n=2}_{\text{av}, b}$ &
    $d^{n=3}_{\text{av}, a}$ & $d^{n=3}_{\text{av}, b}$ \\
    \hline
    $32\times 18$ & $22.98 \times 25.80$ & 199 & 3.0 & 0 & 20.6 & 0.5 & 33.2 & 0.5 \\
    $41\times 14$ & $29.44 \times 20.06$ & 396 & 2.5 & 0 & 6.6 & 0.5 & 11.9 & 0.8 \\
    $50\times 13$ & $35.91 \times 18.63$ & 419 & 1.0 & 1.4 & 3.3 & 3.6 & 5.5 & 5.3 \\
    $54\times 13$ & $38.78 \times 18.63$ & 687 & 0.6 & 0.4 & 1.9 & 1.5 & 2.9 & 3.6 \\
    \hline
  \end{tabular}
\end{table}

\begin{table}
  \caption{Percentage of hits in the $a$ and $b$ directions for FOB-SH simulations of length 900 fs at 200 K using different active cell sizes. $N_\text{traj}$ denotes the total number of FOB-SH trajectories. Hits are counted using $d^{n}_{\text{inst}, x}(t)$ i.e. equation \ref{eq:d_hit_inst}. }
  \label{table:hits_200K_inst_std}
  \begin{tabular}{lllllllll}
      \hline
      \multicolumn{3}{c}{} & \multicolumn{6}{c}{\% hits} \\
    \cline{4-9}
    Active cell & Dimensions $(\text{nm} \times \text{nm})$ & N$_{\text{traj}}$ &   $d^{n=1}_{\text{inst}, a}$ &  $d^{n=1}_{\text{inst}, b}$ & $d^{n=2}_{\text{inst}, a}$ & $d^{n=2}_{\text{inst}, b}$ & $d^{n=3}_{\text{inst}, a}$ & $d^{n=3}_{\text{inst}, b}$ \\
    \hline
    $32\times 18$ & $22.98 \times 25.80$ & 199 & 3.0 & 0 & 34.2 & 0.5 & 59.3 & 1.0 \\
    $41\times 14$ & $29.44 \times 20.06$ & 396 & 1.5 & 0 & 12.6 & 1.0 & 25.3 & 5.1 \\
    $50\times 13$ &  $35.91 \times 18.63$ & 419 & 1.2 & 1.9 & 6.9 & 6.9 & 17.4 & 15.5 \\
    $54\times 13$ & $38.78 \times 18.63$ & 687 & 0.9 & 0.9 & 3.3 & 4.2 & 9.6 & 11.2 \\
    \hline
  \end{tabular}
\end{table}

Simulations with the $32\times18$ cell result in many more hits along the $a$ direction compared to the $b$ direction, indicating that the anisotropy of the cell is not optimal and that the $a$ direction length should be increased, while the $b$ direction length may safely be reduced. This is ameliorated to some extent with the $41\times14$ cell, however there are still significantly more hits along $a$ compared to along $b$. The  $50\times13$ cell yields roughly the same number of hits in each direction, and this remains the case for the $54\times13$ cell, which further reduces the percentage of trajectories which hit the boundary. Such analysis is helpful for a sense of the optimal cell anisotropy, however convergence of the mean-squared-displacement with cell size must be checked explicitly. 

\begin{figure}[!htb]
\centering
\includegraphics[width=1.0\textwidth]{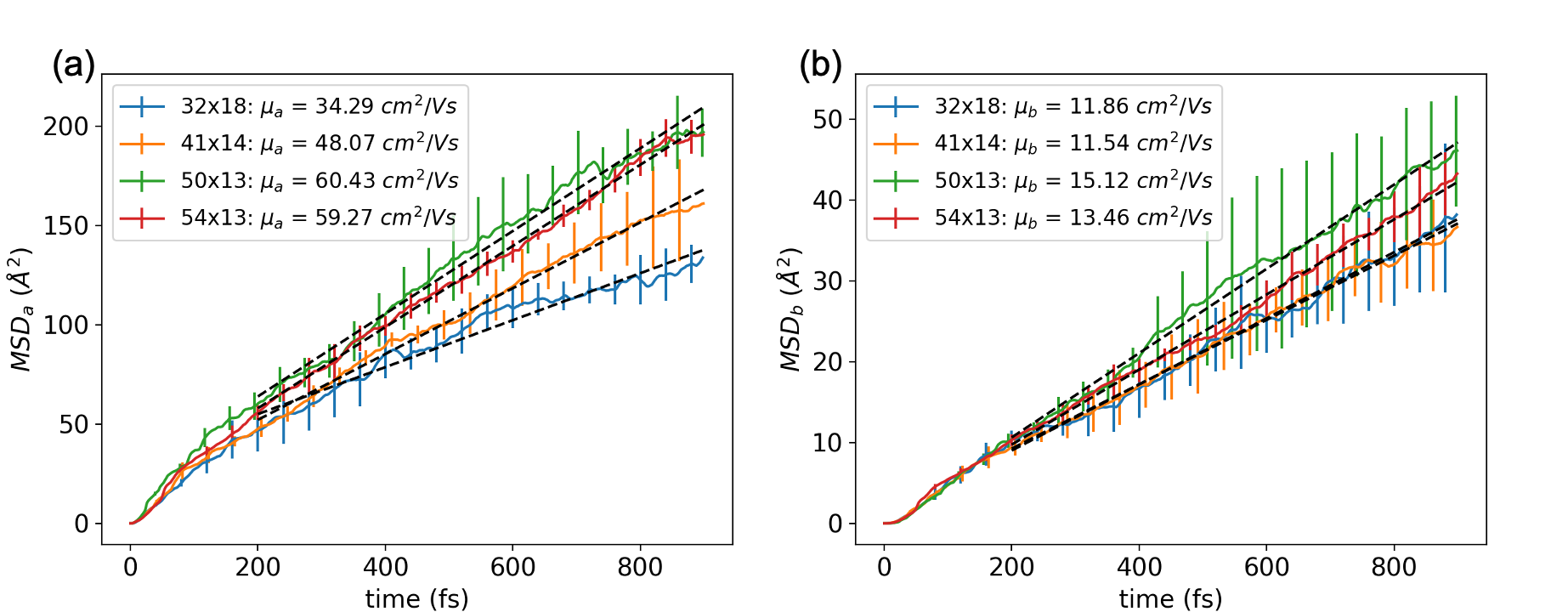}
  \caption{ Mean-squared-displacement along (a) the $a$ direction and (b) the $b$ direction at 200 K against time for different active cell sizes. 
  The good agreement for simulations with an active cell of $50\times13$ and $54\times13$ indicates that the supercell is converged. }
  \label{fig:MSD_conv_200K}
\end{figure}

Figure \ref{fig:MSD_conv_200K} shows the mean-squared-displacement along the $a$ direction against time for the different cell sizes. Error bars are calculated by dividing the trajectories into 5 blocks, and taking the standard deviation over the block averages. It is apparent from Figure \ref{fig:MSD_conv_200K}(a) that cells which are smaller in the high mobility ($a$) direction result in smaller slopes for mean-squared-displacement with time, and hence smaller mobilities. The $50\times13$ cell is converged along the $a$ direction with respect to the larger $54\times13$ cell. The $54\times13$ cell was used for temperatures between 200 K and 250 K. For higher temperatures, the $50 \times 13$ cell was used.

 \clearpage
 
\section{Temperature dependence of MSD and hole mobility}
\label{section:MSD_time}

\begin{figure}[!htb]
\centering
\includegraphics[width=1.0\textwidth]{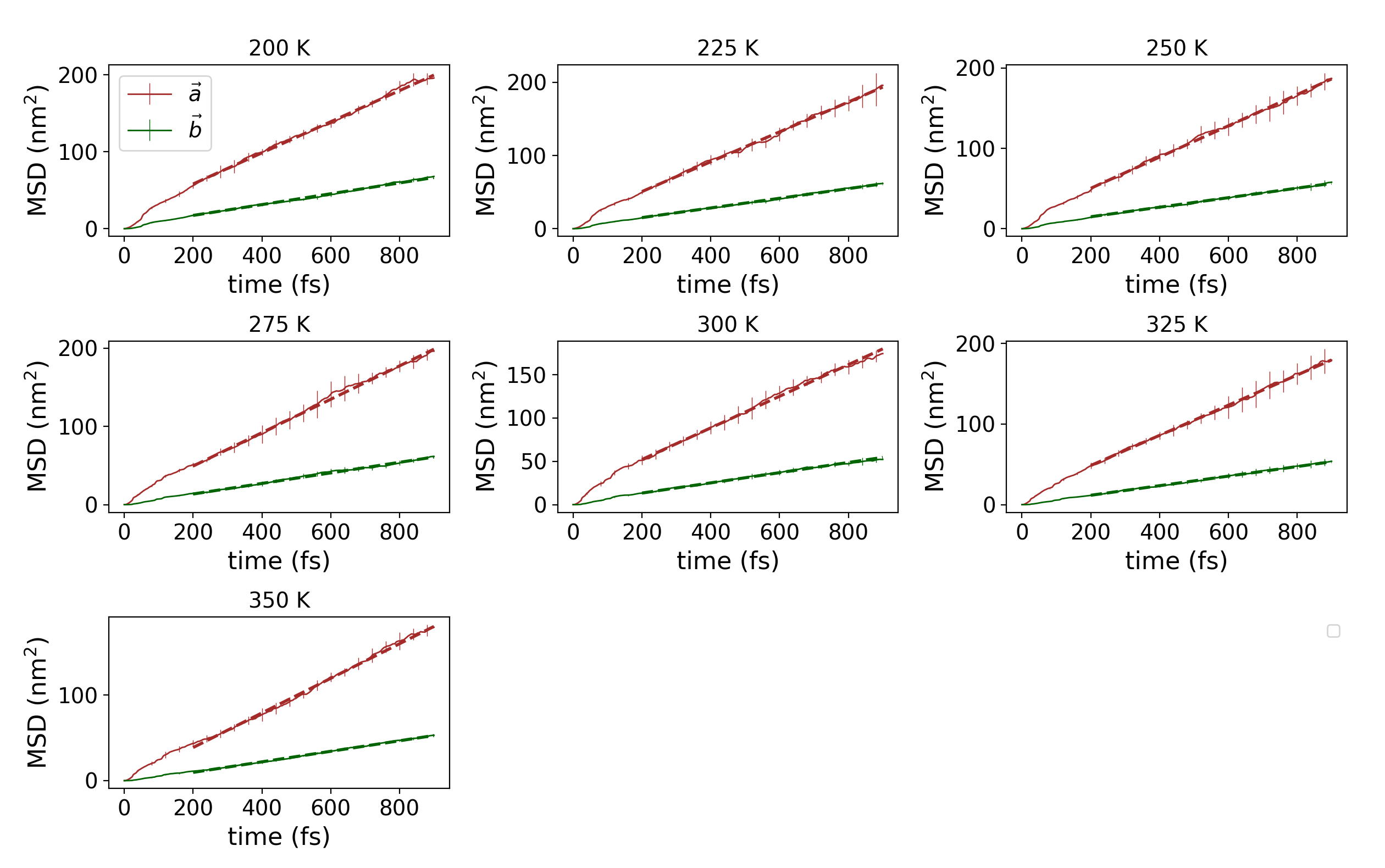}
  \caption{ Mean-squared-displacement against time for all temperatures. Error bars indicate the standard deviation over 5 block averages. In all cases, MSD vs time is linear, indicating diffusive behaviour. }
  \label{fig:MSD_all}
\end{figure}

Figure \ref{fig:MSD_all} shows converged plots of MSD against time for all constant temperature simulations. Error bars were calculated by partitioning the total number of trajectories into 5 blocks, and taking the standard deviation of the block averages. The first 200 fs were neglected in the fit, since the initial part of each trajectory corresponds to quantum relaxation from the initial diabatic state. The active region size, total number of trajectories and resulting values for 
mobility, are listed in Table~\ref{table:mobility_vs_temp}. 

\begin{table}[H]
  \caption{FOB-SH mobilities and standard error along the $a$ and $b$ crystallographic directions of rubrene, for simulations at temperatures between 200 K and 350 K. 
  The active region size, as well as number of trajectories is indicated. }
  \label{table:mobility_vs_temp}
  \begin{tabular}{lllll}
      \hline
    Temperature (K) & Active region & N$_{\text{traj}}$ & $\mu_a$ $(\text{cm}^2/Vs)$&  $\mu_b$ $(\text{cm}^2/Vs)$\\
    \hline
    200 & $54 \times 13$ & 699 & 59.3 $\pm$ 13.1 & 13.2 $\pm$ 0.9\\
    225 & $54 \times 13$ & 698 & 51.5 $\pm$ 4.6 & 12.0 $\pm$ 0.4\\
    250 & $54 \times 13$ & 699 & 44.1 $\pm$ 5.6 & 8.4 $\pm$ 1.2\\
    275 & $50 \times 13$ & 690 & 44.9 $\pm$ 3.9 & 8.5 $\pm$ 1.0\\
    300 & $50 \times 13$ & 686 & 34.8 $\pm$ 3.5 & 6.5 $\pm$ 0.8 \\
    300 & $50 \times 7$ & 400 & 33.0 $\pm$ 5.1 &  \\
    325 & $50 \times 13$ & 888 & 33.4 $\pm$ 2.5 & 6.8 $\pm$ 1.2 \\
    350 & $50 \times 13$ & 888 & 33.8 $\pm$ 5.2 & 6.5 $\pm$ 0.5\\
    \hline
  \end{tabular}
\end{table}

 \clearpage

\section{Temperature dependence of DOS, Valence band energy, IPR}
\label{section:IPR_T}

\begin{figure}[!htb]
\centering
\includegraphics[width=1.0\textwidth]{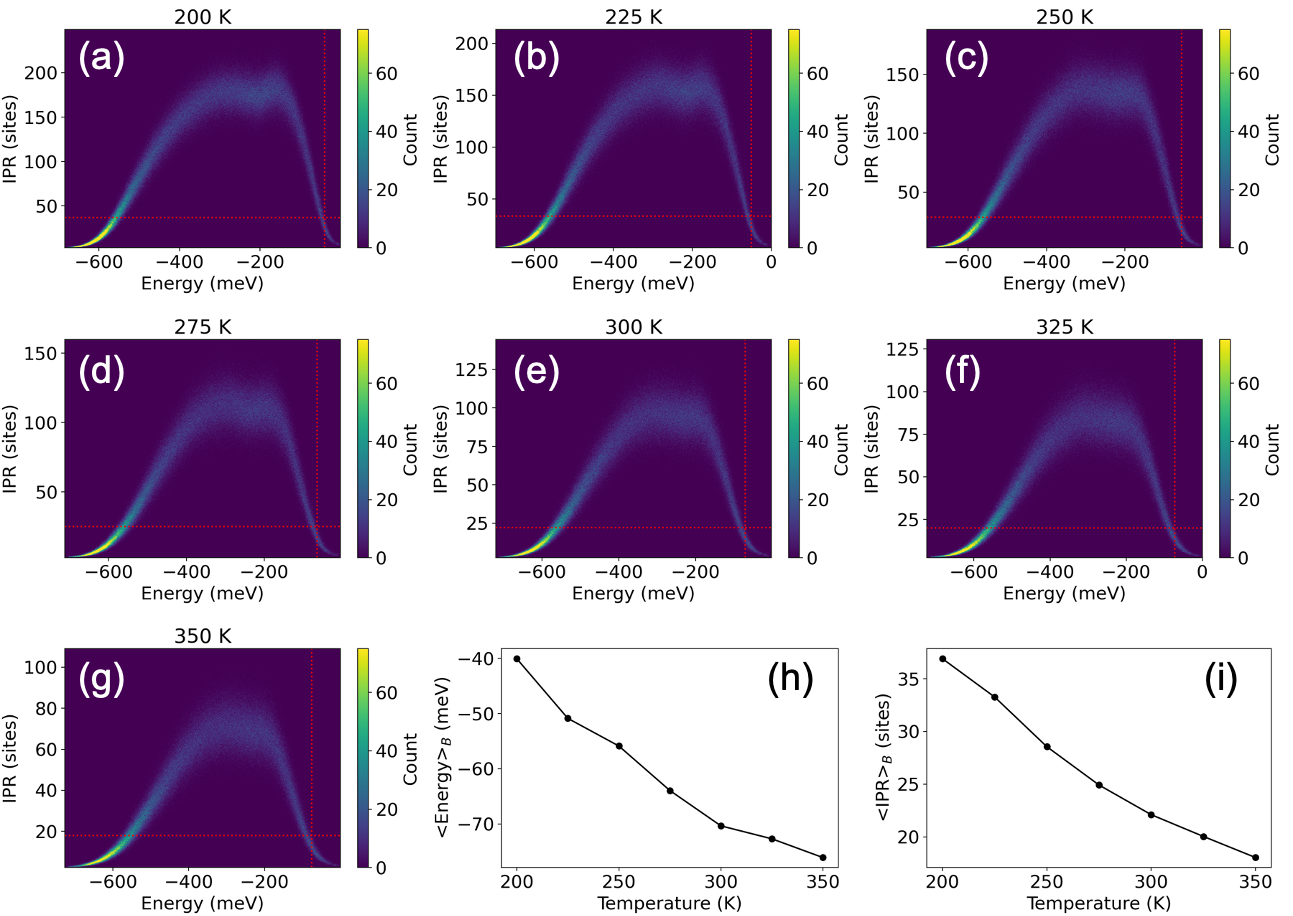}
  \caption{ Energy and IPR-resolved density of valence band states of rubrene. (a)-(g) show 2D histograms of IPR against energy, where the heat-map axis indicates the number of states counted. 
  Red dashed vertical and horizontal lines indicate the Boltzmann averaged energy and IPR, respectively. The quantities are plotted against temperature in panels (h) and (i).}
  \label{fig:IPR_E_all}
\end{figure}

The effect of increasing electronic disorder with temperature (see Table 1 of the main text) is to cause localisation of the eigenstates (i.e. adiabatic states) of the electronic Hamiltonian. Figure \ref{fig:IPR_E_all} shows the energy-resolved IPR of the valence band (also termed adiabatic) states, averaged over snapshots from 20 trajectories for each temperature. The heat-map axis indicates the number of states counted at a particular energy and IPR value. The energy of the valence band maximum was set to zero for each snapshot. 
The Boltzmann averages of the valence band energy and IPR are given by equations \ref{eq:E_Boltz} and \ref{eq:IPR_Boltz}, respectively. 

\begin{equation}
 \langle E \rangle^B = \frac{\sum_k E_k \exp{(E_k / k_B T)}}{\sum_k \exp{(E_k / k_B T)}}, \label{eq:E_Boltz}   
\end{equation}

\begin{equation}
 \langle \text{IPR} \rangle^B = \frac{\sum_k \text{IPR}_k \exp{(E_k / k_B T)}}{\sum_k \exp{(E_k / k_B T)}}, \label{eq:IPR_Boltz}   
\end{equation}
where $k$ sums over the adiabatic states, $k_B$ is the Boltzmann constant and $T$ is temperature. These quantities are plotted as a function of temperature in panels \ref{eq:IPR_Boltz}(h) and \ref{eq:IPR_Boltz}(i). 

In all cases, states are localised at the band edges and become increasingly delocalised towards the middle of the band. As temperature increases, the IPR of the adiabatic states at a given energy becomes smaller (most clearly illustrated in Figure 2(d) of the main text), indicating increasing localisation due to dynamic disorder. This can also be seen by considering the Boltzmann average of the IPR with temperature, shown in Figure \ref{fig:IPR_E_all}(i). An additional effect of increased temperature is an increase in the Boltzmann averaged energy with temperature, shown in Figure \ref{fig:IPR_E_all}(h). This is due to increased thermal energy with temperature, allowing the charge carrier to explore states deeper within the valence band where states are more delocalised. However, from panel \ref{fig:IPR_E_all}(i), clearly the effect of increasingly localised states with temperature is more prominent and overall the states occupied by a charge carrier will be more localised at higher temperatures\cite{Giannini23}. 

\clearpage

\section{Transient Delocalisation Mechanism for Constant Temperature Simulations}
\label{section:TD_mechanism_const_T}

\begin{figure}[!htb]
\centering
\includegraphics[width=1.0\textwidth]{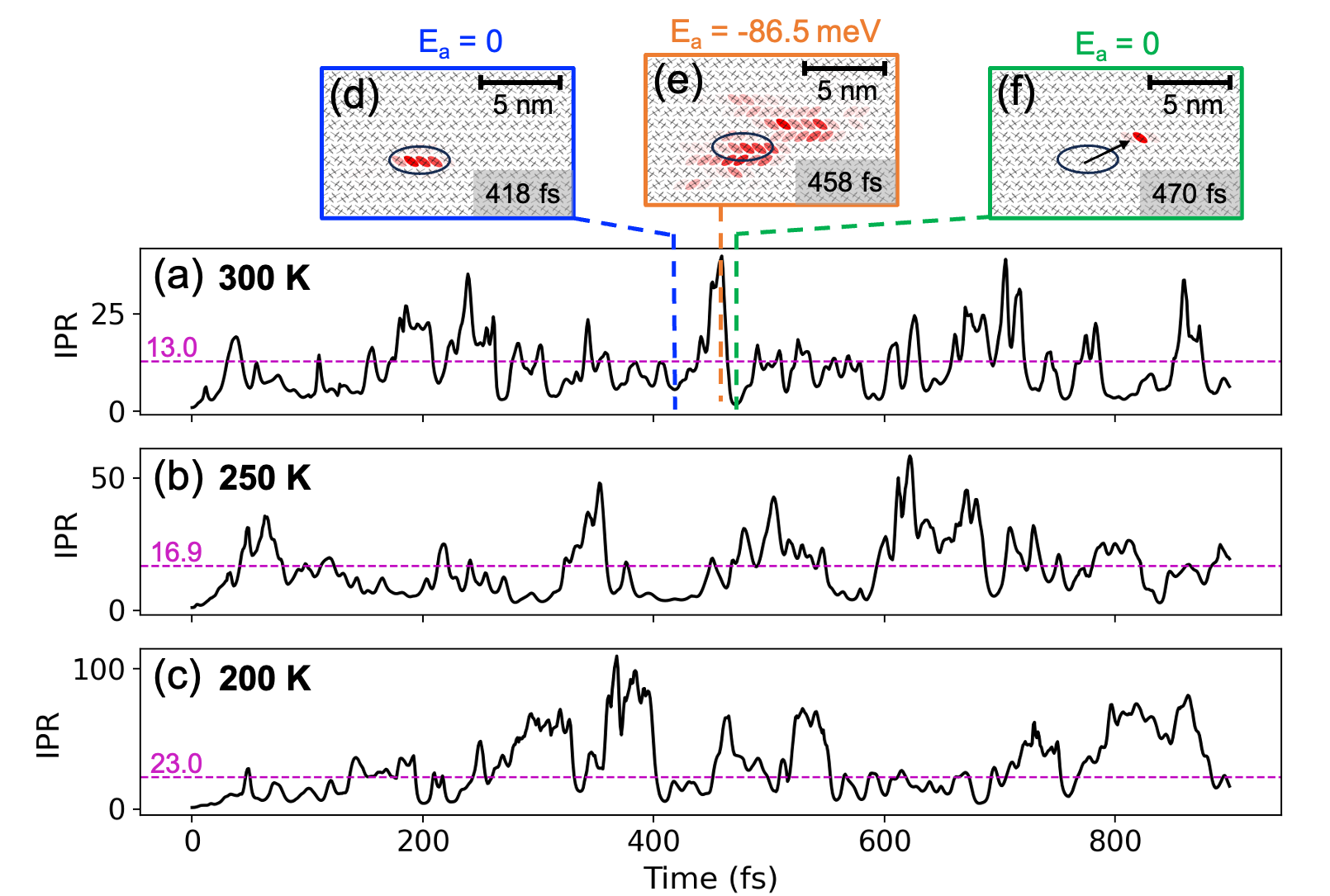}
\caption{IPR of the charge carrier wavefunction (Eq. 11 of the main text) for individual trajectories at (a) 300 K, (b) 250 K and (c) 200 K. The average IPR (averaging over $t > 200$ fs and all trajectories) is indicated by the magenta dashed line in each case. Both the average and root-mean-square fluctuations of IPR become larger with decreasing temperature, see Table~1 of the main text. The mechanism of transient delocalisation (TD) is illustrated for the trajectory at 300 K in panels (d)--(f). 
The wavefunction is represented by superposing red ellipses onto each molecular site with opaqueness proportional to the site population, $|u_k|^2$. The energy of current active adiabatic state, $E_a$, is indicated in each case. 
Such transient expansions of the wavefunction are driven by surface hops to delocalized excited states (see Fig.~2(b) of the main text) and significantly contribute to the overall diffusivity\cite{giannini2020flickering, giannini2022charge}.
  }
  \label{fig:IPR_traj}
\end{figure}

Figure~\ref{fig:IPR_traj}(a), (b) and (c) show inverse participation ratio (IPR) of the charge carrier wavefunction as a function of time along a single FOB-SH trajectory at 300~K, 250~K and 200~K, respectively.
The IPR exhibits significant fluctuations about the mean (indicated by the dashed line in magenta), which increases with decreasing temperature.
Panels (d), (e) and (f) depict, respectively, the charge carrier wavefunction before, during and after an event of transient delocalization (TD). The charge carrier wavefunction is represented by superposing red ellipses on each molecular site (corresponding to fragment molecular orbital basis functions), with opaqueness proportional to the wavefunction site population $|u_k|^2$.

The TD mechanism, illustrated in Figure~\ref{fig:IPR_traj}(d)--(f), allows for significant displacement of the charge carrier over distances up to the extent of the transiently delocalized state\cite{giannini2020flickering,giannini2022charge}. 
In the example shown, at $t = 418$~fs the charge carrier is delocalized over 5--6 molecules (IPR = 5.6) and nuclear dynamics is propagated on the ground state (i.e. top of the valence band, $E_a = 0$). Subsequently, a series of surface hops to excited valence band states take place such that the active adiabatic state at $t = 458$~fs is the 43\textsuperscript{rd} valence band state, $E_a = - 86.5$~meV. This is concomitant with TD of the wavefunction, which expands to cover approximately 39 molecules (IPR = 38.9) at $t = 458$~fs, a factor of 3 larger than the average value.
Finally, following a series of surface hops back to the ground state, the wavefunction contracts and at $t = 470$~fs is momentarily delocalized over just 1--2 molecular sites (IPR = 1.8). Over the course of the TD event, the centre-of-charge of the wavefunction is displaced by approximately 3.5 nm. In general, displacements following a TD event may be larger or smaller and are limited only by the extent of the transiently delocalized state. 
Such TD events are driven by nonadiabatic transitions (i.e. surface hops) to excited valence band states, which become more delocalized towards the centre of the band (see Figure~2(b) of the main text). The extent of delocalization of the hole wavefunction in the TD state reflects the extent of delocalization of the active adiabatic state on which the nuclear dynamics is propagated. 

\clearpage
 
\section{Transient Localization Theory Calculations}
\label{section:TLT_calc}

Transient localization theory (TLT) was used to calculate mobility for the different temperatures considered (Figure~2 of main text, magenta diamonds). 
The transient localization mobility is given by\cite{ciuchi2011transient, fratini2016transient}

\begin{equation}
  \mu_{x(y)} = \frac{e}{k_BT}\frac{\bar{L^2}_{x(y)}(\tau)}{2\tau} \label{eq:TLT_mobility}  
\end{equation}
where $e$ is the elementary charge, $k_B$ is the Boltzmann constant, $T$ is temperature, $\tau$ is the timescale of 
lattice vibrations driving transient localization events, $L$ is the so-called transient localization length and $\bar{L^2}$ denotes the average of $L^2$ over multiple realisations of disorder. The squared transient localization length is given by 

\begin{equation}
 L^2_{x(y)}(\tau) = \frac{1}{Z} \sum_{n,m}e^{\beta E_n} |\langle n|\hat{j}_{x(y)}|m\rangle|^2 \frac{2}{(\hbar/\tau)^2 + (E_m - E_n)^2} \label{eq:TLT_Lsq}   
\end{equation}
where $Z$ is the partition function, $\hat{j}$ is the current operator and ($|{n}\rangle$, $E_n$) refer to the eigenstates and eigenvalues of a Hamiltonian corresponding to a particular realisation of disorder. Note, a positive sign is used in the Boltzmann factor since we consider hole transport. In practice, disordered Hamiltonians are constructed by sampling from specified distributions of electronic couplings for a given supercell size. The average squared localization length, $\bar{L^2}$ is obtained by averaging over the transient localization lengths corresponding to all Hamiltonians considered. Finally, mobility is calculated using Eq. \ref{eq:TLT_mobility}. TLT mobility calculations for each temperature were carried out using freely available code \cite{nematiaram2019practical} (https://github.com/CiuK1469/TransLoc, code version 0.4, downloaded 14th December 2020). We used a $39\times26$ supercell under periodic boundary conditions, which gave well-converged results in all cases, and we averaged over the transient localization lengths obtained from 50 distinct disordered Hamiltonians with matrix elements sampled from the temperature dependent distributions of electronic couplings listed in Table~1 of the main text.
The timescale $\tau$ was calculated by averaging over the ab initio power spectrum for $J_a$ fluctuations, $S_a$ (Fig. 1(c), cyan), $\tau \!=\! [\int \text{d}\omega \, \omega (S_{a}(\omega)/\omega) / \int \text{d}\omega (S_{a}(\omega)/\omega)]^{-1} = 0.43$ ps. We note that the dependence of mobility on $\tau$ in equation \ref{eq:TLT_mobility} is relatively weak due to the dependence of $L^2_{x(y)}$ on $\tau$, compensating the denominator \cite{nematiaram2019practical}.
Numerical values for TLT mobility at different temperatures are listed in Table~\ref{table:TLT}.

\begin{table}
  \caption{TLT mobilities along the $a$ and $b$ crystallographic directions for temperature between 200 K and 350 K. }
  \label{table:TLT}
  \begin{tabular}{lll}
      \hline
    Temperature (K) & $\mu_a$ $(\text{cm}^2/Vs)$&  $\mu_b$ $(\text{cm}^2/Vs)$ \\
    \hline
    200 & 62.5 & 16.4 \\
    225 & 55.1 & 13.9 \\
    250 & 50.3 & 12.4 \\
    275 & 41.4 & 9.9 \\
    300 & 39.5 & 9.1 \\
    325 & 35.0 & 7.9 \\
    350 & 31.1 & 6.8\\
    \hline
  \end{tabular}
\end{table}

 \clearpage

\section{Preparation of simulation cell with temperature gradient}
\label{section:Simulating_grad_T}

\begin{figure}[!htb]
\centering
\includegraphics[width=1.0\textwidth]{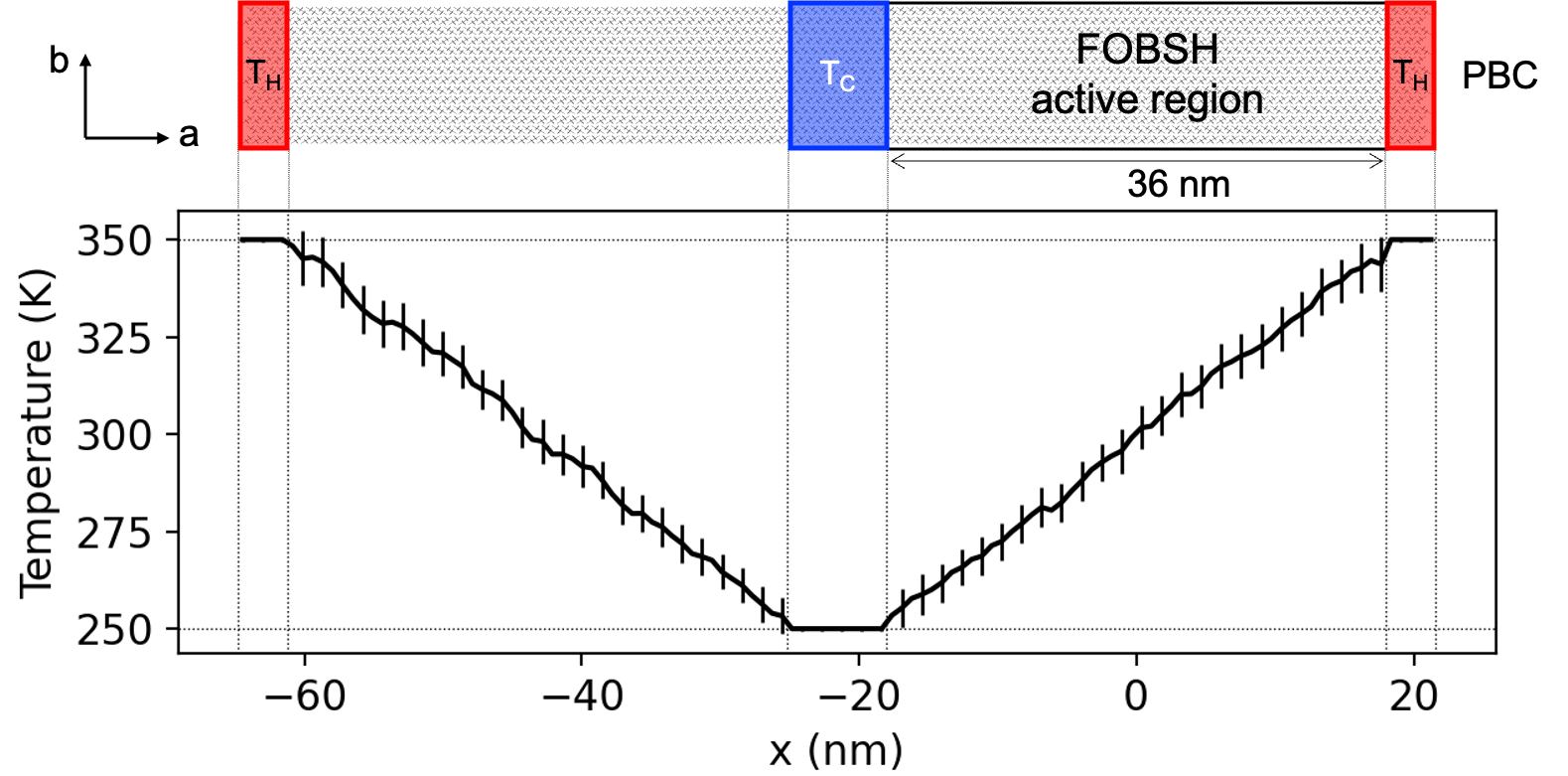}
  \caption{Full supercell for simulations under a temperature gradient and temperature profile, averaged over 200 ps of MD. Error bars represent the root-mean-square fluctuations in local kinetic temperature including all atoms within a slab of dimensions $1\times7\times1$ unit cells. Thermal bath regions, of size $10 \times 7 \times 1$ unit cells, are maintained at temperatures $T_C = 250$ K and $T_H = 350$ K through a velocity rescaling procedure. The thermal bath regions are separated by an active region (later used for hole propagation in FOB-SH), where the temperature varies with a uniform gradient. The supercell shown is periodically replicated in 3D and the active region in FOB-SH simulations ($50 \times 7$ unit cells in the $a-b$ plane, 36.01 nm $\times$ 10.12 nm) 
  is defined over one of the linear parts of the temperature profile, as indicated, and is not periodically replicated.}
  \label{fig:TEMP_GRAD}
\end{figure}

Figure \ref{fig:TEMP_GRAD} shows the simulation box and temperature profile, averaged over 200 ps of MD. The supercell contains $120\times7\times1$ unit cells (4 molecules per unit cell) and is periodic in all dimensions. The temperature gradient is along $x$ which is parallel to the $a$ crystallographic direction. Since the overall cell is 3D periodic, the temperature at either side of the box along  $x$ must be the same, therefore a saw-tooth temperature profile is required. The FOB-SH active region, which is not periodic, is defined over one of the linear portions of the temperature profile ($50\times7\times1/2$ unit cells), as indicated. 

\begin{figure}[!htb]
\centering
\includegraphics[width=1.0\textwidth]{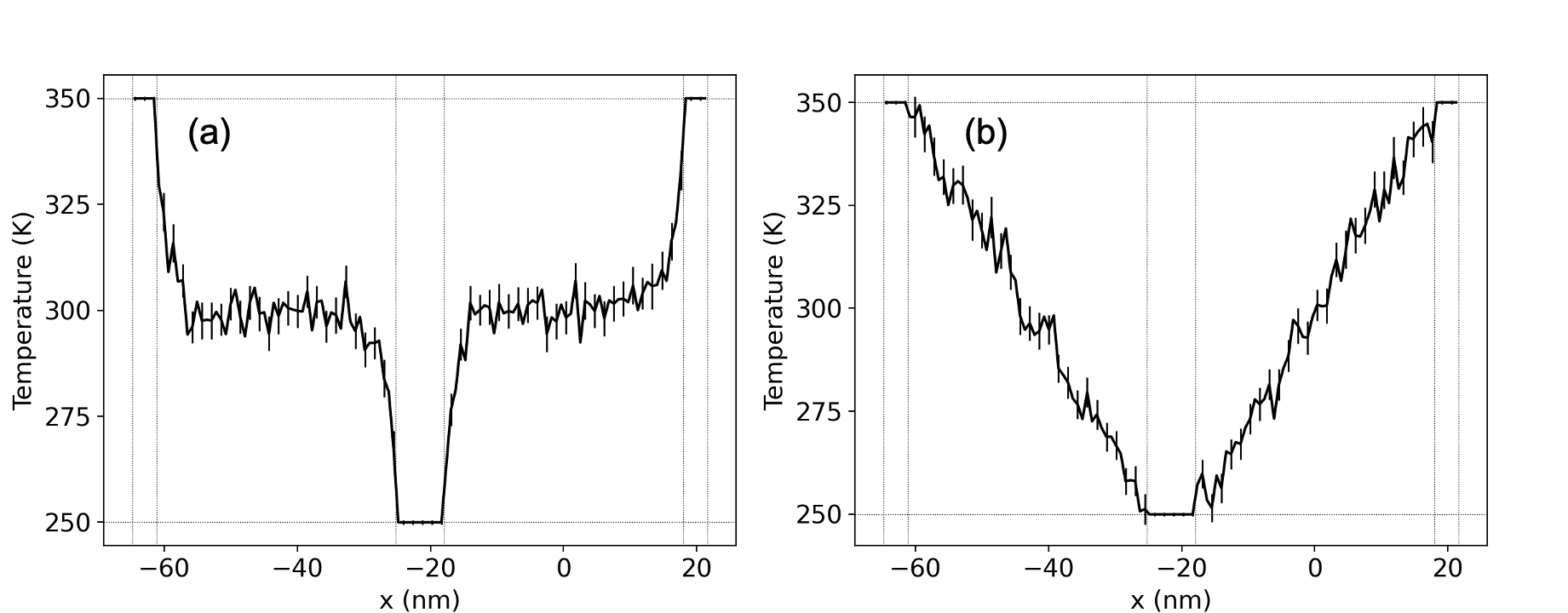}
  \caption{Temperature profile averaged over 1 ps using different relaxation procedures. (a) 50 ps of molecular dynamics with velocity rescaling in the thermal bath regions only. The target temperatures are 250 K and 350 K in the cold and hot baths, respectively and the temperature tolerance for rescaling was set to 1 K. (b) Relaxation using sequential molecular dynamics runs with decreasing thermostatting applied across the entire system, after which thermostatting outside of the bath regions is turned off entirely.  }
  \label{fig:temp_profile_1ps}
\end{figure}

The temperature profile is achieved by defining thermal bath regions at 250 K and 350 K ($10\times7\times1$ unit cells, 280 molecules, 19600 atoms), as indicated in Figure \ref{fig:TEMP_GRAD}, which are pinned to their respective temperatures through a velocity rescaling procedure. Velocity rescaling to the target temperature occurs whenever the difference between the instantaneous kinetic temperature, $T_\text{inst}$, and the target temperature, $T_\text{target}$, exceeds some defined tolerance, $|T_\text{inst} - T_\text{target}| > T_\text{tol}$. $T_\text{tol}$ was set to 1 K in the bath regions. 
Figure \ref{fig:temp_profile_1ps}(a) shows the temperature profile, averaged over 1 ps, after 50 ps of MD. The temperature profile is highly non-linear, indicating that a long time scale is needed for relaxation to the steady state using this approach. To speed up convergence of the temperature profile, we initially apply velocity rescaling over the full simulation cell. To do so we 
defined in the active region consecutive slabs with dimensions $1\times7\times1$ 
unit cells (28 molecules, 1960 atoms) and set the target temperature in each slab to the temperature expected from a linear temperature gradient. The temperature tolerance in the slabs  
was set to $T_\text{tol} = \{1, 5, 20\}$ K for sequential molecular dynamics runs of length 2 ps whilst a tolerance of $T_\text{tol} = 1$ K was applied to the thermal bath regions 
throughout the relaxation procedure. Hence, thermostatting of the active region was sequentially made weaker. Following this initial relaxation, thermostatting in the active region was turned off 
entirely keeping only the thermal bath regions thermostatted. The temperature profile achieved in this manner is linear and stable over long time scales ($\gg$ 200 ps, the time scale averaged over in Fig. \ref{fig:TEMP_GRAD}), see Figure \ref{fig:temp_profile_1ps}(b) where the profile was averaged over 1 ps, after 50 ps of MD. The FOB-SH simulations with temperature gradient were initialized from snapshots 
taken from this trajectory keeping only the thermal bath regions (and not the active region) thermostatted by velocity rescaling.

\begin{figure}[H]
\centering
\includegraphics[width=0.62\textwidth]{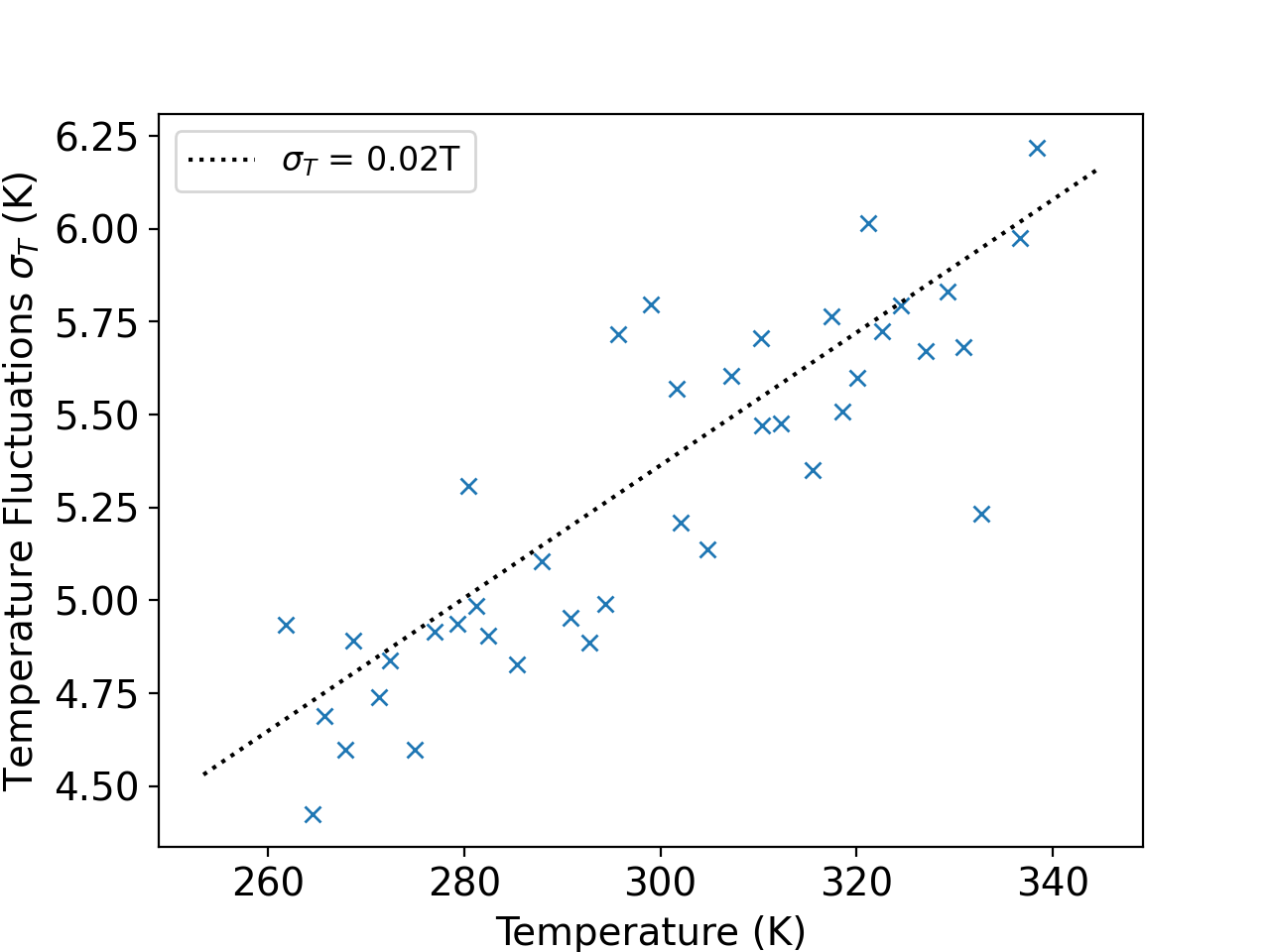}
  \caption{Root-mean-square temperature fluctuations (i.e. error bars in Fig.~\ref{fig:TEMP_GRAD}) as a function of local temperature over bins of size $1\times7$ unit cells, showing a weak linear dependence on the local temperature.}
  \label{fig:temp_fluctuations}
\end{figure}

The root-mean-square fluctuations in temperature within a given slab (i.e. error bars in Fig. \ref{fig:TEMP_GRAD}) are of the order 5 K, which determines the local temperature resolution in our simulations. 
The magnitude of temperature fluctuations shows a slight linear increase with local temperature, see Figure~\ref{fig:temp_fluctuations}. This is consistent with the expected temperature dependence of temperature fluctuations in the NVT ensemble from statistical mechanics, $\sqrt{\langle \Delta T^2 \rangle} = \sqrt{\frac{k_B}{C}}T \sim \frac{T}{\sqrt{N}}$, where $C$ is the heat capacity \cite{chui1992temperature}. However, direct comparisons with properties of equilibrium ensembles should be approached with caution given the nonequilibrium nature of the system under consideration.

 \clearpage

\section{Analysis of drift velocity, $\langle v_x\rangle$}
\label{section:Velocity_distr}

The instantaneous drift velocity at each time step and over each trajectory is calculated from the difference in centre-of-charge between time increments of 2 fs i.e. $v^t = (\langle x \rangle^{t+\delta t} - \langle x \rangle^{t}) / \delta t$, with $\delta t = 2$ fs. We use bins of length 4 unit cells (2.88 nm) along $x$ (parallel to the $a$-crystallographic direction) and allocate each velocity to the bin which includes the initial centre-of-charge position, $\langle x \rangle^{t}$. The temperature gradient is 2.78 K/nm, therefore the temperature difference between the centre of adjacent bins is 8~K. The Seebeck coefficient is determined from the average velocity in the central bin (i.e. the bin centred at the position where $T\!=\!300$ K). 

\begin{figure}[!htb]
\centering
\includegraphics[width=1.0\textwidth]{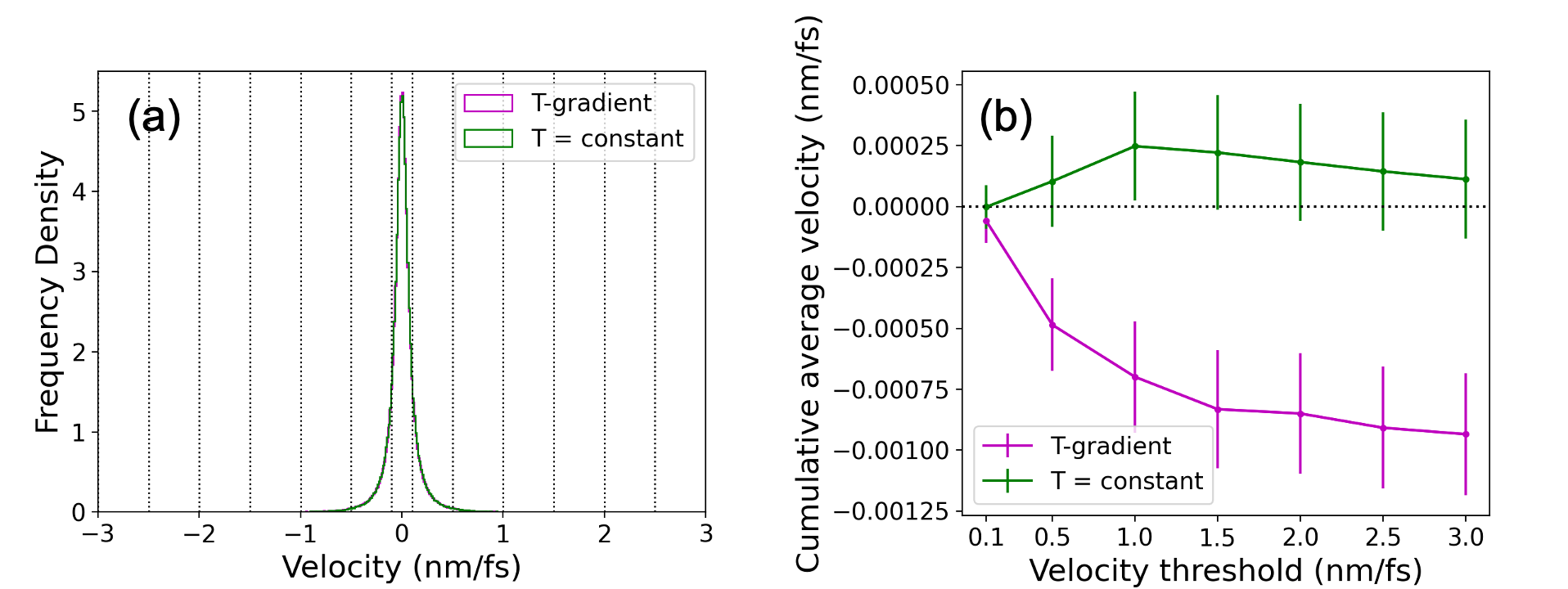}
  \caption{(a) Histogram of velocities in the central 4 unit cells for simulations under the temperature gradient and at constant T = 300 K. (b) Cumulative average velocity, taking into account only velocities whose absolute value is smaller than a threshold.  
  }
  \label{fig:velocities}
\end{figure}

Due to fluctuations in local temperature ($\approx 5$ K, Figure~\ref{fig:temp_fluctuations}), a temperature difference between two points in space is only well-resolved for distances larger than $\approx$ 1 nm. Therefore when velocities are very small (i.e. the change in centre-of-charge after a time increment is significantly smaller than 1 nm), there should be little contribution to the Seebeck coefficient since locally the temperature is indistinguishable on this length scale. For large velocities, the temperature profile is clearly resolved and there is a significant contribution to the Seebeck coefficient. Large velocities generally occur after a successful hop to a new active state at some distance from the old active state. The charge carrier wavefunction spatially follows the active state on average, due to the decoherence correction which damps the non-active-surface adiabatic coefficients (see main text {\it Methods}). As explained in the main text, the temperature gradient induces a symmetry breaking which causes the probability of surface hopping to states on the cold side to be larger than the probability of surface hopping to states on the hot side. Figure \ref{fig:velocities}(a) shows a histogram of the centre-of-charge velocities, only counting velocities where the centre-of-charge lies within the central 4 unit cells along $x$ for constant temperature simulations at $T\!=\!300$ K and simulations under a temperature gradient, including all time steps and all trajectories. 

Visually, the distributions of drift velocities for simulations at constant temperature and under a temperature gradient appear very similar. However, the mean velocity for the constant temperature case 
is $0.10 \pm 0.24$ nm/ps, 
while the mean velocity for simulations under a temperature gradient is $-0.89 \pm 0.25$ nm/ps. Figure \ref{fig:velocities}(b) shows the cumulative drift 
velocity obtained by averaging over signed instantaneous drift velocities with absolute value smaller than some threshold. In the case of constant temperature simulations (green), this remains close to 0, within the margin of statistical error. For simulations under a temperature gradient, the drift velocity becomes more negative as the threshold is increased, corresponding to the net motion from 
hot to cold, and it is well converged when all instantaneous velocities with absolute value smaller than $\approx$ 1.5 nm/fs are included. Velocities larger than this are extremely rare and therefore 
do not affect the average significantly. Notice that the average drift velocity (i.e. the cumulative drift velocity after all velocities are included) is 2-3 orders of magnitude smaller than typical 
instantaneous drift velocities as positive and negative instantaneous drift velocities cancel to a large extent resulting in only relatively small net drift velocity, $\langle v_x\rangle$.    

\clearpage

\section{Analysis of thermoelectric motion}
\label{section:Analaysis_adiabatic_states}

\begin{figure}[!htb]
\centering
\includegraphics[width=0.8\textwidth]{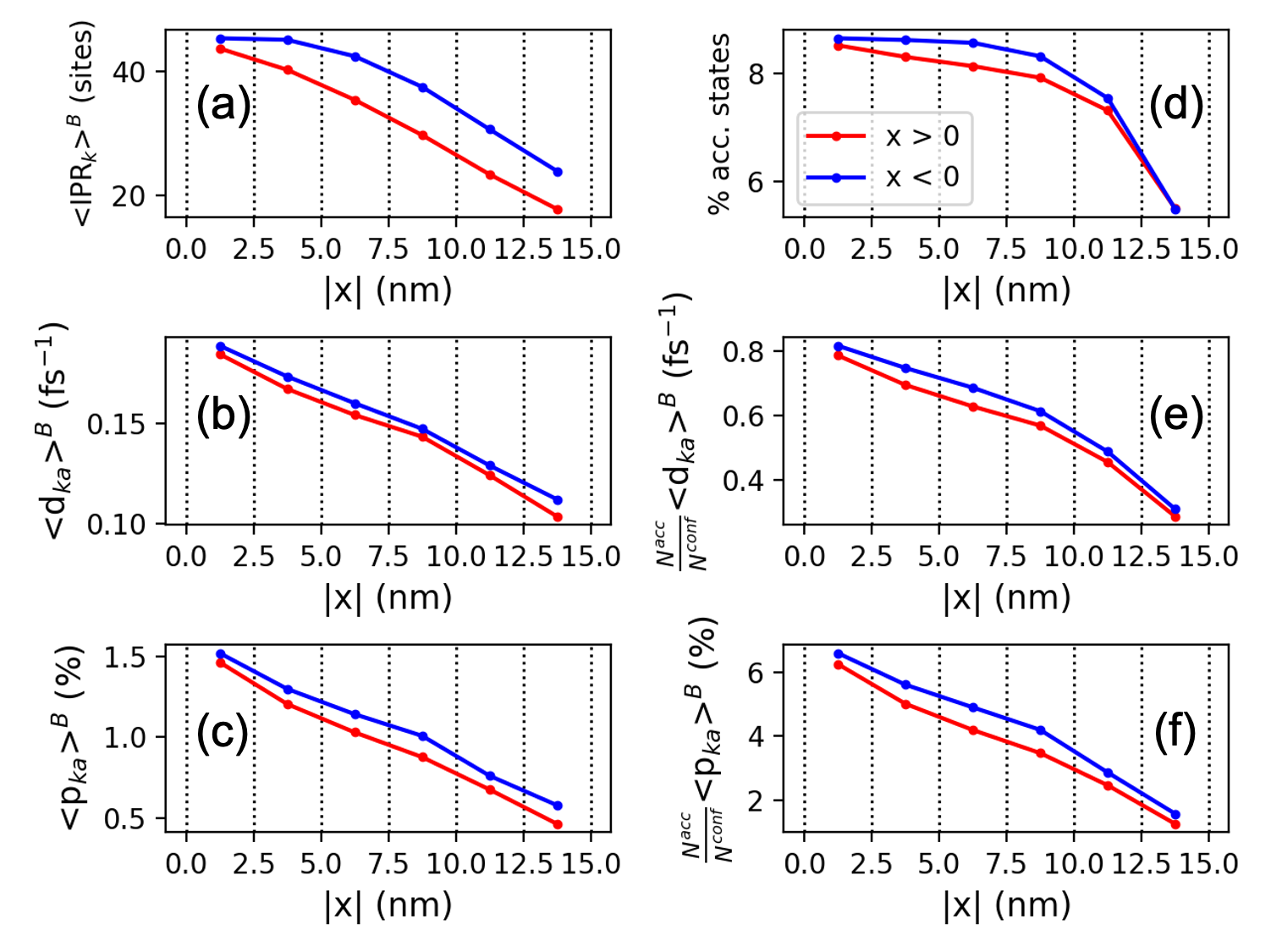}
  \caption{Position-resolved properties of valence band states $\psi_k$ when a temperature gradient is present. Properties for states located towards the cold side and hot side, 
  relative to the location of the current active state, are represented in red and blue, respectively. (a) Boltzmann averaged IPR, (b) Boltzmann averaged NACE, (c) Boltzmann averaged hopping probability, 
  (d) percentage of thermally accessible states, (e) sum of Boltzmann weighted NACEs within a given position bin, (f) sum of Boltzmann weighted hopping probabilities within a given position bin. 
  }
  \label{fig:grad_adiab_analysis}
\end{figure}

\begin{figure}[!htb]
\centering
\includegraphics[width=0.8\textwidth]{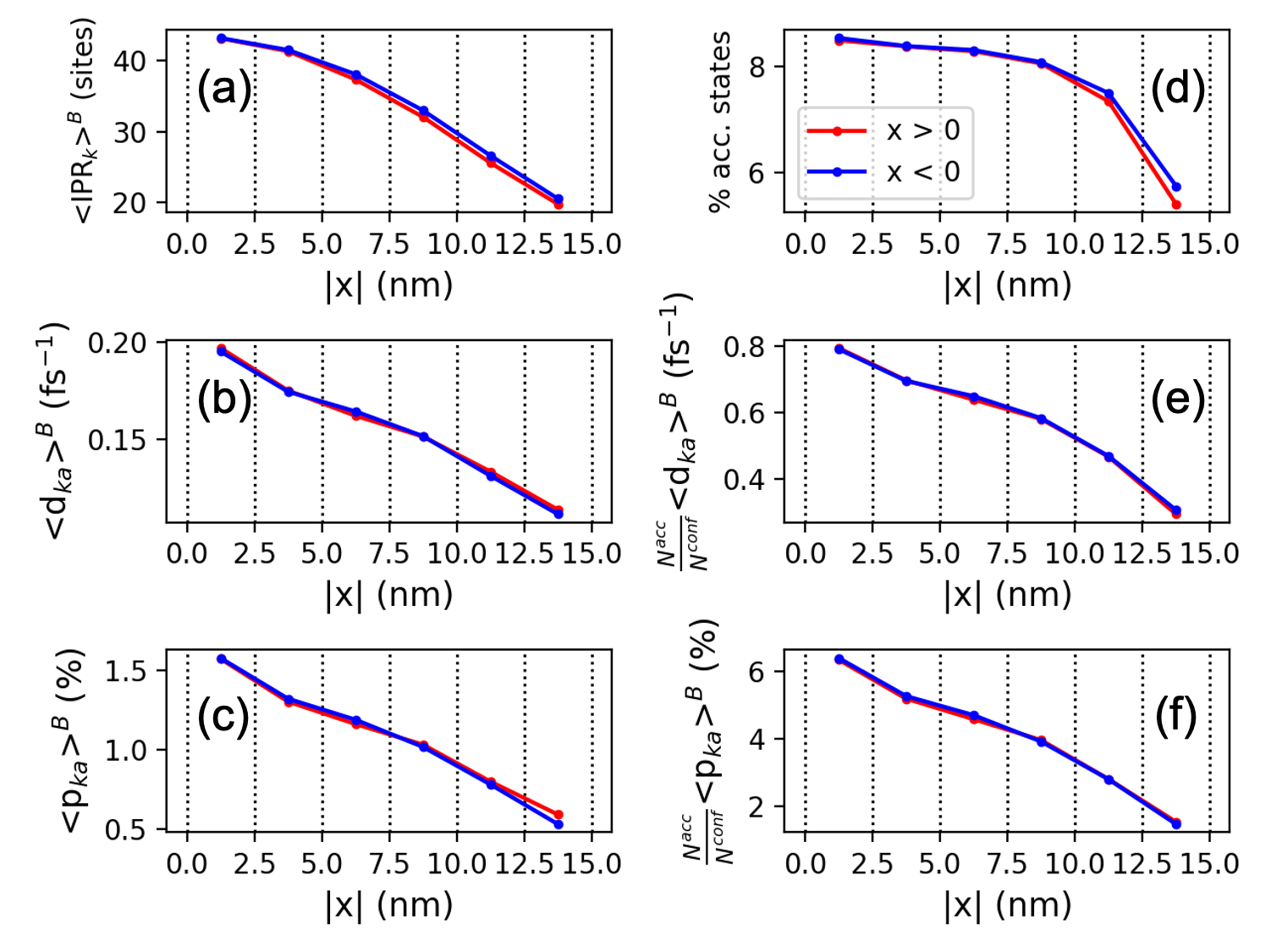}
  \caption{Position-resolved properties of valence band states $\psi_k$ for simulations at constant $T = 300$ K. Properties for states located towards the cold side and hot side, relative to the location of the current active state, are represented in red and blue, respectively. (a) Boltzmann averaged IPR, (b) Boltzmann averaged NACE, (c) Boltzmann averaged hopping probability, (d) percentage of thermally accessible states, (e) sum of Boltzmann weighted NACEs within a given position 
  bin, (f) sum of Boltzmann weighted hopping probabilities within a given position bin.
  }
  \label{fig:const_adiab_analysis}
\end{figure}

Figures \ref{fig:grad_adiab_analysis} and \ref{fig:const_adiab_analysis} show distance resolved properties of the valence band states $\psi_k$ for simulations with a temperature gradient and 
at constant $T = 300$ K, respectively. The Boltzmann average, indicated by $\langle ... \rangle^\text{B}$, is defined in section {\it Methods} of the main text, Eq. 16. 
$N^\text{conf}$ refers to the total number of configurations (i.e. Hamiltonians) selected for the analysis, according to the criteria that a successful hop occurs with active state located in the central 10 unit cells of the simulation box along $x$. Panels \ref{fig:grad_adiab_analysis}(b) and \ref{fig:grad_adiab_analysis}(d) show the same information as panels 4(b) and 4(d) of the main text, while panels \ref{fig:const_adiab_analysis}(b) and \ref{fig:const_adiab_analysis}(d) show the same information as panels 4(c) and 4(e) of the main text. 

It is evident that in the case of constant temperature simulations (Figure \ref{fig:const_adiab_analysis}), there is no difference between states with $\Delta \text{COC}_{ka} < 0$ compared to states with $\Delta \text{COC}_{ka} > 0$ (blue vs red lines). When a temperature gradient is applied (Figure \ref{fig:grad_adiab_analysis}), an asymmetry arises causing a clear splitting between quantities corresponding to states on the cold side compared to the hot side. $\langle d_{ka} \rangle^\text{B}$ is larger for states towards the cold side ($\Delta \text{COC}_{ka} < 0$), which results in a larger Boltzmann averaged surface hopping probability, $\langle p_{ka} \rangle^\text{B}$, for states towards the cold side. Additionally, there is a greater number of thermally accessible states towards the cold side. These two effects result in a larger overall probability of surface hopping to any state at distance $\Delta \text{COC}_{ka} \in x$, determined by the product $\frac{N^\text{acc}}{N^\text{conf}} \times \langle p_{ka}\rangle^\text{B}$. 

\begin{figure}[H]
\centering
\includegraphics[width=1.0\textwidth]{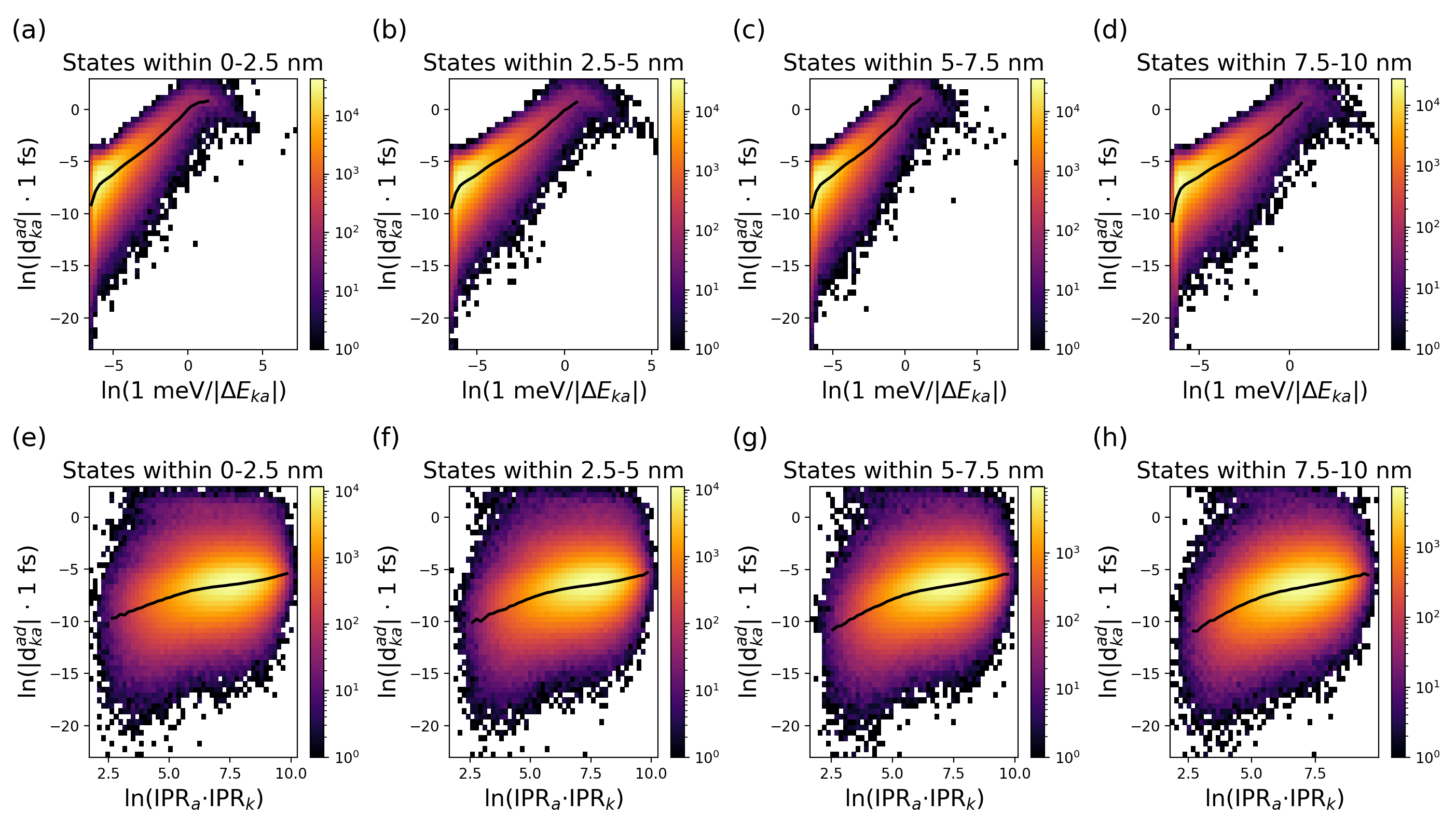}
  \caption{2D histograms plotting the natural logarithm of the unsigned NACE, $\ln(|d^\text{ad}_{ka}|)$, against $\ln(1/|\Delta E_{ka}|)$ (a-d), and $\ln(\text{IPR}_a\text{IPR}_k)$ (e-h) including states centred at different distance bins from the position of the active state (same binning procedure as used in Fig.~5 of the main text). The color scale indicates the number of states in the relevant 2D bin. Black lines in bold indicate the average value, $\langle\ln(|d^\text{ad}_{ka}|)\rangle$. 
  There is a clear correlation between $\ln(|d^\text{ad}_{ka}|)$ and $\ln(1/\Delta E_{ka})$, expected due to the explicit dependence of the NACE on $1/\Delta E_{ka}$, equation~\ref{eq:NACE_explicit_formula}. This results in the asymmetry in average NACE to states located on the cold side compared to the hot side (Fig.~5(b) of the main text) due to the increased likelihood of encountering small $\Delta E$ values as the density of accessible states increases from hot to cold along the temperature gradient. 
  The correlation between the NACE and the product $\text{IPR}_a\text{IPR}_k$ is smaller (albeit non-zero), indicated that state delocalization is a less significant factor determining the magnitude of the NACE.
  }
  \label{fig:NACE_correlations}
\end{figure}

To assess the factors contributing to the higher average NACE for states on the cold side compared to those on the hot side, we present 2D histograms in Figure \ref{fig:NACE_correlations} correlating $|d^\text{ad}_{ka}|$ with either the inverse energy difference between states, $1/|\Delta E_{ka}|$ (panels a-d), or the product of the IPR of the active state $a$ and state $k$, IPR$_a$IPR$_k$ (panels (e-h)). The NACE varies by orders of magnitude, therefore the natural logarithm of each quantity is taken. All states located in a given distance bin (towards hot and cold) relative to the position of the active state are included in the same plot, using the same binning procedure described in Figure~5 of the main text.

The non-adiabatic coupling element between adiabatic states $k$ and $l$, $d^\text{ad}_{kl} = \bra{\psi_k}\ket{\dot{\psi_l}}$ is related to the non-adiabatic coupling vector, $\mathbf{D}^\text{ad}_{kl}$, through the chain rule, $d^\text{ad}_{kl} = \mathbf{D}_{kl} \cdot \frac{\partial \mathbf{R}}{\partial t}$ where\cite{baer2002non}

\begin{align}
\mathbf{D}^\text{ad}_{kl} &= \bra{\psi_k}\partial_{\mathbf{R}}\ket{\psi_l} \\
                 &= \frac{\bra{\psi_k}\partial_\mathbf{R}H\ket{\psi_l}}{\Delta E_{lm}}.
\label{eq:NACE_explicit_formula}
\end{align}
Here, $\partial_\mathbf{R}$ denotes the derivative with respect to coordinates, $H$ is the electronic Hamiltonian and $\Delta E_{lm} = E_l - E_m$ is the energy difference between states $l$ and $m$. Equation \ref{eq:NACE_explicit_formula} shows that the non-adiabatic coupling becomes large when states become close in energy, evident in Figure~\ref{fig:NACE_correlations}(a-d), where scatter is due to variation in the numerator.

\clearpage

\section{Convergence of the kinetic contribution to Seebeck coefficient, $\alpha_v$}
\label{section:Converge_Seebeck}

\begin{figure}[!htb]
\centering
\includegraphics[width=1.0\textwidth]{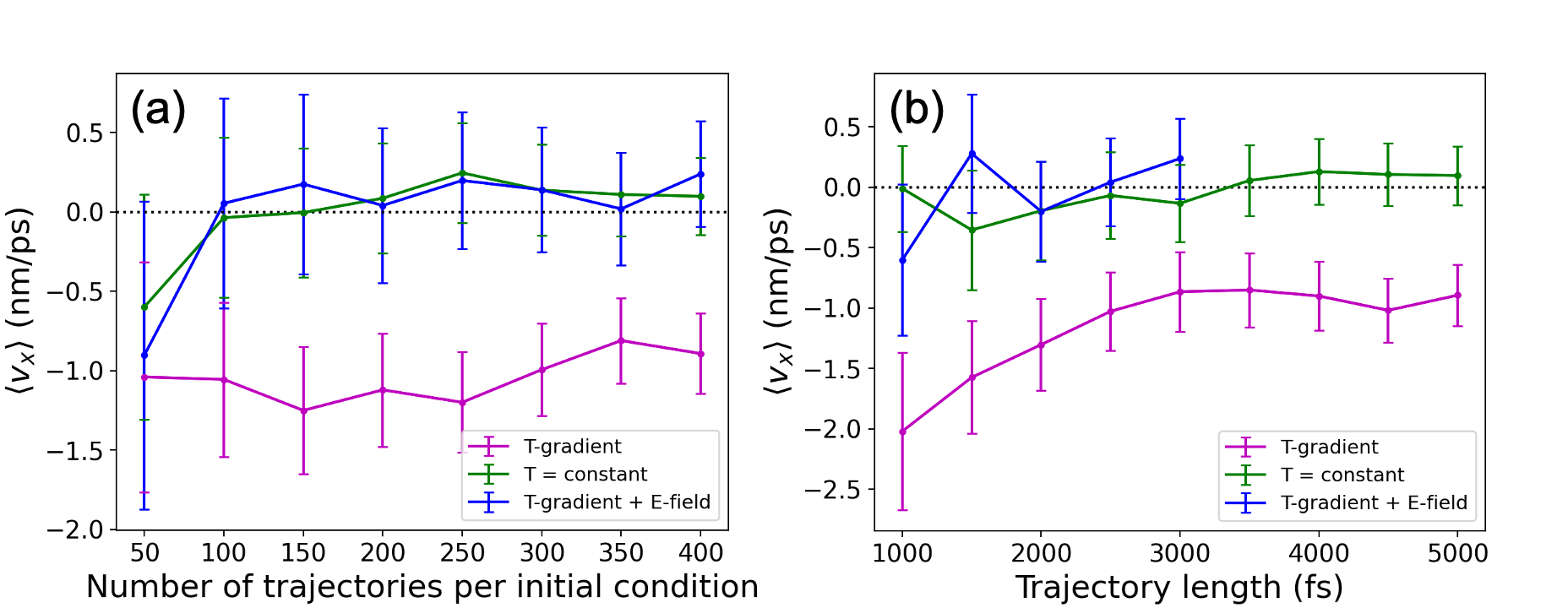}
  \caption{ 
  Convergence of the average drift velocity in the central position bin, $\langle v_x \rangle$, for simulations employing a temperature gradient with no external $E$-field (magenta), constant temperature $T = 300$~K (green) and a temperature gradient with an external $E$-field chosen to cancel the kinetic contribution to the Seebeck coefficient (blue), i.e. same colour scheme as Figure~3 of the main text. 
  (a) Convergence with number of trajectories per initial position of the hole wavefunction. (b) Convergence with trajectory length using all 2000 trajectories. For $T\text{-gradient}$ and $T = \text{constant}$ simulations, trajectories were run to 5~ps, whereas for $T\text{-gradient} + E\text{-field}$ simulations, trajectories were run to 3~ps. 
  }
  \label{fig:S_converge}
\end{figure}

Simulations were carried out with a temperature gradient and no external field (denoted $T\text{-gradient}$), at constant temperature $T = 300$~K (i.e. the control simulations, denoted $T= \text{constant}$) and with a temperature gradient and an external field (denoted $T\text{-gradient}+ E\text{-field}$). In each case, we ran a total of 2000 FOB-SH trajectories. For $T\text{-gradient}$ and $T= \text{constant}$ simulations, trajectories were of length 5~ps (i.e. a total of 10~ns of dynamics), whereas for the $T\text{-gradient}+ E\text{-field}$, trajectories were run to 3~ps (i.e. a total of 6~ns of dynamics).
In each case, the wavefunction was initialised from 5 different positions spread uniformly along the $x$-direction of the active region in FOB-SH simulations, i.e., 400 trajectories per initial condition. 
The kinetic contribution to the Seebeck coefficient, $\alpha_v$, is calculated from the average drift velocity of the charge carrier in the central position bin, $\alpha_v = -\frac{\langle v_x \rangle}{\mu \partial_xT}$ (first term on the RHS of Eq.~2 in the main text).
Figure~\ref{fig:S_converge} shows the convergence of $\langle v_x \rangle$ for the different simulations with (a) number of trajectories (including their full length) and (b) trajectory length using all 2000 trajectories. Error bars represent the standard error of the velocity distribution in the central bin, $\sigma_v/\sqrt{N}$, where $\sigma_v$ is the standard deviation of velocities observed in the central bin and $N$ is the number of data points. 
For the $T$-gradient simulations (magenta), using just 50 trajectories per initial condition 
(i.e. 250 trajectories overall) already yields a mean value close to the final result, however the relatively small amount of data leads to a large error in the mean. Using trajectories of length shorter than 3 ps leads to an overly negative value for $\langle v_x \rangle$ (i.e. an overestimation of $\alpha_v$). This is because trajectories which start from the hot side reach the central bin 
(where drift velocities are used in the average) faster than trajectories starting from the cold end, leading to a small amount of bias. $\alpha_v$ is well converged if trajectories have length 3 ps or longer. 
For the $T = \text{constant}$ simulations (green) and $T\text{-gradient} + E\text{-field}$ simulations (blue), the average velocity in the central bin is approximately converged using around 100 trajectories per initial condition (i.e. 500 trajectories overall). 
As described in the main text, the electric field is chosen to compensate for the net drift velocity observed in the $T$-gradient simulations such that open-circuit conditions are achieved. This is clearly illustrated in Figure~\ref{fig:S_converge}; when the electric field is included, the drift velocity in the central bin becomes close to zero (approximately matching the situation observed for $T = \text{constant}$ simulations). 

\clearpage

 \section{Chemical Potential Contribution to Seebeck coefficient, $\alpha_c$}
\label{section:Chemical_potential_contribution}

As described in the main text {\it Methods}, the chemical potential contribution to the Seebeck coefficient (second term on the RHS of equation 2 of the main text) is given by
\begin{equation}
 \alpha_c = -\frac{1}{q}\frac{\partial_x \mu_c}{\partial_x T} = -\frac{1}{q}\frac{\partial \mu_c}{\partial T}, \label{eq:alpha_muc_contribution}    
\end{equation}
where we have used the chain rule in the second equation of equation \ref{eq:alpha_muc_contribution}. 
The chemical potential at temperature $T$ and reference carrier concentration $n^\text{ref}$ is given by the change in free energy upon insertion of a charge carrier into the band (Eq.~18 of the main text):
\begin{align}
    \mu_c^\text{ref}(T, n^\text{ref}) &= F_\text{hole}(T, n^\text{ref}) - F_\text{neutral}(T, n^\text{ref}) \tag{S9} \\
    &= -k_B T [\ln(Z_\text{hole}) - \ln(Z_\text{neutral})] \tag{S10} \\
    &= -k_B T \ln \frac{\int d\mathbf{R} \sum_i^{\text{vb}} e^{+\beta E_i(\mathbf{R})}}{\int d\mathbf{R} e^{-\beta E_\text{neutral}(\mathbf{R})}} \tag{S11} \\
    &= -k_B T \ln \frac{\int d\mathbf{R} \sum_i^{\text{vb}} e^{+\beta [E_i(\mathbf{R}) + E_\text{neutral}(\mathbf{R})]} e^{-\beta E_\text{neutral}(\mathbf{R})}}{\int d\mathbf{R} e^{-\beta E_\text{neutral}(\mathbf{R})}} \tag{S12} \\ 
    &= -k_B T \ln \Big \langle \sum_i^{\text{vb}} e^{+\beta [E_i(\mathbf{R}) + E_\text{neutral}(\mathbf{R})}  \Big \rangle_{E_\text{neutral}(\mathbf{R})}^{n^\text{ref}}, \tag{S13} \label{eq:mu_c_si_ref_si}
\end{align} where $F$ denotes free energy, $Z$ is the partition function, $\mathbf{R}$ denotes nuclear coordinates, $E_i(\mathbf{R})$ is the $i^\text{th}$ eigenstate of the FOB-SH Hamiltonian describing an excess hole at nuclear geometry $\mathbf{R}$ and $E_\text{neutral}$ is the energy of the neutral system at nuclear geometry $\mathbf{R}$. Note the $+$ sign in the Boltzmann weight over valence band (vb) states, indicating that increasing excitation corresponds to decreased energy. The brackets denote taking the thermal average, which can be obtained by sampling nuclear configurations from a molecular dynamics trajectory of the system in the charge-neutral 
state at carrier density $n^\text{ref}$.
The chemical potential at temperature $T$ and general carrier density $n$ is given by equation 17 of the main text, reproduced here for clarity:
\[  \mu_c(T, n) = \mu_c^\text{ref}(T, n^\text{ref}) + k_B T \ln\frac{n}{n^\text{ref}}. \label{eq:mu_c_si} \tag{S14} \]
Taking the derivative with respect to temperature: 
\[  \frac{\partial \mu_c(T, n)}{\partial T} = \frac{\partial \mu_c^\text{ref}(T, n^\text{ref})}{\partial T} + k_B \ln\frac{n}{n^\text{ref}}, \label{eq:dmu_c_dT} \tag{S15} \] from which $\alpha_c$ may be calculated using equation \ref{eq:alpha_muc_contribution}:
\[ \alpha_c (T, n) = -\frac{1}{q}\frac{\partial \mu_c}{\partial T} = -\frac{1}{q}\Bigg[ \frac{\partial \mu_c^\text{ref}}{\partial T} + k_B \ln\frac{n}{n^\text{ref}}\Bigg]. \label{eq:alpha_c_si} \tag{S16} \] 
To obtain the first term on the RHS of equation~\ref{eq:alpha_c_si}, $\alpha_c^\text{ref}\!=\!-\frac{1}{q}\frac{\partial \mu_c^\text{ref}(T, n^\text{ref})}{\partial T}$ at 300 K, the chemical potential 
$\mu_c^\text{ref}$ was calculated for a given reference concentration $n^\text{ref}$ (see below) at temperatures $T = 275, 300, 325$ K according to equation~\ref{eq:mu_c_si} using MD simulation on the neutral rubrene crystal with a simulation box of $50 \times 7 \times 1$ unit cells
(the same size used for FOB-SH simulations with a temperature gradient).
The temperature derivative was then obtained from the best slope fit. In order to verify that the chemical potential (Eq.~\ref{eq:mu_c_si}) and Seebeck coefficient (Eq.~\ref{eq:alpha_c_si}) are independent of the chosen value for $n^\text{ref}$, $\alpha_c^\text{ref}$  was calculated at 6 different values of $n^\text{ref} = 1 / A$, where $A$ is the area of the active region parallel to the $a$-$b$ crystallographic plane. The active region is the part of the supercell for which the valence band electronic Hamiltonian (Eq.~4 main text) is constructed. 
The values of $n^\text{ref}$ used, along with the corresponding active region sizes and results for $\mu_c^\text{ref}$ and $\alpha_c^\text{ref}$ are listed in Table~\ref{table:chemical_pot}.
The uncertainty in $\alpha_c^{\text{ref}}$ represents one standard deviation, calculated from the square root of the relevant diagonal value of the covariance matrix for the fit parameters. 

\begin{figure}[!htb]
\centering
\includegraphics[width=0.8\textwidth]
{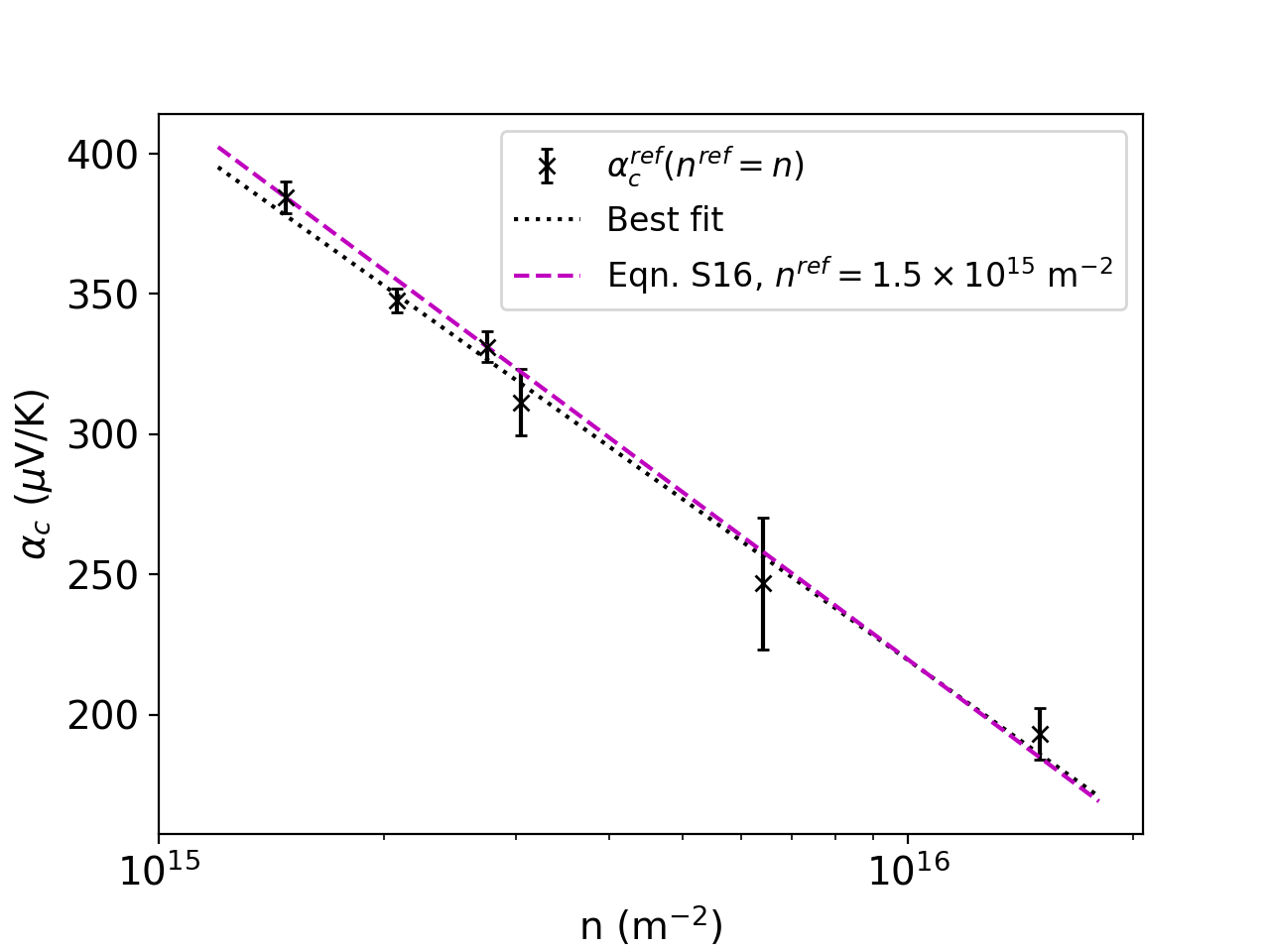}
  \caption{Chemical potential contribution to the Seebeck coefficient, $\alpha_c$, as a function of carrier concentration, $n$. $\alpha_c^\text{ref}$ calculated for the concentrations shown on the x-axis, i.e. $n^{\text{ref}}\!=\!n$, are shown as black crosses (values taken from Table S8), with the best linear fit of $\alpha_c^\text{ref}$ to $\ln(n)$ indicated with the dotted black line. The analytic dependence of $\alpha_c$ on concentration (Eq.~\ref{eq:alpha_c_si}) 
  is shown in dashed magenta, where a single value for $\alpha_c^\text{ref}$ at $n^\text{ref} = 2.74\times 10^{15}$\,m$^{-2}$ was used. See text for details. 
  The explicitly calculated data points (black crosses) align extremely well with the analytic expression given by equation \ref{eq:alpha_c_si} across the entire concentration range. The slopes $\partial \alpha_c/\partial \ln(n)$ for the best fit line and equation \ref{eq:alpha_c_si} are $-82.9$ $\mu$V/K and $-\frac{k_B}{q} = -86.2$ $\mu$V/K, respectively, validating the concentration dependence in equation~\ref{eq:alpha_c_si} and showing that $\alpha_c(T, n)$ does not depend on the chosen reference density~$n^\text{ref}$.
  }
  \label{fig:alpha_c_vs_n}
\end{figure}

Figure \ref{fig:alpha_c_vs_n} plots the calculated values of $\alpha_c (T\!=\!300 \text{ K}, n)$, equation \ref{eq:alpha_c_si}, for the different definitions of $n^\text{ref}$ listed in Table~\ref{table:chemical_pot}, at the 
concentration  $n^\text{ref}$, i.e., $\alpha_c (T\!=\!300 \text{K}, n\!=\!n^\text{ref}) \!=\!  -\frac{1}{q}\Bigg[ \frac{\partial \mu_c^\text{ref}}{\partial T} + k_B \ln\frac{n^\text{ref}}{n^\text{ref}}\Bigg]
\!=\!-\frac{1}{q}\Bigg[ \frac{\partial \mu_c^\text{ref}}{\partial T} \Bigg]\!=\! \alpha_c^\text{ref}$ (black crosses). The best linear fit of $\alpha_c^\text{ref}$ to $\ln(n)$ is indicated with the dotted back line.
The magenta dashed line shows the carrier concentration dependence of $\alpha_c$ from equation \ref{eq:alpha_c_si} when a value of $n^\text{ref} = 2.74\times 10^{15}$\,m$^{-2}$ is chosen 
(i.e. a simulation box of $50\times7\times1$ unit cells).
The explicitly calculated data points (black crosses) coincide very well with the analytic expression equation \ref{eq:alpha_c_si} (within a single error bar). The slope of the best fit line (dotted black) is $\partial \alpha_c/\partial \ln(n)=-82.9$ $\mu$V/K, in very close agreement with the expected slope from equation \ref{eq:alpha_c_si}, $-\frac{k_B}{q} = -86.2$ $\mu$V/K, implying that the thermodynamic theory and simulation provide a fully consistent description of the concentration dependence of~$\alpha_c$ which does not depend on the chosen reference density $n^\text{ref}$.

\begin{table}[H]
  \caption{Calculated values for the chemical potential $\mu_c^\text{ref}$ at different values of $n^\text{ref}$ corresponding to different active region sizes for 
  temperatures $T = 275, 300, 325$ K and the resulting value for $\alpha_c^\text{ref}$. }
  \label{table:chemical_pot}
  \begin{tabular}{llllllll}
      \hline
    $n^\text{ref}$ & Active Region & $\mu_c^\text{ref}|_{275 \text{K}}$ & $\mu_c^\text{ref}|_{300 \text{K}}$  & $\mu_c^\text{ref}|_{325 \text{K}}$ & $\alpha_c^\text{ref}$\\

     (m$^{-2}$) & (unit cells) & (meV) & (meV) & (meV) & ($\mu\text{V}/\text{K}$)\\
    \hline
    $1.5\times 10^{15}$ & $50\times13$ & -315.1 & -324.5 & -334.3 & 384.5 $\pm$ 5.5 \\
    $2.1\times 10^{15}$ & $42\times11$ & -306.5 & -315.1 & -323.9 & 347.7 $\pm$ 4.4 \\
    $2.7\times 10^{15}$ & $50\times7$ & -300.0 & -308.0 & -316.5 & 331.3 $\pm$ 5.6 \\
    $3.0\times 10^{15}$ & $35\times9$ & -297.3 & -304.6 & -312.9 & 311.3 $\pm$ 11.9 \\  
    $6.4\times 10^{15}$ & $25\times6$ & -279.1 & -284.2 & -291.4 & 246.8 $\pm$ 23.6 \\
    $1.5\times 10^{16}$ & $16\times4$ & -255.7 & -260.1 & -265.3 & 193.2 $\pm$ 9.1 \\
    \hline
  \end{tabular}
\end{table}

\clearpage

\clearpage

\section{Experimental Determination of Mobility and Seebeck Coefficient}
\label{sec:Experimental Determination of Mobility and Seebeck Coefficient}
\begin{figure}
    \centering
    \includegraphics[width=\textwidth]{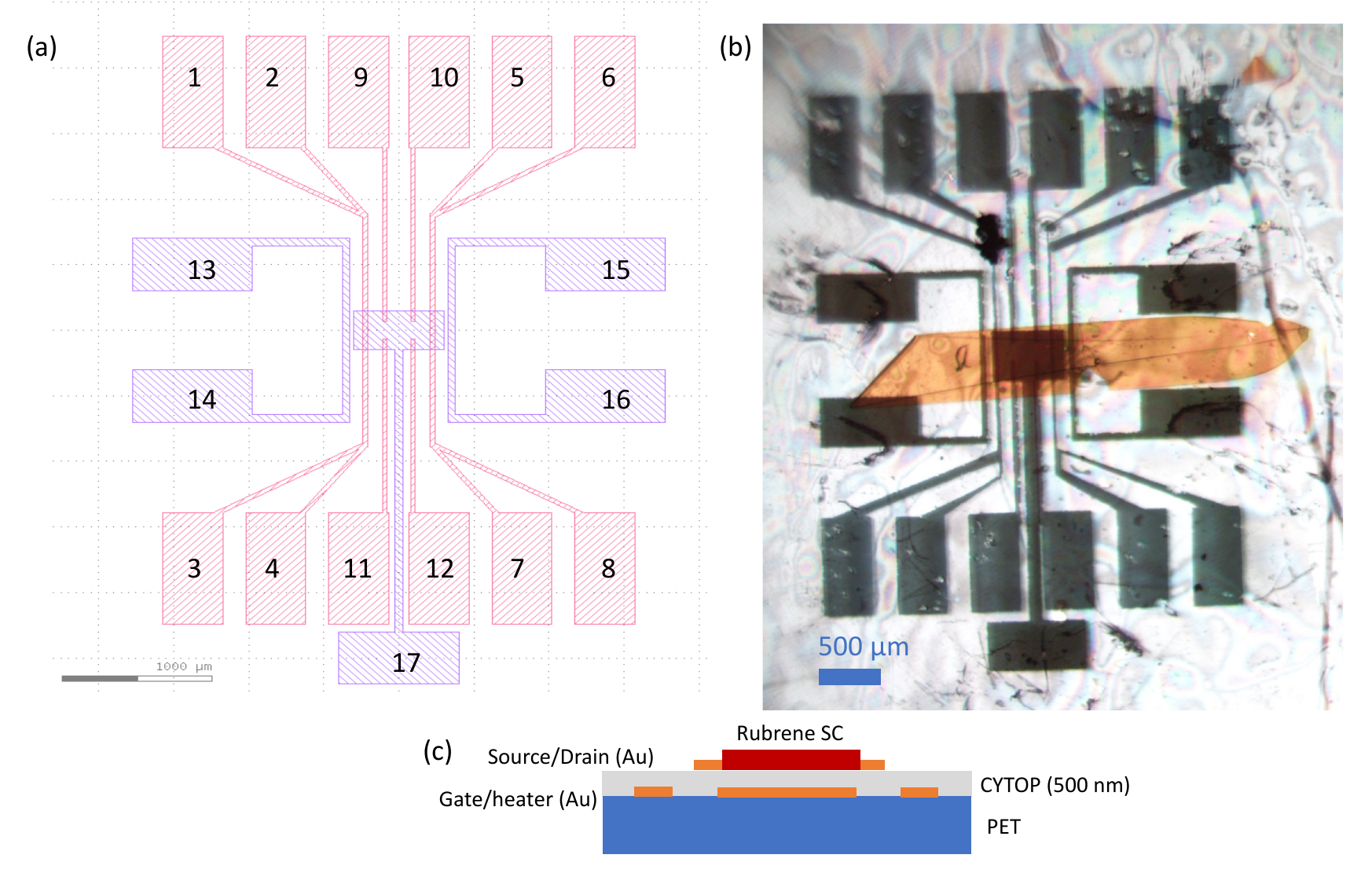}
    \caption{(a) Shadow mask pattern used for measuring mobility and Seebeck coefficient in rubrene single crystals. The electrodes in blue are evaporated directly on the PET, while those in red are evaporated on the CYTOP dielectric, as shown in panel (c). Contacts 1--4 connect to the source/hot thermometer; 5--8 to the drain/cold thermometer; 9--12 are used as the voltage probes for the 4-point probe mobility measurements; 13--16 allow either heater wire to be used; 17 is the gate electrode. (b) Micrograph of the finished device with rubrene single crystal. (c) Side view showing the layers of transistor architecture.}
    \label{fig:device}
\end{figure}

A schematic of the electrode pattern used, and an image of the actual device are given in Figure~\ref{fig:device}.

The mobility of the rubrene single crystals is determined via a 4-point probe method in FET geometry. The crystals show an increasing mobility with decreasing temperature, indicative of band-like transport, as well as a turn on voltage close to 0~V suggesting a very low deep trap density. 

\begin{figure}[!htb]
\centering
\includegraphics[width=\textwidth]{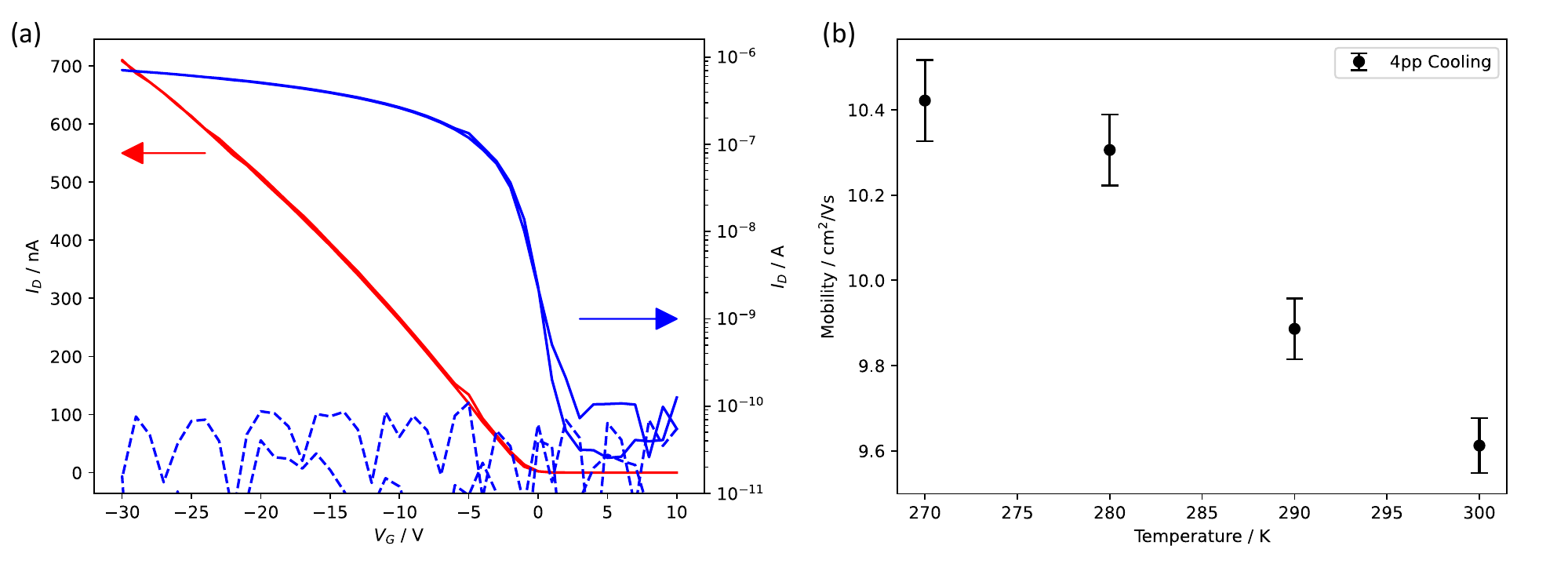}
\caption{Transfer properties of the rubrene single crystal (a) The measured transfer curve in both linear (red) and log (blue) current scale. Solid lines show the source-drain current while dashed line shows the source-gate current. Here $V_{SD} = -1$~V. (b) The experimental mobility vs. temperature curve measured \textit{via} 4-point probe method.}
\end{figure}

As we cool down the temperature from 300~K to 270~K, the threshold voltage shifts from near 0~V to -1~V indicating the existence of some trap states.

\begin{figure}[!htb]
\centering
\includegraphics[width=\textwidth]
{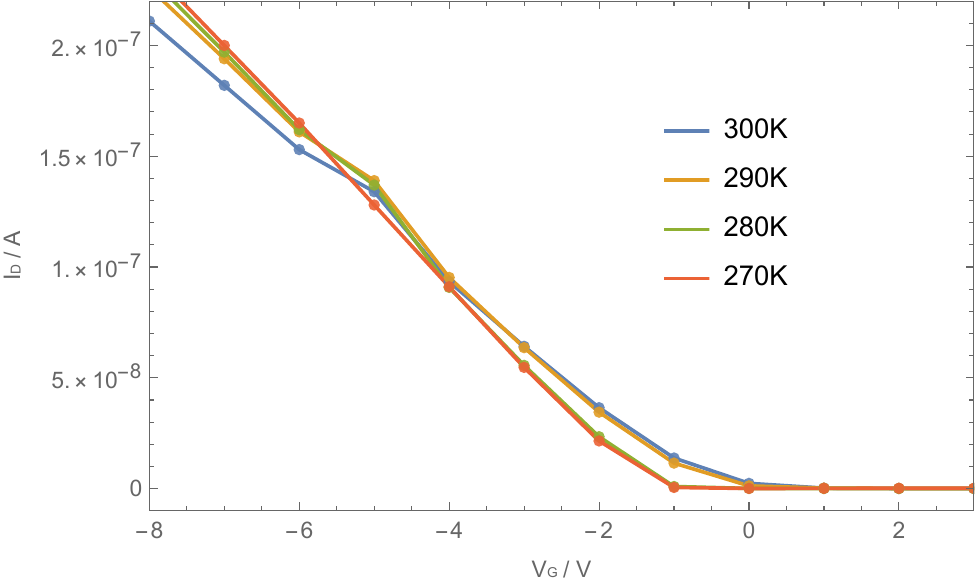}
\caption{Linear regime transfer characteristics around the turn on voltage at temperatures from 300~K to 270~K.} 
\end{figure}

In order to calculate the Seebeck coefficient of a material, the generated voltage across a known thermal gradient is measured. As described in previous works \cite{staz2020seebeck}, in this study, the thermal gradient is created via resistive heating by on-device heater electrodes, with applied heater power $W$,

\begin{equation}
    \alpha = \frac{\Delta V}{\Delta T} = \frac{\frac{\Delta V}{\Delta W}}{\frac{\Delta T}{\Delta W}}.
\end{equation}

To find the absolute temperature difference between the two ends, the resistance of the source and drain electrodes are measured by a 4-point probe method. We apply a current across one pair of electrodes (e.g. 1 and 3) and measure the voltage across the other pair (2 and 4), which negates any contact resistance from the probes or resistance from cabling. Because the source and drain are gold -- which has a linear resistivity vs. temperature curve -- the applied temperature at the hot and cold ends can be found by measuring the resistance ($R$) as a function of applied heater power. Finally, each thermometer needs calibrating by measuring its resistance at varying cryostat base temperatures $T$.

In this way:
\begin{equation}
    \frac{\Delta T}{\Delta W} = \frac{dT}{dW}_{hot} - \frac{dT}{dW}_{cold},
\end{equation}
and 
\begin{equation}
    \frac{dT}{dW}_{hot, cold} = \frac{\frac{dR}{dW}_{hot, cold}}{\frac{dR}{dT}_{hot, cold}}.
\end{equation}

\begin{figure}[!htb]
\centering
\includegraphics[width=\textwidth]{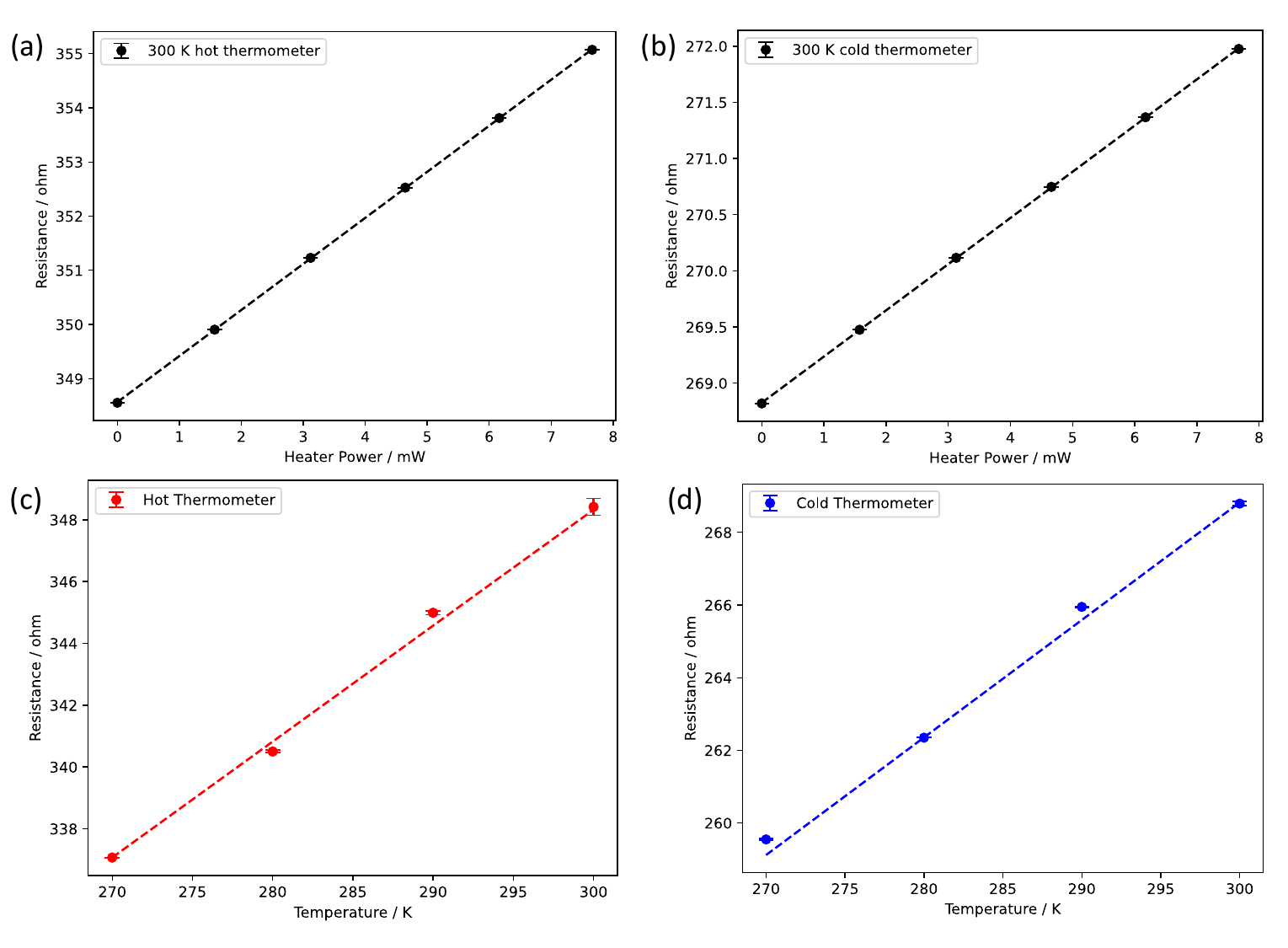}
\caption{Hot and cold sensor calibrations for determining the temperature difference across the device. (a) and (b) show the $R$ vs. $W$ curves at 300~K, while (c) and (d) show the sensor wire resistances as a function of cryostat temperature.}
\end{figure}

\begin{figure}[!htb]
\centering
\includegraphics[width=\textwidth]{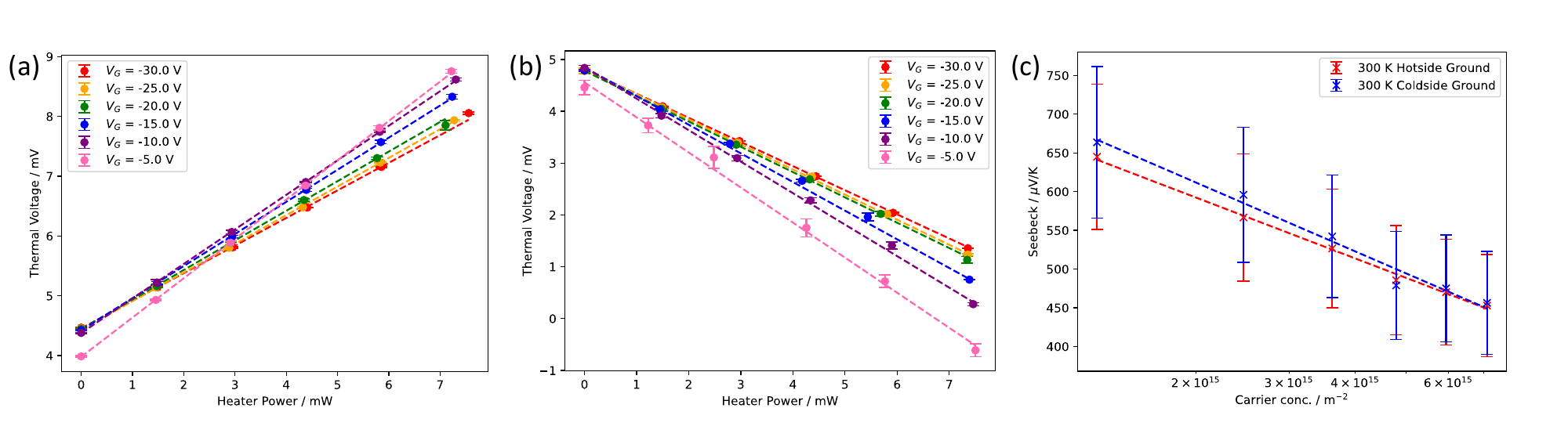}
\caption{Thermal voltage vs. heater power for the rubrene crystal with both the hot side grounded (a) and the cold side grounded (b). As expected for devices with no leakage both give the same magnitude of gradient except inverted, and so the extracted Seebeck coefficient is the same (c). Note that the voltage offset in (a) and (b) is due to the pre-amplifier used with the measuring SMU.}
\end{figure}

\clearpage

\end{document}